\documentclass[%
prx,twocolumn,showpacs,superscriptaddress,nobibnotes,
 amsmath,amssymb,nofootinbib,
 aps,
]{revtex4-2}

\usepackage{graphicx}
\usepackage{dcolumn}
\usepackage{bm}
\usepackage[hidelinks]{hyperref}

\newcommand{\nocontentsline}[3]{}
\newcommand{\toclesslab}[3]{\vspace{0.1in}\bgroup\let\addcontentsline=\nocontentsline#1{#2\label{#3}}\egroup\vspace{-0.1in}}

\usepackage{xcolor}
\usepackage[sectionbib]{bibunits}

\begin{document}

\title{Dynamics, scaling behavior, and control of nuclear wrinkling}



\author{Jonathan A. Jackson}
\altaffiliation[These authors contributed equally]{}
\affiliation{
Department of Biology, Massachusetts Institute of Technology
}
\affiliation{Graduate Program in Biophysics, Harvard University}
 
\author{Nicolas Romeo}%
\altaffiliation[These authors contributed equally]{}
\affiliation{%
Department of Mathematics, Massachusetts Institute of Technology
}%
\affiliation{%
Department of Physics, Massachusetts Institute of Technology
}%
\author{Alexander Mietke}
\affiliation{%
Department of Mathematics, Massachusetts Institute of Technology
}%
\affiliation{%
School of Mathematics, University of Bristol
}%
\author{Keaton J. Burns}
\author{Jan~F.~Totz}
\affiliation{%
Department of Mathematics, Massachusetts Institute of Technology
}%
\author{Adam C. Martin}
\affiliation{
Department of Biology, Massachusetts Institute of Technology
}%
\author{J\"orn Dunkel}
\email{dunkel@mit.edu}
\affiliation{%
Department of Mathematics, Massachusetts Institute of Technology
}%

\author{Jasmin Imran Alsous}
\email{jalsous@flatironinstitute.org}
\affiliation{%
 Center for Computational Biology, Flatiron Institute, Simons Foundation
}%

\date{\today}

\begin{abstract}
The cell nucleus is enveloped by a complex membrane, whose
wrinkling has been implicated in disease and cellular aging. 
The biophysical dynamics and spectral evolution of nuclear wrinkling during multicellular development remain poorly understood due to a lack of direct quantitative measurements. Here, we combine live-imaging experiments, theory, and simulations to characterize the onset and dynamics of nuclear wrinkling during egg development in the fruit fly, \emph{Drosophila melanogaster}, when nurse cell nuclei increase in size and display stereotypical wrinkling behavior. A spectral analysis of three-dimensional high-resolution data from several hundred nuclei reveals a robust asymptotic power-law scaling of angular fluctuations consistent with renormalization and scaling predictions from a nonlinear elastic shell model. We further demonstrate that nuclear wrinkling can be reversed through osmotic shock and suppressed by microtubule disruption, providing tunable physical and biological control parameters for probing mechanical properties of the nuclear envelope. Our findings advance the biophysical understanding of nuclear membrane fluctuations during early multicellular development.
\end{abstract}

\maketitle

Wrinkling and flickering of flexible sheet-like structures essentially determine mechanics and transport in a wide range of physical and biological systems, from graphene \cite{blees_graphene_2015,los_mechanics_2017} and DNA origami \cite{yoo_situ_2013} to nuclear envelopes (NEs) \cite{kalukula_mechanics_2022,lomakin_nucleus_2020,almonacid_active_2019,biedzinski_microtubules_2020} and cell membranes \cite{brochard_frequency_1975,betz_atp-dependent_2009}. Over the last decade, much progress has been made through experimental and theoretical work in understanding the effects of environmental fluctuations on the bending behaviors of carbon-based monolayers \cite{bowick_non-hookean_2017} and the shape deformations of lipid bilayer membranes of vesicles \cite{kantsler_vesicle_2007,kokot_spontaneous_2022,honerkamp_membrane_2013} and cells \cite{ben-isaac_effective_2011, turlier_equilibrium_2016}. In contrast, the emergence and dynamical evolution of surface deformations in NEs~\cite{chu_origin_2017,almonacid_active_2019,biedzinski_microtubules_2020} at different length- and timescales, to which we refer throughout this paper simply as `wrinkling',
 still pose fundamental open questions, as performing three-dimensional (3D) observations at high spatio-temporal resolution remains challenging 
under natural physiological and developmental growth conditions. Specifically, it is unclear
how NE wrinkle formation proceeds during cellular development,
which biophysical processes govern wrinkle
morphology, and whether there exist characteristic scaling
laws for NE surface 
fluctuations \cite{lomakin_nucleus_2020, venturini_nucleus_2020, almonacid_active_2019}. Addressing these questions through quantitative measurements promises insights into the physics of complex membranes and can clarify the biological and biomedical implications of NE deformations that have been linked to gene expression~\cite{almonacid_active_2019}, cellular aging \cite{scaffidi_lamin_2006}, and diseases like progeria syndrome~\cite{mounkes_progeroid_2003,kalukula_mechanics_2022}. 

\begin{figure*}
\includegraphics{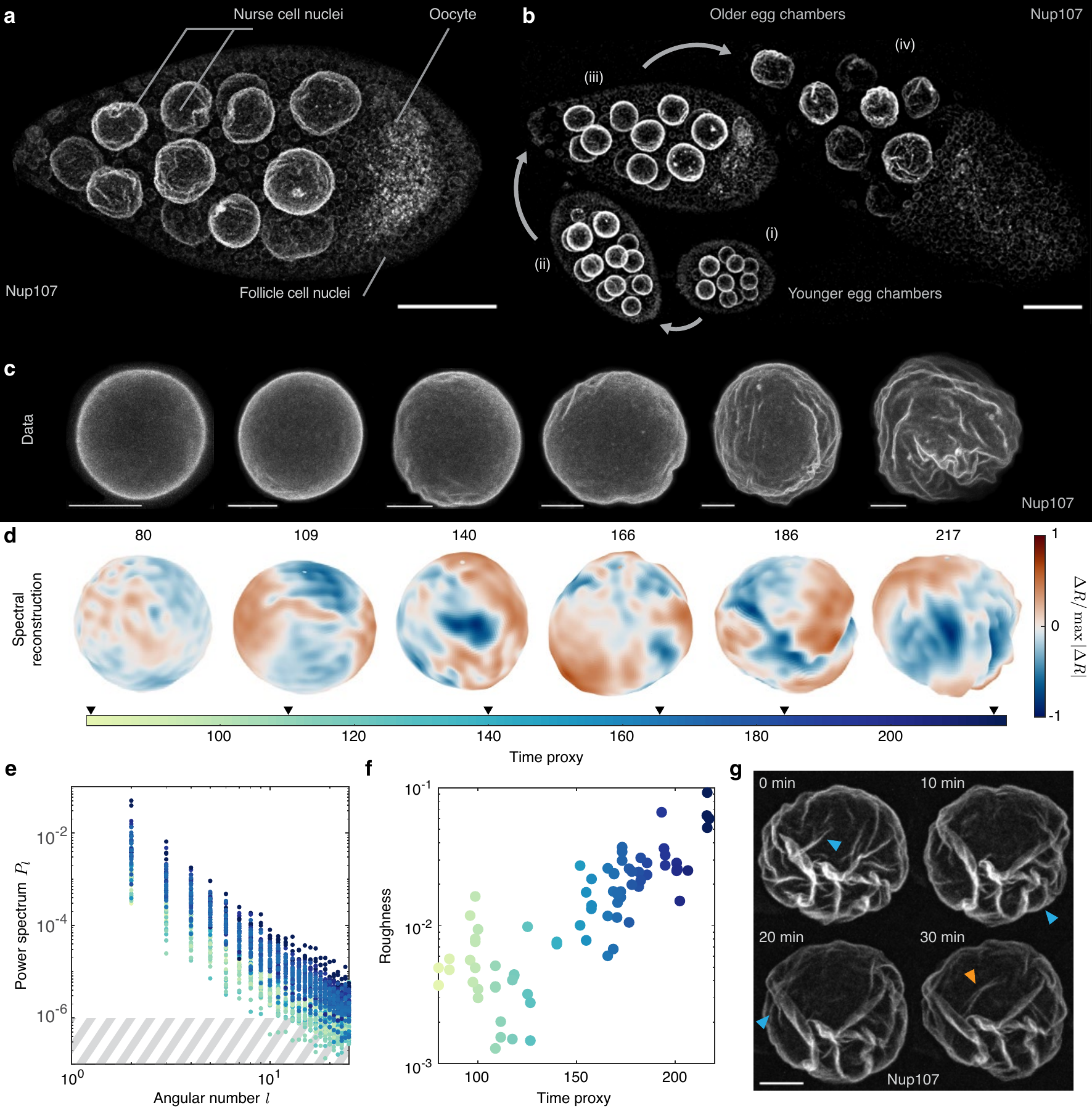}
\caption{\label{fig:fig1}
\textbf{Dynamic wrinkling of nurse cell nuclear envelopes during \emph{Drosophila} egg development.} \textbf{a,} Maximum-intensity projection (MIP) of a 3D image of an egg chamber expressing GFP-labeled Nup107, a component of the nuclear pore complex. The wrinkled nuclei of the 15 nurse cells are substantially larger than those of the surrounding follicle cells. 
\textbf{b,} MIP of four egg chambers showing an increase in nurse cell nuclear size and nuclear surface deformation as egg chambers age. Curved arrows indicate developmental progression from youngest (i) to oldest (iv). 
\textbf{c,} MIPs of individual nurse cell nuclei from six egg chambers spanning all ages included in our dataset, showing an increase in nuclear radius and NE wrinkling with age. Note that scale bar is the same size for each image; oldest nucleus shown is about 2.3 times the diameter of youngest shown. 
\textbf{d,} Spectral reconstruction of NE surfaces shown in \textbf{c} from 3D microscopy data using spherical harmonics with an angular number up to $l_\text{max} =25$ (Eq.~(\ref{eq:Ylm}), Methods). Time proxy values for each nucleus are included above the reconstructions.
\textbf{e,} Power spectra normalized by average radius for $N=78$ nuclei from 39 egg chambers in nurse cells directly connected to the oocyte (results are qualitatively similar for nuclei farther away from the oocyte; Supp. Fig.~\ref{fig:sfig_layers}). Hashed area indicates approximate noise threshold for young nuclei; color indicates the time proxy (corresponding to the color bar in \textbf{d}) as defined in the text and detailed in SI~Sec.~\ref{sec:si_devcoord}. 
\textbf{f,} NE roughness $\mathcal{R}=\sum_{l\ge 3}(2l+1)P_l$ for the same nuclei as in \textbf{e} increases exponentially with time proxy (see also Supp. Fig.~\ref{fig:sfig_layers}).
\textbf{g,} Snapshots of the same nucleus at four different time points illustrate that NE wrinkling is a dynamic process (Supplementary Video 2). Blue and orange arrowheads point to wrinkles that disappear and appear, respectively, between subsequent frames.
Scale bars: 50\,$\mu$m (\textbf{a, b}), 10\,$\mu$m~(\textbf{c, g}).
}
\end{figure*}
\par
Here, we combine 3D confocal microscopy, theoretical analysis, and simulations to characterize the wrinkling morphology and dynamics of nuclear surfaces in fruit fly egg chambers. A spectral analysis of over 300 nuclei provides evidence for an asymptotic power-law scaling of the surface fluctuations, consistent with predictions from renormalization calculations \cite{nelson_statistical_1989,kosmrlj_statistical_2017} and scaling arguments based on a nonlinear elasticity model for thin shells. Although the scaling is found to be highly robust against physical and biological perturbations, its magnitude (prefactor) can be tuned via osmotic pressure variation and microtubule disruption. These two different control mechanisms enable the tuning and probing of the NE's spectral and mechanical properties, and provide biophysical strategies for suppressing and reversing nuclear wrinkling.
\par

The NE is a double membrane that separates the cell's nuclear interior from the surrounding cytoplasm. The two concentric $\sim$4\,nm-thick lipid bilayers are $\sim$20-50\,nm apart and are supported by the nuclear lamina, a non-contractile meshwork of intermediate filaments that lie adjacent to the inner nuclear membrane, conferring mechanical stability and affecting essential cellular processes through regulation of chromatin organization and gene expression \cite{aebi_nuclear_1986,lammerding_lamin_2004}. Among other proteins, the NE contains nuclear pore complexes, multi-protein channels that primarily regulate passage of macromolecules between the nucleus and the cytoplasm \cite{strambio-de-castillia_nuclear_2010,knockenhauer_nuclear_2016}. Recent \emph{in vitro} studies have provided key insights into the role of lamins, cytoplasmic structures, and the physical environment in affecting NE morphology, as well as evidence for the critical importance of nuclear shape for many cellular and nuclear functions \cite{venturini_nucleus_2020,kalukula_mechanics_2022}, including transcriptional dynamics \cite{almonacid_active_2019}. Despite notable progress, a quantitative understanding of how wrinkling phenomenology and 3D spectral properties of nuclear surfaces evolve in time and during cellular development has remained elusive. 

\par
To investigate the biophysical dynamics, scaling behaviors, and reversibility of nuclear wrinkling, we used the egg chamber of the fruit fly \emph{Drosophila melanogaster}, a powerful system amenable to 3D high-resolution live imaging and targeted biological and physical perturbations \cite{hudson_methods_2014}. The egg chamber contains 15 nurse cells and the oocyte (the immature egg cell), all connected via cytoplasmic bridges and enclosed by a thin layer of hundreds of follicle cells (Fig.~\ref{fig:fig1}a, with schematics in Supp. Fig.~\ref{fig:sfig_devcoord}a, \cite{king_oogenesis_1956}). For most of the $\sim$3 days of oogenesis, the nurse cells supply proteins, mRNAs, and organelles to the oocyte through diffusion and microtubule-mediated directed transport \cite{bratu_drosophila_2015,bastock_drosophila_2008,imran_alsous_dynamics_2021,mahajan-miklos_intercellular_1994}. 
To provide the prodigious amount of material and nutrients that the oocyte needs, each nurse cell replicates its DNA $\sim$10 times without undergoing cell division, thereby notably increasing its nuclear and cell sizes \cite{lin_germline_1993}. In the $\sim$30-hour window studied here, the diameter of nurse cell nuclei in the cells directly connected to the oocyte increases from approximately 16 to about 40 micrometers~\cite{tzur_cell_2009,lin_germline_1993}, accompanied by the progressive appearance of fold-like deformations in the NE, providing an ideal test bed for studying the onset and evolution of NE wrinkling (Fig.~\ref{fig:fig1}b,c).
\par

To compare nurse cell nuclei within the same egg chamber and across different egg chambers, we defined a proxy measurement for developmental time (referred to here as the `time proxy') based on the geometric average of the egg chamber's length and width (Methods, SI~Sec.~\ref{sec:si_devcoord}, Supp. Fig.~\ref{fig:sfig_devcoord}b,c). Since egg chamber geometry correlates closely with developmental progression, adopting this continuous geometric characterization offers finer temporal resolution than the traditional approach of distinguishing 14 discrete morphological stages \cite{bratu_drosophila_2015,bastock_drosophila_2008} (for a comparison between the time proxy and developmental stage, see Supp. Fig.~\ref{fig:sfig_devcoord}c). By time-ordering nuclei according to this metric, we could more accurately determine the time of emergence of nuclear wrinkling and reconstruct its evolution (Fig.~\ref{fig:fig1}b,c). To track the NEs of the nurse cells in space and time, we used a fluorescently-tagged version of the nuclear pore complex protein Nup107 that delineates the nucleus (Supplementary Video 1; qualitatively similar wrinkling patterns were observed using a different labeled protein in the NE and via label-free imaging, see Supp. Fig.~\ref{fig:sfig_otherlabels}); note that this label allows observation only of deformations that include both membranes of the nuclear envelope, but is unlikely to label deformations that include only the inner membrane, such as Type I nucleoplasmic reticula \cite{Malhas2014, fricker_interphase_1997}. Having acquired highly resolved 3D imaging data (Fig.~\ref{fig:fig1}c, Supp. Fig.~\ref{fig:sfig_atlas}), we reconstructed the nuclear surface radius $R(\theta, \phi)$ relative to the geometric center of the nucleus, where $\theta$ and $\phi$ are the spherical polar angles. 
\par
To obtain a compact 3D spectral representation of the nuclear surface deformations, we computed the real spherical harmonic coefficients $f_{lm}$, defined by
\begin{equation}
    R(\theta, \phi) = \sum_{l=0}^{l_\text{max}} \sum_{m=-l}^l f_{lm} Y_{lm}(\theta, \phi) \label{eq:Ylm},
\end{equation}
where $Y_{lm}$ is the spherical harmonic with angular number $l$ and order $m$ (Methods). Equation~\eqref{eq:Ylm} allows for a continuous reconstruction of the NEs (Fig \ref{fig:fig1}d, Supp. Fig.~\ref{fig:sfig_atlas}), with the mode-cutoff $l_\text{max}$ setting the angular resolution of the spectral representation (Methods). The coefficient values $\{f_{lm}\}$ depend on the choice of coordinate system, that is, the orientation of the nuclei. To obtain a rotation-invariant characterization of the surface wrinkles, we consider the power spectrum of radial out-of-plane deformations
\begin{align}
    P_l = \frac{4\pi}{(2l+1)f_{00}^2}\sum_{m=-l}^l f_{lm}^2 \label{eq:roughness}, 
\end{align} 
normalized by the average radius of the shell $\langle R \rangle = f_{00}/\sqrt{4\pi}$. The non-negative numbers $P_l$ measure the average power in a mode of angular wavenumber $l$. A single-valued summary statistic of surface wrinkling can be given in terms of the `roughness' parameter $\mathcal{R}=\sum_{l\ge 3}(2l+1)P_l$, the total power contained in angular numbers $l\ge 3$. By ignoring the long-wavelength modes $l<3$, $\mathcal{R}$ measures the contribution of finer-scale wrinkles to NE deformations. Our analysis of over 300 nurse cell nuclei shows that the power spectrum of NEs maintains an approximately constant shape as development progresses, but with a steadily-increasing amplitude (Fig.~\ref{fig:fig1}e; Supp. Fig.~\ref{fig:sfig_layers}), reflecting the fact that wrinkling becomes more pronounced as nuclei increase in size. $\mathcal{R}$ increases exponentially with the time proxy (Fig.~\ref{fig:fig1}f), suggesting that nurse cell nuclei transition smoothly from an unwrinkled to a wrinkled state.

\par
Nuclear surface wrinkling is a highly dynamic process~\cite{almonacid_active_2019}. By imaging individual nurse cells at $\sim$40\,s intervals, we too observed that NE surface shapes fluctuate substantially, with smaller features appearing and disappearing faster than larger ones (Fig. \ref{fig:fig1}g, Supplementary Video 2). Specifically, power spectra $P_l$ of repeatedly imaged nuclei changed on timescales of minutes or faster~(Supp. Fig.~\ref{fig:sfig_layers} and Supp. Fig.~\ref{fig:sfig_fluct}). The rotational invariance of spectra implies that these fluctuations are not the result of whole body rotations, but instead reflect a rapid shape dynamics of NE surfaces. Experimental limitations prevented quantification of timescales for the entire 3D surface, but our observations are qualitatively consistent with findings that smaller wrinkles typically decay faster \cite{zilman_undulations_1996,turlier_equilibrium_2016}. Furthermore, the fact that the deformation spectrum is monotonically decreasing (Fig.~\ref{fig:fig1}e) implies that there is no preferred wavelength, suggesting that the observed NE shapes do not correspond to fluctuations about the steady-states of buckled shells, but instead reflect dynamic wrinkling across all experimentally resolved angular scales. 

\begin{figure*}
\includegraphics{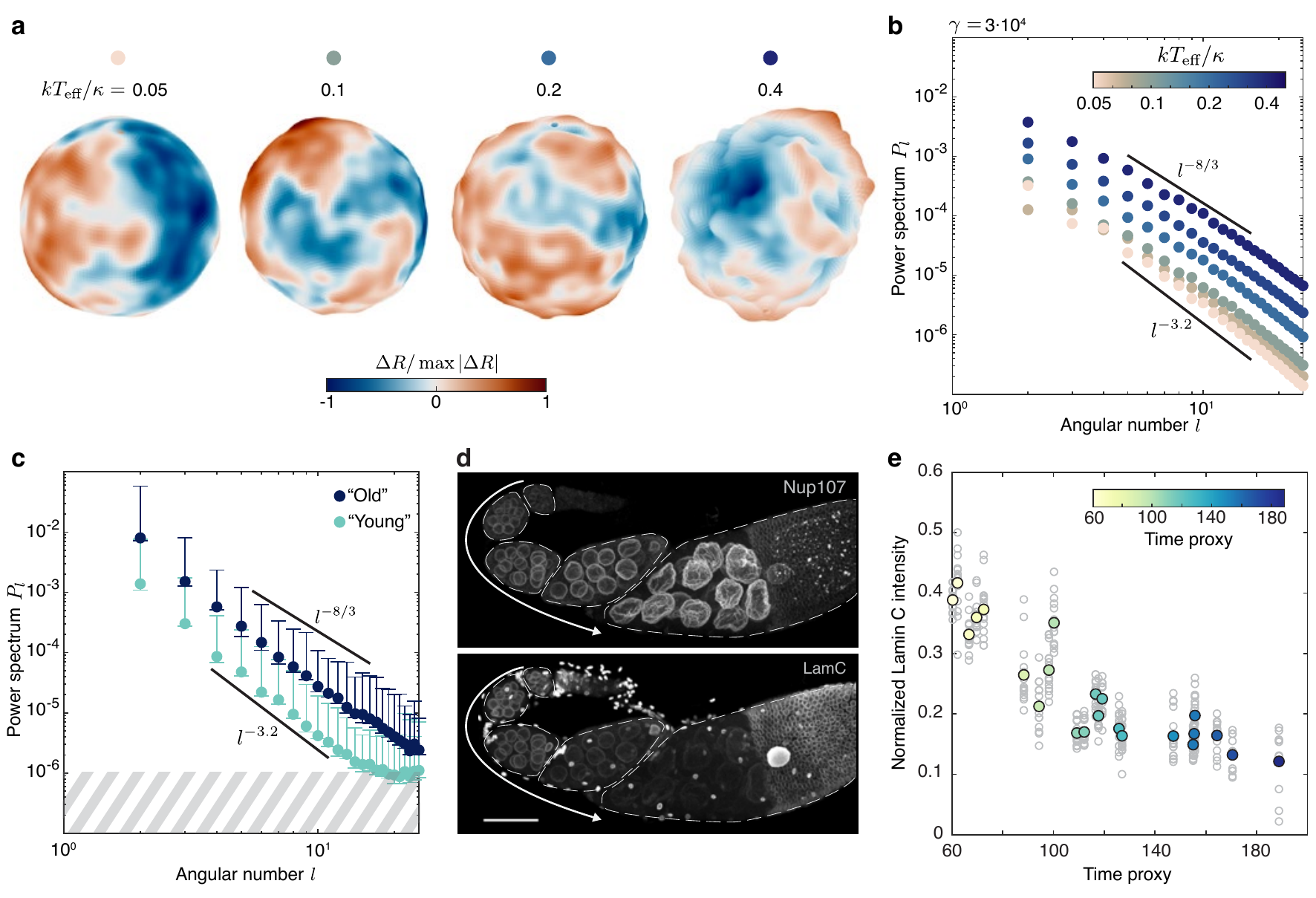}
\caption{\label{fig:fig2} 
\textbf{Fluctuating elastic shell theory predicts a scaling law with exponent $\approx 3$ for the wrinkle power spectrum, in agreement with experiments}. 
\textbf{a,}~Equilibrium simulation snapshots of nuclei at temperature $T_\mathrm{eff} = 10 T_{eq}$, undeformed radius $R = 25$ $\mu$m and $R_c /R = 20$, at fixed FvK number $\gamma =3\times10^4$ for varying elastic moduli controlled by $kT_\text{eff}/\kappa$. Color indicates the normalized deviation of the surface from the mean shell radius.
\textbf{b,}~Time-averaged spectra of simulated NEs of undeformed radius $R=25$ $\mu$m, $R_c/R = 20$, $T_\mathrm{eff} = 10 T_\text{eq}$ for different moduli $\kappa, Y$ at fixed $\gamma = 3\times10^4$, showing the transition from weak nonlinearity to strong nonlinearity as bending rigidity decreases. Color bar matches the dots from \textbf{a}.
\textbf{c,}~Binned averages of spectra from nuclei in nurse cells directly connected to the oocyte reveal that shape fluctuations follow a scaling law with an exponent between $-3.2$ and $-8/3$ that is obeyed throughout development. `Young' nuclei have a time proxy between $80-140$, $N=29$ nuclei, from 12 egg chambers; `Old' nuclei have a time proxy between $160-220$, $N=40$ nuclei, from 22 egg chambers. Bars show extremal values. Hashed area indicates approximate noise threshold for young nuclei. (See Supp. Fig.~\ref{fig:sfig_layers} for comparison between nuclei at different positions in the egg chamber)
\textbf{d,}~Fixed egg chambers expressing Nup107::RFP and stained for Lamin~C, showing a decrease in Lamin~C intensity in nurse cell nuclei as egg chambers increase in age. In contrast, Nup107::RFP intensity stays relatively constant. The same trend is observed in live imaging of \textit{ex vivo} egg chambers expressing LamC::GFP and Nup107::RFP (Supp. Fig.~\ref{fig:sfig_lamC}a). Wrinkling of nuclei in younger egg chambers (all but the rightmost) is a result of fixation and is not observed in live imaging until later stages. Arrows indicate increasing age; egg chamber boundaries are shown in dashed outlines. Scale bar: 50 $\mu$m. 
\textbf{e,}~Normalized Lamin~C fluorescence intensity decreases by approximately 5-fold over time. Normalization details are specified in the Methods. $N= 337$ nuclei from 23 egg chambers; colored dots show means for each egg chamber.
}
\end{figure*}

Both maximum-intensity projections and spectral reconstructions show that NE wrinkles and creases are sharp, with narrow bent regions separated by flatter areas (Fig.~\ref{fig:fig1}). This morphology is reminiscent of the nonlinear stress-focusing characteristic of crumpled elastic sheets and shells such as ordinary paper sheets, which are much more easily bent than stretched \cite{andrejevic_model_2021,blees_graphene_2015, witten_stress_2007}. In particular, these geometric nonlinearities lead to anisotropic responses when point forces are applied to the shell \cite{witten_stress_2007}. To rationalize the experimentally observed wrinkle morphology at spatial scales larger than the NE thickness, we constructed a minimal effective elastic model, describing the NE as a deformed spherical shell (equilibrium radius $R$). In spherical coordinates $\mathbf{r} = (\theta, \phi)$, the shell has an isotropic elastic free energy \cite{paulose_fluctuating_2012,kosmrlj_statistical_2017}
\begin{equation}
    F_\mathrm{shell} = \int \mathrm{d}^2\mathbf{r} \left[ \frac{\kappa}{2} (\nabla^2 f)^2 + \frac{\lambda}{2} \epsilon_{ii}^2 + \mu \epsilon_{ij}^2 \right],\label{eq:free_energy}
\end{equation} where $i,j\in\{\theta, \phi\}$ and using the Einstein summation convention. The energy functional~\eqref{eq:free_energy}, accounts for bending stiffness through a Helfrich-like bending term that penalizes out-of-plane deformation $f$ (positive when pointing inwards), and the stretching of the membrane through the nonlinear strain tensor $\epsilon_{ij}$. The 2D Lam\'e parameters $\lambda, \mu$ are proportional to the 2D Young's modulus~$Y$. The strain combines contributions from $f$ and from the in-plane deformation $\mathbf{u}(\mathbf{r})$ (SI~Sec.~\ref{sec:si_theory}). We also allow for a preferred radius of curvature $R_c$ of the shell mismatched with the radius $R$ of the shell $R_c \geq R$, which in the large-F\"oppl-von K\'arm\'an (FvK) regime leads to a strain tensor $\epsilon_{ij} = \frac{1}{2}\left( \partial_i u_j + \partial_j u_i + \partial_i f \partial_j f \right) - \delta_{ij} f/ R_c$ (SI~Sec.~\ref{sec:si_theory}). 
Previous work \cite{guilak_viscoelastic_2000,funkhouser_mechanical_2013,kim_volume_2015} has shown the NE to be stiffer than most biological membranes and to be well described as a thin membrane of a 3D isotropic elastic material with an effective 3D Young's modulus $E\approx1$\,kPa and thickness of $h\sim 10-100$\,nm (for a more detailed discussion of limitations of fluid membrane models, see SI~Sec.~\ref{sec:si_fluid}), leading to a bending rigidity of $\kappa = 100-300\; kT_\text{eq} \approx 10^{-18}$\,J, where $T_\text{eq}$ is the room temperature, and a stretching rigidity, captured by the 2D Young's modulus, of $Y \approx 10^{-4}\,$N/m \cite{dahl_nuclear_2004}. By construction, these moduli are approximately related through the effective thickness $h \sim \sqrt{\kappa/Y}$ \cite{kosmrlj_statistical_2017}. Note that $Y$ is a factor of $10^3$ smaller than the stretching rigidity of a lipid bilayer, potentially explained by the presence of `area reservoirs' in NEs and by transmembrane protein conformational changes~\cite{enyedi_nuclear_2017}. For a shell of radius~$R$, one can define the FvK number $\gamma = YR^2 / \kappa $ which describes the relative propensity of the material to bend rather than to stretch. Using the above values, we find that the NE has a large FvK number $\gamma \sim 10^4-10^6$, comparable to that of a sheet of paper or graphene \cite{blees_graphene_2015}. Accordingly, the NE is more amenable to bending than to stretching, and deformations are expected to appear as sharp wrinkles and creases, in agreement with our observations (Fig. \ref{fig:fig1}).

\par

To compare the surface shapes and fluctuation predicted by Eq.~\eqref{eq:free_energy} with our experimental data, we simulated the equilibrium Langevin PDE derived from this free energy (see Methods and SI~Sec.~\ref{sec:si_simulations} for simulation details). The simulations account for hydrodynamic coupling and both passive and active fluctuations, which are modeled by an effective temperature $kT_\mathrm{eff}$. Despite the model's minimal character and theoretical limitations of Eq.~\eqref{eq:free_energy} at long wavelengths where $l\rightarrow0$ (SI Sec.~\ref{sec:si_theory}), the numerically obtained shapes (Fig.~\ref{fig:fig2}a) are qualitatively similar to those in the experiments (Fig.~\ref{fig:fig1}d). In the experimentally accessible range of low-to-intermediate angular wave numbers $3\lesssim l\lesssim 11$, the angular spectra extracted from the simulations at different ratios of $kT_\mathrm{eff}/\kappa \in[0.05,0.5]$ (Fig.~\ref{fig:fig2}b) and experimental data (Fig.~\ref{fig:fig1}e) also show an approximately similar decay, suggesting that the minimal elastic shell model in Eq.~\eqref{eq:free_energy} captures relevant features of the NE, providing a basis for further analysis and predictions.
\par

A main feature of the experimentally measured spectra is that both younger and older nuclei exhibit a similar asymptotic power law decay in the limit of small angular numbers $l\le 10$ (Fig.~\ref{fig:fig2}c). To rationalize this observation, we first note that the scaling behavior in our experiments deviates from the basic linear response theory predictions, which is expected because, even for younger nuclei, the radial fluctuations $f$ typically exceed the NE thickness $h\sim 10^{-3} R$ (Fig.~\ref{fig:fig1}c-f). More precisely, for small fluctuations ($f\ll h\ll R$) and small thermodynamic pressure ($p\ll p_c = 4\sqrt{\kappa Y}/R_c^2$, where $p_c$ is the critical buckling pressure of the sphere), linear response theory predicts that the power spectrum $P_l$ exhibits a plateau for $l\le l_c$ and falls of as $l^{-4}$ for $l\gg l_c$ with a crossover value $l_c \approx \gamma^{1/4}\sqrt{R/R_c}$ (SI~Sec.~\ref{sec:si_theory})\cite{kosmrlj_statistical_2017,pecreaux_refined_2004}, which is not seen in our experiments (Figs.~\ref{fig:fig1}e and~\ref{fig:fig2}c). Indeed, classical shell theory~\cite{los_mechanics_2017} states that nonlinear effects become important when the out-of-plane deformations~$f$ become comparable to or exceed the shell thickness $h$, which is generally the case in our data where $h\ll f\ll R$ (Fig.~\ref{fig:fig1}c,d,g). Nonlinear analysis of elastic plates and shells has a long history \cite{landau_theory_2009,nelson_statistical_1989} and has seen major advances in the last decade~\cite{paulose_fluctuating_2012, kosmrlj_statistical_2017}, motivated in part by the discovery of graphene~\cite{novoselov_electric_2004}. As demonstrated above, the FvK number of the NE is comparable to that of graphene, so we can borrow and apply recent theoretical results to understand the fluctuation spectra of the NE. Specifically, a detailed renormalization group (RG) analysis~\cite{kosmrlj_statistical_2017,baumgarten_buckling_2018} of Eq.~\eqref{eq:free_energy} showed that, for sufficiently small plate fluctuations, elastic nonlinearities lead to a modified asymptotic decay of $P_l\propto l^{-3.2}$, consistent with our experimental and simulated data (Figs.~\ref{fig:fig1}e and~\ref{fig:fig2}b,c) and with previous experiments in red blood cell spectrin networks \cite{schmidt_existence_1993}. Notably, earlier studies~\cite{nelson_statistical_1989,kosmrlj_statistical_2017, paulose_fluctuating_2012} also predicted that the interplay of elastic nonlinearities and fluctuations can cause the spontaneous collapse of sufficiently large shells, suggesting a physical mechanism that could contribute to the eventual breakdown of the nurse cell NE when these cells donate their contents to the oocyte~\cite{yalonetskaya_nuclear_2020,imran_alsous_dynamics_2021}.
\par
The previously mentioned RG methods can give rise to divergences in large deformation regimes, where nonlinearities dominate the shell's response (SI~Sec.~IV, Fig.~\ref{fig:sfig_pressure}). To obtain an analytical prediction for the scaling in the larger-deformation regime $h\ll f\ll R< R_c$, relevant to older nuclei, we performed an asymptotic dimensional analysis that provides additional insight into how NE wrinkling can be controlled. To that end, we added to the elastic free energy $F_\mathrm{shell}$ an effective pressure term $F_p = - \int \mathrm{d}^2\mathbf{r}\,p_\text{eff}f$, where $p_\text{eff}$ accounts for a normal load, which may arise from osmotic pressure differences or microtubule-induced local stresses. Denoting by $L$ the characteristic surface variation length scale and omitting numerical prefactors that depend on details of the adopted thin-shell modeling approach (SI~Sec.~\ref{sec:si_theory}), one finds for shells of thickness $h \sim \sqrt{\kappa/Y}$ that the various free-energy components give scaling contributions of the form
\cite{kosmrlj_statistical_2017} 
\begin{equation}
\frac{\delta F}{Y} \sim 
\left(\frac{h}{L}\right)^2 \left(\frac{f}{L}\right)^2 + 
\left(\frac{f}{R_c}\right)^2 +
\frac{f}{R_c} \left(\frac{f}{L}\right)^2+ 
\left(\frac{f}{L}\right)^4 -\frac{p_\text{eff}f}{Y}. \label{eq:scaling_scheme}
\end{equation}
The first term corresponds to bending, and the second and third terms arise from the non-zero curvature of the undeformed shell. The fourth term describes the nonlinear response associated with changes in the Gaussian curvature of the shells. For well-developed wrinkles with $f\gg h$, the first term can be neglected as it is smaller than the fourth term. Considering wrinkle amplitudes $f_l$ at the spatial length scale $L\sim R/l$, where $l$ is the angular wave number, the remaining terms can be recast as
\begin{equation}
    \frac{\delta F_l}{Y} \sim \left(\frac{f_l}{R_c}\right)^2 + \frac{f_l}{R_c}\left(\frac{f_l}{R}\right)^2 l^2 + \left(\frac{f_l}{R}\right)^4 l^4 - \frac{p_\text{eff}}{Y}f_l.\label{eq:scaling_scheme_l}
\end{equation} 
Since $R_c > R$, the first two terms will be dominated by the $l^4$-term implying that, at steady-state, this quartic term and the pressure term must balance out, consistent with a corresponding earlier result for flat plates with $R_c = \infty$ \cite{los_mechanics_2017}. We thus find $f_l \sim (p_\text{eff}/Y)^{1/3} (R/l)^{4/3}$, and hence for the angular power spectrum $P_l \sim (f_l/R)^2$ [see Eq.~(\ref{eq:roughness})] the scaling law
\begin{equation}
P_l \sim \left(\frac{p_\text{eff}R}{Y}\right)^{2/3} l^{-8/3}. \label{eq:scaling}
\end{equation}
In this scaling regime, the surface deformation dynamics is dominated by the shell's resistance to stretching, which causes changes in its Gaussian curvature~\cite{during_strong_2019}. Both our experimental data (Fig. \ref{fig:fig2}c, Supp. Fig.~\ref{fig:sfig_layers}) and spherical shell simulations (SI, Supp. Fig.~\ref{fig:sfig_sims}) show an asymptotic spectral decay $P_l \propto l^{-\alpha}$ with an exponent $\alpha$ in the range $8/3< \alpha < 3.2$, predicted by this scaling analysis and renormalization group calculations.
\par
\begin{figure*}
\includegraphics{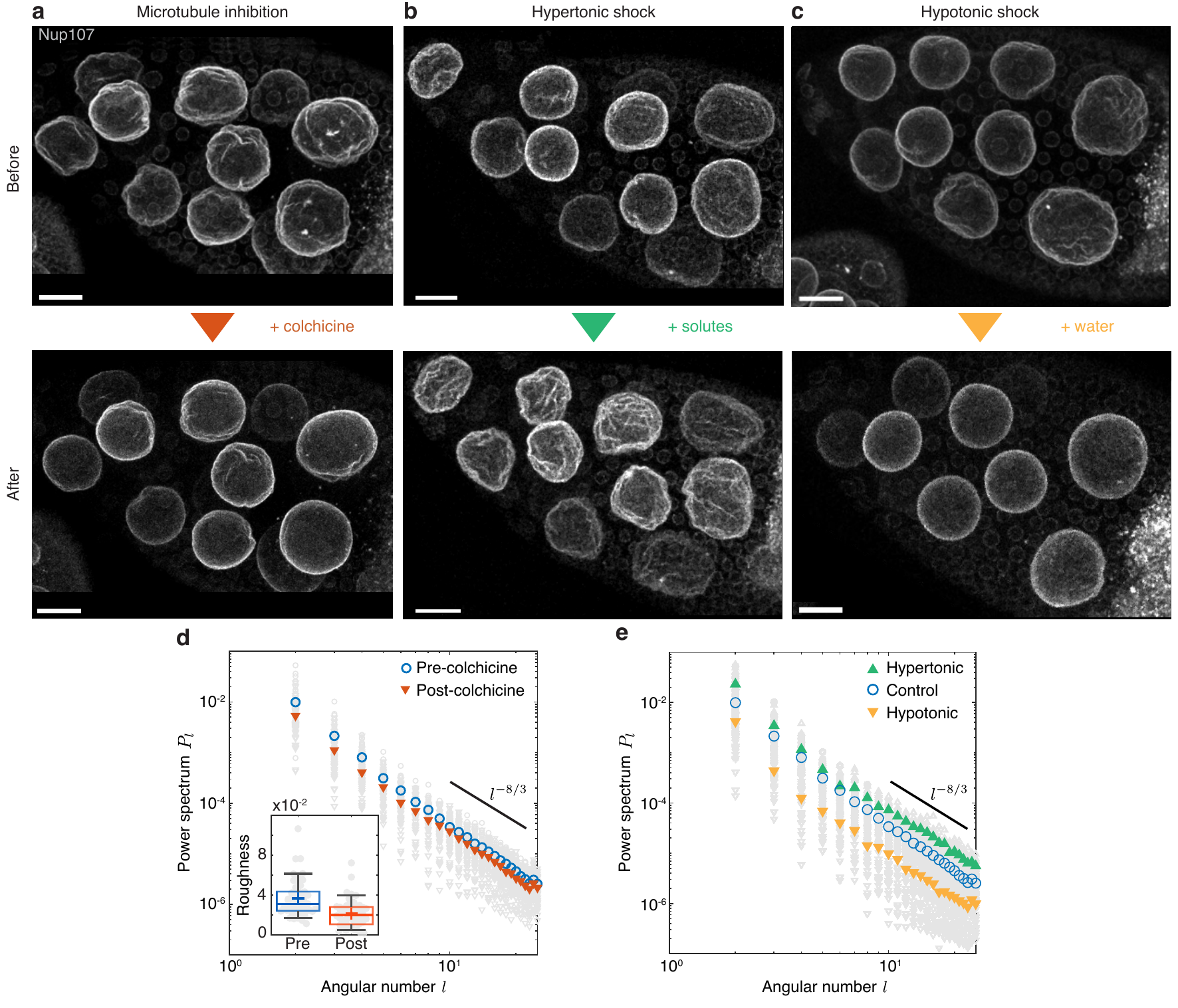}
\caption{\label{fig:fig3}
\textbf{Perturbation experiments confirm robustness of observed scaling laws and reveal NE wrinkling reversal mechanisms.} 
\textbf{a,} MIP of one egg chamber before (top) and after (bottom) inhibition of microtubule polymerization by colchicine, showing that microtubule disruption can reverse wrinkling (Supplementary Video 4). 
\textbf{b,} MIPs before and after hypertonic shock using an external culture medium of 1.5x osmolarity, showing an increase in wrinkling.
\textbf{c,} MIPs before and after hypotonic shock using an external culture medium of 0.5x osmolarity, showing a decrease in wrinkling.
Egg chambers in \textbf{a}, \textbf{b}, and \textbf{c} have time proxies of 171, 174, and 171, respectively. 
\textbf{d,} The power spectrum after microtubule inhibition by colchicine still follows a power law with roughly the same exponent, with a reduction of roughness by a factor of 2 (inset). $N=49$ pre-colchicine and post-colchicine nuclei, from 6 egg chambers, time proxies $169-182$. For box plots, plus signs denote mean, middle line is the median, top and bottom edges of the box are the upper and lower quartiles, and whiskers span from 9\% to 91\% of the data range. 
\textbf{e,} In the presence of increased inwards (hypertonic) or outwards (hypotonic) pressure, the overall shape of the power spectrum remains approximately conserved. Hypotonic shock treatment reduces the wrinkle amplitudes, providing a reversal mechanism for NE wrinkling. Spectra were computed using $49$ control, $15$ hypertonic, and $30$ hypotonic nuclei in the time proxy range $165-185$, from 6, 3, and 6 egg chambers, respectively, using nuclei from all nurse cells rather than only those directly connected to the oocyte. Supp. Fig.~\ref{fig:sfig_osmotic} shows roughness values over a larger range of time proxies. Scale bars: 20 $\mu$m.
}
\end{figure*}

Both Eq.~\eqref{eq:scaling} and the robustness of the experimentally observed scaling behavior in time (Fig. \ref{fig:fig2}c, Supp. Fig.~\ref{fig:sfig_layers}), and under different chemical and physical perturbations (Fig. \ref{fig:fig3}), suggest the emergence of NE wrinkling is primarily controlled by the material properties and the effective pressure $p_\text{eff}$ induced by thermal and by active fluctuations. For Gaussian fluctuations with effective temperature $T_\text{eff}$, previous theoretical work~\cite{paulose_fluctuating_2012,kosmrlj_statistical_2017} showed that $p_\text{eff} \sim p_c (kT_\text{eff}) \sqrt{\gamma} \sim (Y/R)(kT_\text{eff}/\kappa)$, with $p_c = 4 \sqrt{\kappa Y}/R^2$ the critical buckling pressure for a homogeneous spherical shell. If in addition to a fluctuating pressure $p'$, there are uniform loads, such as those caused by osmotic pressure differences, we find  $p_\text{eff} = p' (R/\langle R\rangle)^3$ accounts for excess area contributions to the amplitude (as long as the shell is not stretched taut; SI Sec.~\ref{sec:si_pressure}). Inserting these results into Eq.~\eqref{eq:scaling}, scaling analysis predicts that wrinkle formation can be tuned by changing the bending rigidity~$\kappa$, the activity $kT_\text{eff}$ and the cell's osmotic pressure balance (for comparison with RG predictions, see SI Sec.~\ref{sec:si_Pl_large}). 

To test these predictions and investigate the role of the NE's material structure during wrinkle formation, we performed live-imaging experiments in which we measured the concentration of a structural component that determines nuclear stiffness. Elastic properties of the NE are known to depend strongly on the nuclear lamina \cite{swift_nuclear_2013}, a roughly 10-100\,nm thick meshwork of intermediate filaments that abuts the NE's inner membrane \cite{aebi_nuclear_1986}. \emph{Drosophila} have two lamin proteins, Lamin~C (a developmentally-regulated A-type lamin similar to mammalian Lamin A/C, \cite{schulze_molecular_2005, riemer_expression_1995}) and Lamin Dm0 (a B-type lamin present in most cell types). Both through live imaging of egg chambers simultaneously expressing a fluorescently-labeled nuclear pore complex protein (Nup107) and Lamin~C, and through fixed imaging with an antibody against Lamin~C, we found that, as egg chambers age and nurse cells grow in size, the ratio of Lamin~C to Nup107 decreases (Fig. \ref{fig:fig2}d,e; Supp. Fig.~\ref{fig:sfig_lamC}a-c) while the intensity of Nup107 remains roughly constant (Supp. Fig.~\ref{fig:sfig_lamC}d). Nonetheless, at the spatial scales resolved experimentally, Lamin C continues to appear alongside Nup107 at the sites of wrinkles (Supp. Fig.~\ref{fig:sfig_lamC}e-i). This reduction in Lamin~C concentration might cause softening of the NE and a reduced bending rigidity $\kappa$ that increases wrinkle amplitudes \cite{swift_nuclear_2013} as predicted by Eq.~\eqref{eq:scaling}. However, given the challenge in performing perturbative studies (see Discussion and SI~Sec.~\ref{sec:si_LamC}), it is not possible to assess the functional importance of Lamin C decrease from our data.

In addition to material properties, active fluctuations~\cite{agrawal_active_2022} or hydrodynamic effects~\cite{chakrabarti_flexible_2020, kantsler_vesicle_2007} can substantially affect buckling and pattern formation in shells and membranes~\cite{loubet_effective_2012,vutukuri_active_2020,kokot_spontaneous_2022}. 
To explore how changes in cytoskeleton-mediated intracellular activity \cite{bausch_bottom-up_2006} influence the spectrum of NE deformation, we performed additional perturbation experiments targeting the cytoplasmic microtubule and actin networks. Previous work showed that incoherent microtubule dynamics can cause fluctuations of the NE during cellularization of the \textit{Drosophila} embryo \cite{hampoelz_microtubule-induced_2011}. Consistent with this earlier report and with the predictions of Eq.~\eqref{eq:scaling}, we found that inhibition of microtubule polymerization by the small-molecule inhibitor colchicine notably reduces the amplitude of fluctuations (Fig.~\ref{fig:fig3}a; Supp. Fig.~\ref{fig:sfig_drugaddition}a,b; Supplementary Videos~4,5). Colchicine treatment also decreased nucleus volume by 5\%-20\%; however, as a volume decrease for a similar surface area would lead to a rougher rather than a smoother NE in the absence of other factors, it is unlikely that volume reduction explains the effects seen upon colchicine treatment. Furthermore, colchicine addition reduced the motion of cytoplasmic contents of the cells (Supplementary Videos~6,7), suggesting microtubule-mediated active fluctuations contribute to NE wrinkling. However, colchicine addition in older egg chambers (time proxies roughly over 185) had a less noticeable effect, decreasing rotational motion of the nuclei but not leading to the same extent of observable unwrinkling as in younger egg chambers. In contrast, perturbation of actin by cytochalasin D did not unwrinkle the NE (Supp. Fig.~\ref{fig:sfig_drugaddition}c,d; Supplementary Videos~8,9), suggesting cytoplasmic F-actin is not a major contributor to NE wrinkling during the developmental stages studied here. The observation that inhibition of microtubule polymerization reduces the wrinkle amplitude but does not change the spectral scaling behavior (Fig.~\ref{fig:fig3}d) suggests that, to leading order, non-equilibrium contributions to NE fluctuations arising from microtubule dynamics can be modeled through an effective temperature $kT_\mathrm{eff}$~\cite{betz_atp-dependent_2009,turlier_equilibrium_2016}. We also tested other mechanisms known to contribute to NE deformation, such as sustained impingement by cytoskeletal filaments or changes to chromatin structure (SI~Sec.~\ref{sec:si_otherStuff}), but found Lamin~C decrease to be the dominant factor correlating with NE wrinkle formation (Supp. Fig.~\ref{fig:sfig_cytoskeleton},~\ref{fig:sfig_drugaddition},~\ref{fig:si_chromatinLINC}; Supplementary Video 3). Nonetheless, other factors than the ones addressed in this study could also participate in NE wrinkling, as our model is a simplification of a complex biological process.

\par 
Eq.~\eqref{eq:scaling} also suggests that osmotic pressure variations can, by tuning the available excess area \cite{deviri_balance_2022, lemiere_control_2022}, enhance or reverse wrinkle formation by up- or down-shifting the deformation spectrum without changing its characteristic decay. To test this prediction, we performed osmotic shock experiments and found that adding salt to the ambient culture medium (hypertonic shock) increases the total external pressure on the NE, which drives fluid out of the nucleus and leads to visibly more wrinkled surfaces (Fig.~\ref{fig:fig3}b,e). Conversely, reducing the salt concentration in the culture medium (hypotonic shock) decreases the external pressure and leads to substantial smoothening of wrinkled NE (Fig.~\ref{fig:fig3}c,e, Supp. Fig.~\ref{fig:sfig_osmotic}; see SI~Sec.~\ref{sec:si_osmo} for experimental details and SI Sec.~\ref{sec:si_pressure} for an explanation of the connection between osmotic pressure and wrinkle amplitude). In both cases, NEs maintained their altered morphology for 15-30 minutes before nuclear shapes trended back towards their pre-shock state, presumably through regulatory mechanisms that partially compensate for osmotic changes. In agreement with Eq.~\eqref{eq:scaling}, the spectral slopes remained approximately preserved for both types of shocks. Taken together, these results support the hypothesis that wrinkle morphology and dynamics of deformation of the nurse cells' NE are dominated by a nonlinear elastic response rather than liquid-like behavior.

\par
NE wrinkles have been associated with biological processes including nuclear positioning \cite{almonacid_active_2019}, and as a mechano-sensitive element of the cell, the NE can regulate chromatin dynamics and force-induced transcription factor movement through nuclear pore complexes \cite{cosgrove_nuclear_2021,elosegui-artola_force_2017,makhija_nuclear_2016,chu_origin_2017}. Here we observed an increase in NE wrinkling during egg chamber growth that correlates with an increase in nuclear size along with a decay in Lamin~C concentrations. Although NE wrinkling may affect the nurse cells' chromatin organization and transcriptional states, NE wrinkles may instead simply result from concomitant nucleus growth and Lamin~C density decrease in cells that are fated to die to enable egg development. It is nonetheless tempting to propose that NE wrinkles could act as a tension buffer: tension applied to the NE would initially unfold the wrinkles before leading to significant in-plane strain that might cause NE rupture. Such a two-stage response to tension has indeed been observed during cell spreading, in which a NE stretch-mediated response occurs only after the initially-wrinkled nucleus flattens by a certain amount \cite{lomakin_nucleus_2020,venturini_nucleus_2020}.

Prior work has shown that some nurse cells in stage 5-9 egg chambers have a high level of intranuclear actin, which decreases from stage 10 onwards \cite{kelpsch_fascin_2016}. Additionally, in other contexts such as the \textit{Drosophila} larval muscle, Lamin C mutants can induce formation of intranuclear actin rods and potentially deform the nuclei \cite{dialynas_role_2010}. These findings suggest that changes in intranuclear actin levels or organization may also affect wrinkling; however, whether they increase or decrease wrinkling remains unclear (see SI~Sec.~\ref{sec:si_otherStuff}). Another future prospect 
is to investigate how NE wrinkling changes when Lamin C density is exogenously modified, allowing further comparisons to theory and clarifying whether NE wrinkling has biological function. Due to complications with existing fly reagents for perturbing Lamin C levels in the female germline, increasing or decreasing expression of Lamin C in the nurse cell NEs proved challenging (see SI~Sec.~\ref{sec:si_LamC} for details); therefore, developing such genetic tools would constitute a substantial technical advance. A further limitation of our study is that the continuum model is a simplification of the complex anatomy of the NE and that our measurements cannot distinguish between features that only include a single NE membrane layer or are below the resolution limit of confocal light microscopy.

To conclude, our experimental and theoretical results suggest that essential qualitative aspects of NE wrinkling can be understood within the framework of nonlinear elastic thin-shell mechanics. As NEs have a F\"oppl-von K\'arm\'an number similar to both graphene and paper \cite{blees_graphene_2015}, we expect our theoretical observations to be relevant for these and other similar systems, where fluctuations push the membranes and shells into larger deformation regimes. With the power-law exponent set by the elastic behavior of the shell, the amplitude of wrinkles is controlled by the effective pressure, which we have manipulated here through osmotic shocks and microtubule inhibition. Our findings therefore raise the question of whether cellular control over pressure could be a generic biophysical mechanism for avoiding undesirable consequences of NE wrinkling~\cite{mounkes_progeroid_2003,venturini_nucleus_2020,kalukula_mechanics_2022}.

\textbf{Acknowledgments}
The authors thank the MIT SuperCloud and Lincoln Laboratory Supercomputing Center for providing HPC resources that have contributed to the research results reported within this paper. We thank Mehran Kardar, Roger D. Kamm, Eric Folker, Mary Ann Collins and Douglas P. Holmes for helpful discussions. This work was supported by a MathWorks Science Fellowship (N.R.), NSF Award DMS-1952706 (J.D. and N.R.), Sloan Foundation Grant G-2021-16758 (J.D.), MIT Mathematics Robert E. Collins Distinguished Scholar Fund (J.D.), Feodor Lynen Research Fellowship from the Humboldt foundation (J.F.T.), Jarve Fund MIT grant (A.C.M. and J.D.), and the National Institute of General Medical Sciences of the National Institutes of Health under award number R01GM144115 (A.C.M). N.R. and J.F.T. acknowledge participation in the KITP online workshop `The Physics of Elastic Films: from Biological Membranes to Extreme Mechanics' supported in part by the National Science Foundation under Grant No. NSF PHY-1748958.

\textbf{Data and resource availability.}
The code used for numerical simulations is publicly available at \url{https://github.com/NicoRomeo/d3shell}. Raw data and all fly lines used in the study are available upon request.

\textbf{Competing interests.}
The authors declare no competing interests.
\renewcommand\thesection{\arabic{section}}
\toclesslab\section{Methods}{sec:methods}
Detailed descriptions of experimental methods, image processing, simulations, and theory can be found in the Supplementary Information.
\toclesslab\subsection{Experiments}{sec:methexp}
A list of fly lines used for this study can be found in Table S1. In short, ovaries were removed from well-fed flies and cultured \textit{ex vivo} \cite{prasad_cellular_2007} for 1-4 hours. Images were acquired using laser-scanning confocal microscopy with a 40x/1.2 NA water or 63x/1.4 NA oil objective. For immunofluorescence, ovaries were fixed in 4\,\% (wt/vol) paraformaldehyde and stained with phalloidin-Alexa-568, Hoechst, and antibodies against Lamin~C or tri-methylated histone H3K9. Perturbations were performed by adding NaCl-spiked culture medium, water, or small-molecule inhibitors to the culture medium to the following final conditions: 1.5x original osmolarity (hypertonic shock), 0.5x osmolarity (hypotonic shock), 9\,mg/mL colchicine, or 10\,$\mu$g/mL cytochalasin D.

\toclesslab\subsection{Image processing}{sec:imgprocmeth}
Nuclear envelopes were segmented and their coordinates extracted using FIJI and the MorphoLibJ plugin \cite{schindelin_fiji_2012,legland_morpholibj_2016}. Custom-built Matlab code~\cite{mietke_dynamics_2018} was used to perform least-square fits to determine spherical harmonic coefficients up-to the cutoff mode number $l_\text{max}=25$, which corresponds to an angular scale of roughly $7\,^\circ$, as well as to measure the LamC:Nup107 ratio over developmental time.

\toclesslab\subsection{Simulations}{sec:simmeth}

 We use the pseudo-spectral solver Dedalus 3 \cite{burns_dedalus_2020} to solve the Langevin equation derived from the non-euclidean elastic free energy Eq.~\eqref{eq:free_energy} accounting for hydrodynamic interactions with a viscous environment~\cite{lin_brownian_2004,turlier_equilibrium_2016}; see the Supplementary Information for details and validations. Due to the stiffness of the equations governing the dynamics of plates and shells, simulations of such thin surfaces is numerically expensive. We simulate the partial differential equations on a $512\times 256$ Driscoll-Healy spherical coordinate grid, for $3\times10^5$ timesteps using the MIT Supercloud cluster~\cite{Supercloud}. Each simulation is run on 32 cores for a total of approximately $\sim 360$ hours of CPU time per simulation. Additional processing to study simulation results used the pyshtools library \cite{wieczorek_shtools_2018}.
The fact that the power spectrum $P_l$ does not change qualitatively when the nuclear radius $R$ increases suggests that the preferred curvature radius $R_c$ is rather large. In simulations, we thus set the preferred curvature radius to $R_c = 20 R$ (SI Sec.~{\ref{sec:si_simulations}}). Consistent with experiments (Fig.~\ref{fig:fig2}c), the appearance of a plateau region in the power spectrum $P_l$ at small angular numbers is suppressed for this choice of $R_c$ (SI Sec.~{\ref{sec:si_linresponse}}, Suppl.~Fig.~\ref{fig:sfig_sims}). Indeed, the typical length scale above which finite-curvature effects become visible is $L_{\text{el}}=R \gamma^{-1/4} \sqrt{R_c/R}$, which is for FvK number $\gamma=10^4$ and $R_c = 20 R$ approximately equal to $R/2$. Further details on parameter selections are provided in SI Sec.~{\ref{sec:si_simulations}}.

\let\oldaddcontentsline\addcontentsline
\renewcommand{\addcontentsline}[3]{}
\bibliography{bibliography}
\let\addcontentsline\oldaddcontentsline

\onecolumngrid
\newpage
\centerline{\LARGE\textbf{Supplementary Information}}
\ \vspace{-0.2cm}\\
\begin{center}
\textbf{\large Dynamics, scaling behavior and control of nuclear wrinkling}\\
\vspace{0.3cm}
Jonathan A. Jackson,$^{1,\,2,\,*}$ Nicolas Romeo,$^{3,\,4,\,*}$ Alexander Mietke,$^{3,\,5}$ Keaton J. Burns,$^3$ \\Jan F. Totz,$^3$ Adam C. Martin,$^1$ J\"orn Dunkel,$^{3,\,\text{†}}$  and Jasmin Imran Alsous$^{6,\,\text{‡}}$\\
\vspace{0.15cm}
\textit{
$^1$Department of Biology, Massachusetts Institute of Technology\\
$^2$Graduate Program in Biophysics, Harvard University\\
$^3$Department of Mathematics, Massachusetts Institute of Technology\\
$^4$Department of Physics, Massachusetts Institute of Technology\\
$^5$School of Mathematics, University of Bristol\\
$^6$Center for Computational Biology, Flatiron Institute, Simons Foundation}
\end{center}

\setcounter{page}{1}
\renewcommand\thefigure{S\arabic{figure}}    
\setcounter{figure}{0}
\renewcommand\thetable{S\,\arabic{table}}    
\setcounter{table}{0}
\renewcommand\theequation{S\arabic{equation}}    
\setcounter{equation}{0}
\setcounter{section}{0}

\renewcommand\thesection{\Roman{section}}
\renewcommand\thesubsection{\arabic{subsection}}
\renewcommand\thesubsubsection{\arabic{subsection}.\arabic{subsubsection}}
  
\tableofcontents

\makeatletter
\renewcommand*{\@seccntformat}[1]{\csname the#1\endcsname\hspace{0.15cm}}
\makeatother

\vspace{0.3cm}

\begin{bibunit}[apsrev4-2]
\section{Experiments}
\subsection{Fly maintenance and dissection} 
Fly strains used in this study are listed in Table \ref{tab:flylines}. All flies were raised at room temperature ($\sim$22\,°C) and dissected following an established protocol \cite{prasad_cellular_2007}, with the exception of the crosses used for RNA interference, in which the first offspring (F1) generation was raised at 27\,°C. For all experiments, flies of 4-8 days of age were transferred to a fresh vial of food containing a thin layer of dry yeast one day prior to dissection.

\begin{footnotesize}
\begin{table}[!h]
\renewcommand{\arraystretch}{1.3}
\resizebox{\textwidth}{!}{\begin{tabular}{ |l|l|l| }
 \hline
 \textbf{Fly line} & \textbf{Source} & \textbf{Used in} \\ \hline\hline
 w[*]; Nup107::GFP & BDSC (35514) & Fig. 1-3; Supp. Fig. S1, S10-S12 \\ \hline
 y,w[*];; koi$^{CB04483}$ & BDSC (51525) & Supp. Fig. S2 \\
 \hline
 Oregon R & Lab stocks & Supp. Fig. S2, S12 \\ \hline
  w[*]; Nup107$^\text{E8}$, LamC$^\text{G00158}$ ttv$^\text{G00158}$/CyO; Nup107::RFP & BDSC (35516) & Supp. Fig. S6 \\ \hline
 alphaTub67C-Gal4::VP16, UtrABD::GFP/CyO; Nup107::RFP/TM3 & This study & Supp. Fig. S9 \\ \hline
 w[*]; wg[Sp-1]/CyO; Nup107::RFP & BDSC (35517) & Supp. Fig. S12 \\ \hline
 alphaTub67C-Gal4::VP16 & Lab stocks & Supp. Fig. S12 \\ \hline
 alphaTub67C-Gal4::VP16/CyO; Nup107::RFP/TM3 & This study & Supp. Fig. S12 \\ \hline
 y[1] sc[*] v[1] sev[21]; P\{y[+t7.7] v[+t1.8]=TRiP.HMS01612\}attP2 (Klar shRNA) & BDSC (36721) & Supp. Fig. S12 \\ \hline
y[1] sc[*] v[1] sev[21]; P\{y[+t7.7] v[+t1.8]=TRiP.HMS02172\}attP40 (Koi shRNA) & BDSC (40924) & Supp. Fig. S12 \\ \hline
 w[*];; His2Av::mRFP & BDSC (23650) & Supp. Fig. S12 \\ \hline
 Jup::GFP & Gift from Vladimir Gelfand & Confirming colchicine efficacy \\ \hline
 UAS-mCherry::MoeABD & Gift from Todd Blankenship & Confirming cytochalasin D efficacy \\ \hline
 y[1] v[1]; P\{y[+t7.7] v[+t1.8]=TRiP.JF01406\}attP2 (LamC RNAi) & BDSC (31621) & Attempt to perturb LamC levels \\ \hline
 w[*]; P\{w[+mC]=GAL4-nanos.NGT\}40/CyO; P\{y[+t7.7] w[+mC]=UAS-3xFLAG.dCas9.VPR\}attP2 & BDSC (67052) & Attempt to perturb LamC levels \\ \hline
 y[1] sc[*] v[1] sev[21]; P\{y[+t7.7] v[+t1.8]=TOE.GS01246\}attP40 (LamC sgRNAs) & BDSC (79670) & Attempt to perturb LamC levels \\ \hline
 UASt-LamC & Gift from Lori Wallrath & Attempt to perturb LamC levels \\ \hline
  Traffic jam-Gal4 & Lab stocks & Attempt to perturb LamC levels \\ \hline
\end{tabular}}
\caption{\textbf{Fly lines used in this study}. BDSC = Bloomington Drosophila Stock Center; parentheses = stock number} \label{tab:flylines}
\end{table}
\end{footnotesize}

\subsection{Microscopy} 
All imaging was performed using a Zeiss LSM710 laser-scanning confocal microscope and the Zen Black software, with either a 40x/1.2 NA Apochromat water immersion objective or a 63x/1.4 NA Plan Apochromat oil immersion objective. Lasers used include 405\,nm and 561\,nm diode lasers, an argon 488\,nm laser, and a HeNe 633\,nm laser. Reflection microscopy was performed using the same setup with the 488\,nm laser, but with a mirror in the beam path in place of the dichroic beam-splitter and with the detection wavelength set to the excitation wavelength.

\begin{figure}
    \centering
    \includegraphics[scale=0.95]{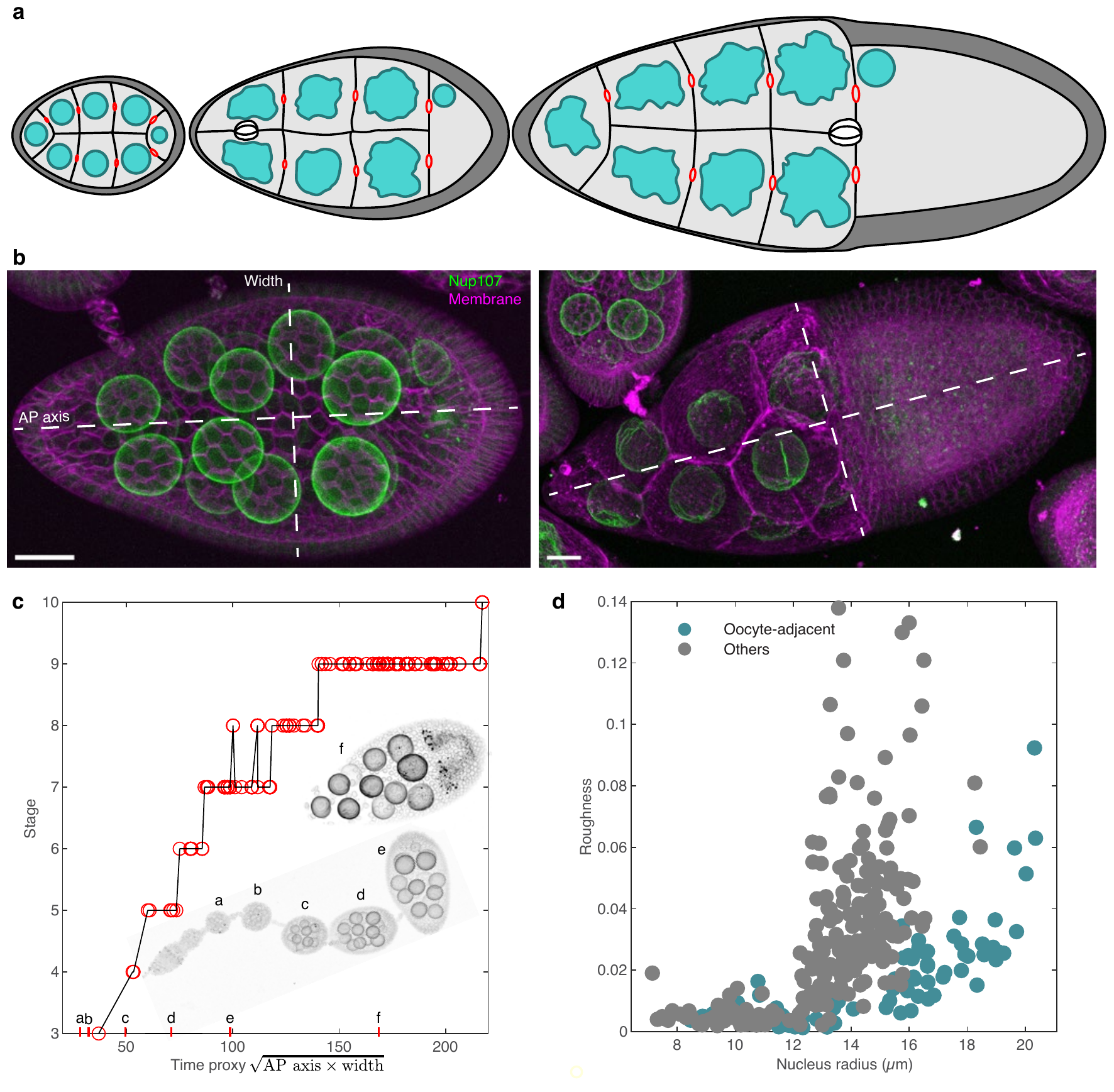}
    \caption{\textbf{Time proxy and relationship of nucleus size to wrinkling.}  \textbf{a,}~Schematics of the midplane of the egg chamber at three different stages, corresponding to the youngest egg chamber included in the roughness measurements (left; approximately stage 6), the stage during which wrinkling begins to become visible (middle; early stage 9), and the oldest egg chamber included in the measurements (right; approximately stage 10A). Dark gray: follicle cell layer; light gray: germline cells, oocyte at right (posterior); blue: nuclei; red: intracellular bridges connecting germline cells' cytoplasm; white: border cell cluster. \textbf{b,} Two egg chambers with dashed lines as used to measure anterior-posterior (AP) axis and width shown in white. Scale bars: 20\,$\mu$m; left egg chamber time proxy: 127, right egg chamber time proxy: 216. \textbf{c,} Developmental stage plotted against the time proxy for 39 egg chambers, showing a nearly monotonic relationship. Inset: ovariole (expressing Nup107::GFP) with six egg chambers, from stage 2 to 9, with corresponding time proxies shown in red tick marks on the x-axis. \textbf{d,} Roughness plotted against radius for 302 individual nuclei from 44 egg chambers; blue points (78 nuclei) represent nurse cells directly connected to the oocyte (`Oocyte-adjacent'), while gray points (224 nuclei) represent all other nurse cells, as measured by counting the number of intercellular bridges separating the oocyte and the nurse cell.}
    \label{fig:sfig_devcoord}
\end{figure}

\subsection{Immunofluorescence and antibodies} 
Ovaries were dissected from 10-15 flies into Schneider's \textit{Drosophila} medium (Thermo-Fisher, 21720001) and fixed in 4\,\% (wt/vol) paraformaldehyde in phosphate-buffered saline (PBS) at room temperature for 15-20 minutes. Ovaries were washed 3x in PBS with 1\% Triton X-100 by volume (PBT), after which ovarioles were separated using tungsten needles and transferred to blocking buffer containing primary antibodies (mouse anti-LamC (LC28.26) from the Developmental Studies Hybridoma Bank, diluted to 1:500; or rabbit anti-trimethylated H3K9, a gift from the Yamashita lab (MIT), also at 1:500 dilution) and incubated overnight at 4\,°C. After washes, egg chambers were incubated for two hours at room temperature with a secondary antibody (goat anti-mouse Alexa-647 or goat anti-rabbit Alexa-488) diluted to 1:500. When included, Hoechst (Invitrogen (H3570), diluted to 30\,$\mu$g/mL) and/or phalloidin-Alexa-568 (Invitrogen (A12380), diluted to 1:1000) were added alongside the secondaries. Egg chambers were mounted in a 1:1 mixture of RapiClear 1.47 (SunJin Laboratory Co.) and Aqua-Poly/Mount (Polysciences, Inc.) and imaged using the 40x/1.2 NA water objective (for Lamin C levels) or 63x/1.4 NA objective (for actin and H3K9 images).

\subsection{Live imaging}
Egg chambers were dissected and cultured \textit{ex vivo} following a modified version of an established protocol \cite{prasad_cellular_2007}. Briefly, ovaries were dissected from 1-4 flies into Schneider’s \textit{Drosophila} medium and individual egg chambers of stages 4-10 were separated using tungsten probes and forceps. Egg chambers to be imaged were transferred to a glass-bottomed dish (MatTek, P35G-1.5-14-C) containing approximately 200\,$\mu$L of fresh Schneider's medium. Imaging was performed over 1-4 hours. To outline cell membranes, when necessary, CellMask Deep Red plasma membrane stain (Invitrogen, C10046) was added to the imaging medium at a 1:1000 – 1:500 dilution. To visualize microtubules, egg chambers were incubated in Schneider's medium at room temperature for approximately 1 hour with Spirochrome SPY555-tubulin live-cell dye (Cytoskeleton, Inc.), diluted 1000x according to the manufacturer’s instructions. Live images were acquired using the 40x/1.2 NA objective, except for those in Supplementary Video 7, which were acquired using the 63x/1.4 NA objective. 

\begin{figure}
    \centering
    \includegraphics[scale=0.95]{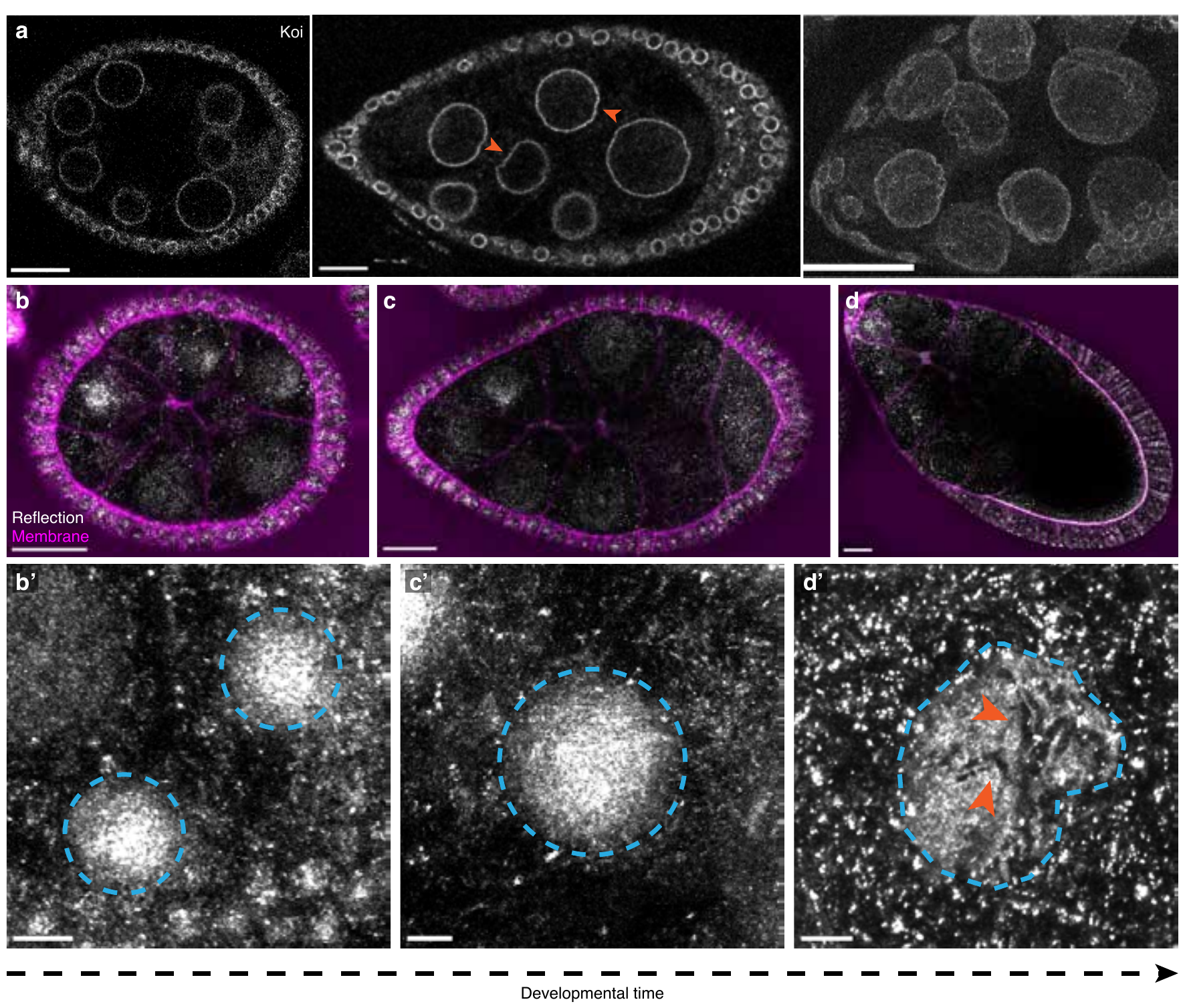}
    \caption{\textbf{NE wrinkling is observed using a label other than Nup107::GFP and through label-free imaging}. \textbf{a,}~Left: single optical section of a stage 6-7 egg chamber (time proxy 83) expressing endogenous levels of GFP-tagged Klaroid (Koi), showing nuclei without visible wrinkles. Middle: single section of an early stage 9 egg chamber (time proxy 148) showing the beginnings of visible wrinkles (orange arrowheads) as with the Nup107::GFP label that is used for analysis throughout this study. Right: projection through $\sim$45\,$\mu$m of a late stage 9 chamber (time proxy 197), displaying more obvious wrinkles; scale bars: 20\,$\mu$m (left and middle), 50\,$\mu$m (right). \textbf{b-d,} Single optical sections from wild-type egg chambers imaged via reflected light, with cell membranes imaged using a membrane-specific fluorescent stain; scale bars: 20\,$\mu$m. \textbf{b’-d’,} Maximum-intensity projections of the reflected light signal at the tops of selected nuclei (outlined by dashed lines) from the egg chambers in \textbf{b-d}. Nuclei are not visibly wrinkled in the first two egg chambers (\textbf{b}, \textbf{c}), while \textbf{d'} shows ridge-like structures (arrowheads) in the nucleus from the egg chamber in \textbf{d}. Scale bars: 5\,$\mu$m; projection depths: $\sim$4\,$\mu$m (\textbf{b'}), $\sim$5.5\,$\mu$m (\textbf{c'}), $\sim$6\,$\mu$m (\textbf{d'}). Developmental time proxies are 79 (\textbf{b}), 128 (\textbf{c}), and 196 (\textbf{d}). Arrow at bottom shows increasing developmental time.}
    \label{fig:sfig_otherlabels}
\end{figure}

\subsection{Osmotic shocks and small-molecule inhibitor addition} 
For all osmotic shock and small-molecule inhibitor experiments, Nup107::GFP egg chambers were stained with CellMask dye to assess cell health and membrane integrity over time.

\textbf{Osmotic shocks.} \label{sec:si_osmo} Because egg chambers appear to recover partially from osmotic shock after approximately 30 minutes, no images were acquired more than 15 minutes after shock.
For hypertonic shock (Fig.~\ref{fig:fig3}b), 3x-osmolarity Schneider's medium was prepared by adding 300\,mM of NaCl to Schneider's medium (which is approximately 300\,mM osmolarity). After acquiring initial images in 240\,$\mu$L of standard Schneider's medium for comparison, 80\,$\mu$L of the 3x-osmolarity medium was added, resulting in a final osmolarity of $\sim$1.5x. Occasionally, intracellular plasma membrane protrusions appeared in the cytoplasm of the nurse cells, likely resulting from excess surface area left after volume loss due to hypertonic shock. After any protrusions retracted, images of the same nuclei were then acquired again.
For hypotonic shock (Fig.~\ref{fig:fig3}c), experiments were performed as described above, but with addition of 200\,$\mu$L of deionized water to 200\,$\mu$L of Schneider's medium, resulting in 0.5x osmolarity medium. Although the egg chambers visibly increased in volume as a result, no protrusions or visible disruptions of plasma membrane integrity appeared.

The NE is studded with nuclear pore complexes, which allow free passage of ions but not of large macromolecules \cite{paine_nuclear_1975}. As a result, osmotically shocking the egg chamber is not likely to create a difference in ion concentration across the NE the way it would across the cell membrane. Indeed, in isolated nuclei, changes in ionic concentration of the surrounding medium does not affect the nuclear volume \cite{finan_nonlinear_2009}. However, those same nuclei inside cells do respond to changes in ion concentration by changing their volume. We propose the nurse cell nuclei studied here display changes in surface roughness in response to the changes osmotic shocks cause to the cytoplasm of the nurse cells. Hypertonic shock leads to water efflux from the cell, concentrating the macromolecules in the cytoplasm and therefore subsequently removing water from the nucleus, leading to volume loss (and vice versa for hypotonic shock). Plotting surface area and volume from reconstructions of nuclei in hypotonic, normal, and hypertonic conditions shows that volume increases (decreases) upon hypotonic (hypertonic) shock (data not shown), supporting this model in our system.

\textbf{Small-molecule inhibitors.} For colchicine addition, images of Nup107::GFP egg chambers in 200\,$\mu$L of Schneider's medium were acquired for reference, during which colchicine (Sigma-Aldrich, C9754) was added to fresh Schneider's medium to a final concentration of 45\,$\mu$g/$\mu$L and the solution allowed to come to room temperature. This colchicine-in-Schneider's solution was prepared fresh each day of experiments. 50\,$\mu$L of colchicine-containing medium was then added to the original 200\,$\mu$L of Schneider's medium, resulting in a final concentration of 9\,$\mu$g/$\mu$L \cite{viktorinova_microtubule_2013}. Colchicine efficacy was assessed by verifying that drug addition to stage 10 egg chambers stopped ooplasmic streaming (not shown), which is known to be dependent on microtubules \cite{theurkauf_premature_1994}, and by imaging egg chambers expressing a GFP-tagged version of the microtubule-associating protein Jupiter.
For movies of colchicine effects (Supplementary Videos 4, 7), images were acquired immediately following drug addition. For higher-resolution images of individual nuclei before and after colchicine addition (Fig.~\ref{fig:fig3}a, Supp. Fig.~\ref{fig:sfig_drugaddition}a-a$'''$), images were taken before and $\geq$30 minutes after colchicine addition. Recovery from colchicine, if at all present, was not clearly visible in the first 60 minutes of imaging.
To disrupt actin dynamics, Cytochalasin D (Enzo Life Sciences (BML-T109-0001)) dissolved in DMSO was added to the imaging medium as described above for colchicine, but to a final concentration of 10\,$\mu$g/mL cytochalasin D and 1\% DMSO (Supp. Fig.~\ref{fig:sfig_drugaddition}c-c$'''$; (Supplementary Video 8). Cytochalasin D efficacy at this concentration was assessed by addition of 10\,$\mu$g/mL to stage 11 egg chambers expressing the fluorescently-tagged actin-binding domain of moesin. Cortical actin decreased in intensity and the cortical actin network fragmented within 15-30 minutes (not shown), in accord with the known actin-depolymerizing effects of the molecule.

\subsection{Attempts to perturb Lamin C levels} \label{sec:si_LamC}

Prior work has established that increasing levels of Lamin A/C increases the stiffness of the nuclear envelope in different systems \cite{schape_influence_2009}, while expressing mutant forms of Lamin A/C reduces stiffness and results in misshapen nuclei \cite{pajerowski_physical_2007, houben_role_2007}. To decouple nuclear size increase and Lamin C decrease in the nurse cells, we attempted to perturb Lamin C levels using existing fly lines. None of our attempts noticeably affected levels of Lamin C, likely because existing fly lines are not optimized for the female germline; nonetheless, the details are presented here for completeness.
\begin{itemize}
    \item We first drove expression of LamC shRNA (BDSC line 31621) in the germline using mat67-Gal4, which did not appear to reduce Lamin C RNA levels when assessed by reverse transcription PCR (not shown). We suspect this is because the vector used to generate this line (VALIUM1) is not typically efficient for expression in the female germline.
    \item We next attempted to overexpress Lamin C in the germline using mat67-Gal4 to drive UASt-LamC (a gift from Dr. Lori Wallrath). When fixed and stained for Lamin C alongside Traffic jam (Tj)-Gal4 driven UASt-LamC (to overexpress Lamin C in the follicle cells), nurse cell nuclear membranes appeared similar in both conditions (however, follicle cell nuclear membranes in the Tj-Gal4 egg chambers did appear brighter than those in the mat67-Gal4 chambers (not shown) - as expected). Again, this is likely due to complications with ectopic expression using UASt sequences in the female germline \cite{deluca_efficient_2018}.
    \item We also tried to overexpress Lamin C using nanos-Gal4 driven expression of a nuclease-dead CRISPR enzyme fused to an activator (BDSC line 67052), crossed to a fly expressing sgRNAs for LamC (BDSC line 79670). Again, egg chambers were fixed and stained with the anti-Lamin C antibody alongside Nup107::GFP flies as a control. No clear differences were visible, although as with the UASt-LamC line, data were not quantified.
\end{itemize}

\section{Excluding other potential drivers of wrinkling as major factors} \label{sec:si_otherStuff}
Previous studies have identified several factors, some residing in the nucleus, others in the cytoplasm, that can exert fluctuating forces on the NE. These factors include impingement by either filamentous actin or microtubules \cite{jorgens_deep_2016,makhija_nuclear_2016,almonacid_active_2019,brandt_developmental_2006,biedzinski_microtubules_2020,chu_origin_2017}, changes in chromatin structure \cite{chu_origin_2017}, or force transduction from the external environment through the Linker of Nucleoskeleton and Cytoskeleton (LINC) complex \cite{versaevel_super-resolution_2015,horn_linc_2014}. Below we address these potential factors.\\

\textbf{Impingement by cytoskeletal filaments.} Impingement by microtubules has been implicated in nuclear shape change during cellularization of the \textit{Drosophila} embryo  and in human hematopoetic stem cells \cite{brandt_developmental_2006,biedzinski_microtubules_2020}. However, while live imaging revealed the presence of microtubules in the vicinity of nurse cell nuclei, these microtubules were present across all earlier stages of development as well, and there was no obvious increase in the density of microtubules surrounding the nucleus around the stages wrinkles first became visible (Supp. Fig.~\ref{fig:sfig_cytoskeleton}a). Once wrinkling did become apparent, there was no clear association between coherent microtubule structures and indentations in the NE (Supp. Fig.~\ref{fig:sfig_cytoskeleton}b-d; Supplementary Video 3). Furthermore, the nurse cells do not contain centrosomes or microtubule organizing centers at this stage \cite{januschke_centrosome-nucleus_2006}. The reduction in NE surface roughness observed upon addition of colchicine argues that microtubules do play a critical role in NE wrinkling in nurse cells; however, it appears microtubules are important for wrinkling via a different mechanism than directed protrusion of growing microtubules into the NE. Instead, microtubules could lead to NE wrinkling through increased bulk flow in the cytoplasm, causing the microtubules or organelles to impact the NE. Indeed, dynein-driven bulk cytoplasmic flow has recently been demonstrated in the nurse cells in stage 9, approximately the same time at which wrinkling becomes clearly visible, and is responsible for transport of other organelles including Golgi bodies and mitochondria \cite{lu_novel_2022} (see in particular Videos 4, 7, and 8 of this reference). Accordingly, our own live imaging of nurse cell nuclei using reflection microscopy reveals fluctuations in the cytoplasmic contents of the nurse cells coupled to deformations of the NE (Supplementary Video 6), both of which are substantially reduced following colchicine addition (Supplementary Video 7).

Impingement by filamentous actin (F-actin) cables and forces from actomyosin contractility have also been shown to deform and indent nuclei in a number of contexts in cell culture and mouse oocytes \cite{jorgens_deep_2016,makhija_nuclear_2016,almonacid_active_2019}, as well as in the final stages of \emph{Drosophila} oogenesis, when actin cables emanate from the nurse cell cortex and towards the nucleus, holding nuclei in place to prevent their dislodging during NC dumping \cite{huelsmann_filopodia-like_2013,yalonetskaya_nuclear_2020}. The dynamic wrinkling described here occurs $\sim$15 hours prior to formation of the actin bundles, suggesting it is independent from the actin bundle-driven crumpling seen in older egg chambers. Furthermore, visualization of F-actin, both live via the actin-binding domain of Utrophin (UtrABD) and fixed using labeled phalloidin, did not reveal noticeable actin structures protruding into the nucleus around the time of wrinkling onset or during its progression (Supp. Fig.~\ref{fig:sfig_cytoskeleton}e,f), and treatment with cytochalasin D to depolymerize F-actin did not lead to reduction in wrinkling, as observed with colchicine addition (Supp. Fig.~\ref{fig:sfig_drugaddition}a,c; Supplementary Video 8). Basket-like actin structures associated with the ring canals connecting the nurse cells to each other and to the oocyte \cite{nicolas_dual_2009} are often visible using labeled phalloidin (see rightmost side of Supp. Fig.~\ref{fig:sfig_cytoskeleton}f); however, wrinkling does not preferentially appear near the baskets. Occasionally, large divots in nuclei were observed in egg chambers prior to stage 9, particularly in cells adjacent to the oocyte, despite an absence of visible wrinkles anywhere else. These divots likely result from the actin baskets pressing into the nuclei, but because these divots were not present in the majority of egg chambers and appeared as single local deformations rather than global wrinkling, it is unlikely the actin baskets account for the observed wrinkling patterns.

\textbf{Changes to chromatin structure and organization.} Within the nucleus, the genome is organized in a hierarchical and complex manner, and changes in its organization can result in wrinkling of the NE \cite{chu_origin_2017}, particularly through interactions with lamins \cite{aebi_nuclear_1986}, which are major regulators of chromatin architecture. Nurse cell chromatin is highly organized and is known to undergo a well-defined transition from a blob-like to a more homogeneous dispersed organization \cite{dej_endocycle_1999}; however, as this re-organisation occurs approximately 20 hours prior to visible NE wrinkling onset (Supp. Fig.~\ref{fig:si_chromatinLINC}a), it is unlikely to explain the observed NE wrinkling dynamics. Similarly, there are no noticeable changes in heterochromatin amount (assessed using tri-methylated H3K9) around the time of visible NE wrinkling onset (Supp. Fig.~\ref{fig:si_chromatinLINC}b-e).

\textbf{Force transduction to the nucleus through the LINC complex.} Linking the cytoskeleton to the nuclear lamina is the Linker of Nucleoskeleton and Cytoskeleton (LINC) complex, which transmits mechanical stress and can mediate interactions that affect NE shape \cite{versaevel_super-resolution_2015,horn_linc_2014}. A previous study has shown that null mutants of Klaroid, the sole \textit{Drosophila} SUN-domain protein connecting the lamina to the rest of the LINC complex in nurse cells, as well as both KASH-domain proteins in \textit{Drosophila} (Klarsicht and MSP-300), do not block oogenesis, and that nurse cell nuclei expressing a dominant-negative form of MSP-300 still develop wrinkled NEs \cite{technau_drosophila_2008} (See Fig. 2 in this reference). Furthermore, we performed siRNA-mediated knockdown of Klaroid and Klarsicht, and found that neither seemed to significantly affect the level of NE wrinkling in the nurse cells (Supp. Fig.~\ref{fig:si_chromatinLINC}f-j). These results suggest that forces transduced to the nucleus through the LINC complex are not the primary cause of the observed NE wrinkling dynamics.

\textbf{Intranuclear actin.} Previous work \cite{kelpsch_fascin_2016} has shown that intranuclear actin is visible in some nurse cells, especially in stages 5-9; from stage 10 and onwards, which is already after the time when nurse cell nuclear wrinkling becomes apparent, infrequent and only unstructured nuclear actin is observed in the nurse cells. Previous work \cite{bohnsack_selective_2006} noted that nuclear actin and the actin rods observed upon overexpressing GFP-actin inversely correlate with invagination of the NE in nurse cells, suggesting that nuclear actin may play a supportive role, resisting the forces that deform the NE (also consistent with results in the Xenopus oocyte nucleus). However, in our own experiments, we observed increased roughness of the NE in stages 6-9 upon expression of Utrophin-RFP (not shown), which has also been shown to cause accumulation or stabilization of nuclear actin if expressed sufficiently strongly~\cite{spracklen_pros_2014}. The concentrations of cytochalasin D used in our experiments were likely insufficient to significantly disrupt any potential intranuclear actin structures, as even higher concentrations of have been shown to not enter the nucleus (\cite{munter_actin_2006,sankaran_gene_2019}). Furthermore, in other contexts such as the \textit{Drosophila} larval muscle, Lamin C mutants have been shown to induce formation of intranuclear actin rods \cite{dialynas_role_2010}, which seem to deform the nucleus. Taken together, these results suggest that intranuclear actin likely plays a role in nuclear shape changes, although possibly in conflicting ways and often under conditions of cellular stress or mutations rather than in normal development.

\section{Image processing and analysis of experimental data}
Analysis was performed using FIJI \cite{schindelin_fiji_2012} (to obtain coordinates of nuclear membranes in three-dimensional (3D) space) and custom-built Matlab code. For purposes of display, images in figures were processed using FIJI’s `Subtract background…', `Gaussian blur…', and/or `Remove outliers…' functions and adjusted in brightness/contrast to make nuclear membrane features more obvious. In Fig.~\ref{fig:fig1}b (main text), one egg chamber from a different ovariole that was damaged during dissection was manually removed to reduce clutter (in FIJI, a region of interest was drawn around the damaged egg chamber and the intensity inside the region set to zero). In other figures, portions of other egg chambers extending into the field of view were similarly manually removed.

\begin{figure*}
    \centering
    \includegraphics[scale=0.95]{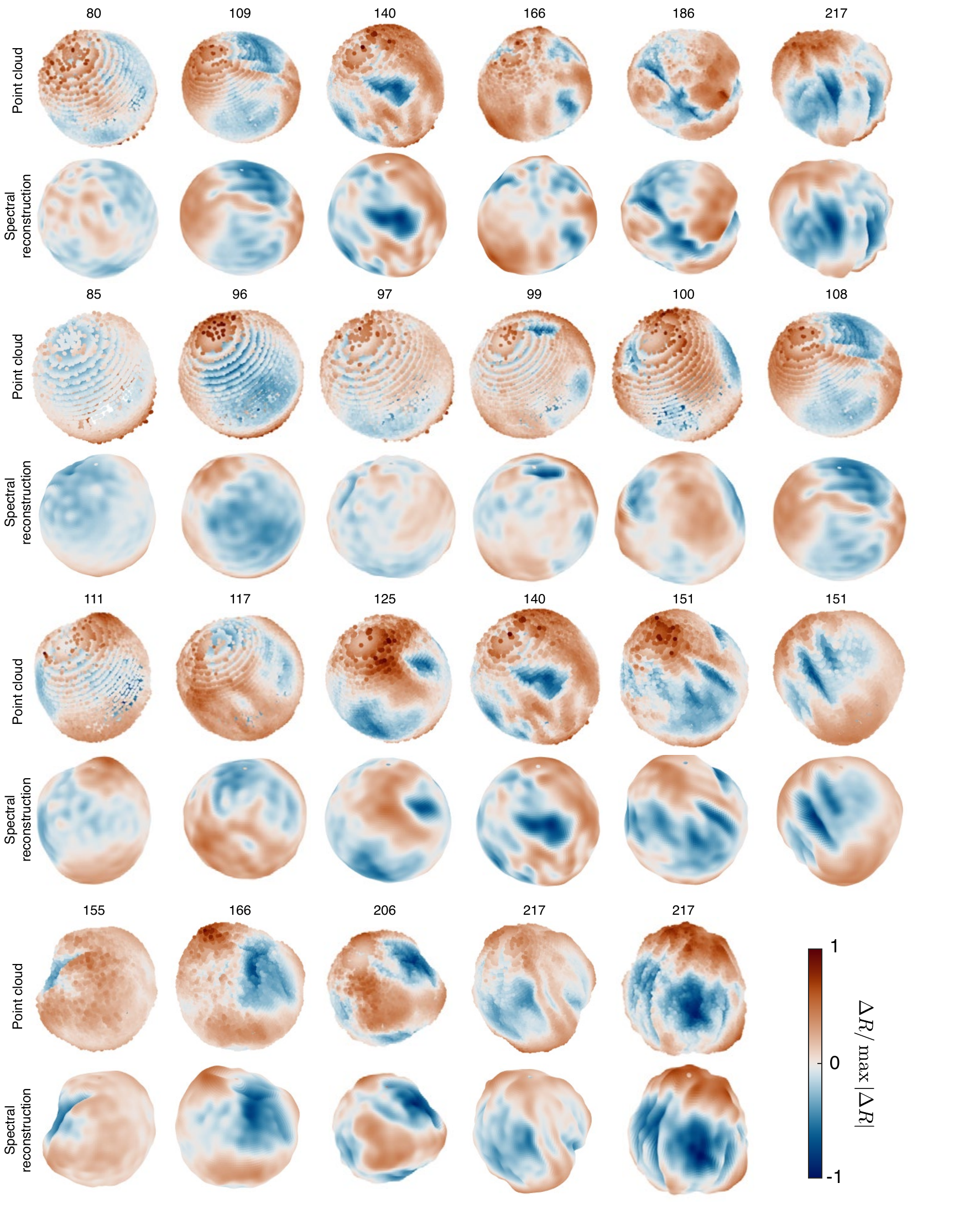}
    \caption{\textbf{Additional examples of nurse cell NE segmentation and reconstructions}. Representative examples of reconstructed surfaces (bottom row) from segmented point clouds (top row) obtained from 3D imaging. Numbers indicate time proxy, while color indicates deviation $\Delta R$ from the mean radius, normalized to maximum absolute value $\max{|\Delta R|}$. The first two rows of nuclei are those in Fig.~\ref{fig:fig1}c,d; other nuclei appear in order of ascending time proxy.}
    \label{fig:sfig_atlas}
\end{figure*}

\subsection{Time proxy: Using egg chamber geometry as proxy for developmental time} 
\label{sec:si_devcoord}
Egg chamber age, or progression through oogenesis, has traditionally been determined using broad morphological features, with chambers grouped into 14 stages, each of which encompasses several hours \cite{bastock_drosophila_2008,bratu_drosophila_2015,lin_germline_1993}. However, these stages provide a discrete time measurement with insufficient resolution for describing the continuous process of wrinkling progression, which occurs over much smaller time scales. Therefore, to determine if a readily-measurable and continuous-valued metric is suited for age comparisons and ranking egg chambers' progression through development, we plotted manually-determined egg chamber stages against several different size and morphological characteristics. One such metric, which we term the `time proxy', is the geometric average of the egg chamber width and anterior-posterior axis length. This time proxy technically has units of distance ($\mu$m), but is left unitless on plots as the measure is used solely for arranging egg chambers in time relative to one another. When egg chambers were aligned with their long axis horizontal to the objective (their midplane aligned with the imaging plane), linear measurements were made using maximum intensity projections (Supp. Fig.~\ref{fig:sfig_devcoord}b); when egg chambers were tilted, axis measurements were made in 3D space using the `3D\_Distance\_Tool' FIJI macro. Stages showed nearly monotonic increase with respect to the time proxy described above (Supp. Fig.~\ref{fig:sfig_devcoord}c), unlike with other metrics such as follicle cell height, egg chamber length or width, or oocyte length (not shown). Notably, this time proxy is similar to egg chamber cross-sectional area and to germline area, previously determined by others to be a good proxy for age \cite{jia_automatic_2016,weichselberger_eya-controlled_2022}. Due to changes in egg chamber size and shape resulting from fixation, the time proxy was not calculated for fixed images, except where it was estimated for Supp. Fig.~\ref{fig:si_chromatinLINC}e to allow approximate comparison to the roughness data.

Nuclei in nurse cells directly connected to the oocyte are larger than those in the more anterior nurse cells during the egg chamber ages considered in this study \cite{imran_alsous_collective_2017}. Because of this size difference within a given egg chamber, plotting roughness against effective nucleus radius reveals separate trends for the two populations of nurse cell nuclei; nucleus radius is thus not an effective metric for comparing egg chambers across time (Supp. Fig.~\ref{fig:sfig_devcoord}d). This size difference also means nurse cells nearer the anterior have higher roughness values for the same nucleus radius than cells nearer the oocyte, suggesting that age increase alone is unlikely to explain the patterns of wrinkling observed in the data.

\subsection{Extraction of nuclear membrane coordinates} 
Images of Nup107::GFP nuclei were preprocessed using FIJI’s built-in rolling-ball background subtraction method with a radius of 50 pixels, followed by applying a Gaussian blur with a width of 2 pixels. Segmentation was performed for still images using the `Interactive Marker-controlled Watershed’ algorithm of the MorphoLibJ library \cite{legland_morpholibj_2016} (Supplementary Video 10), and live images were segmented using the `Marker-controlled Watershed’ algorithm in a custom-built FIJI macro. In both cases, one seed point was used for each nucleus and another was used for the background. Individual nuclei were then isolated in FIJI, if necessary, and 3D coordinates saved for import to Matlab.

\begin{figure*}
    \centering
    \includegraphics[scale=0.95]{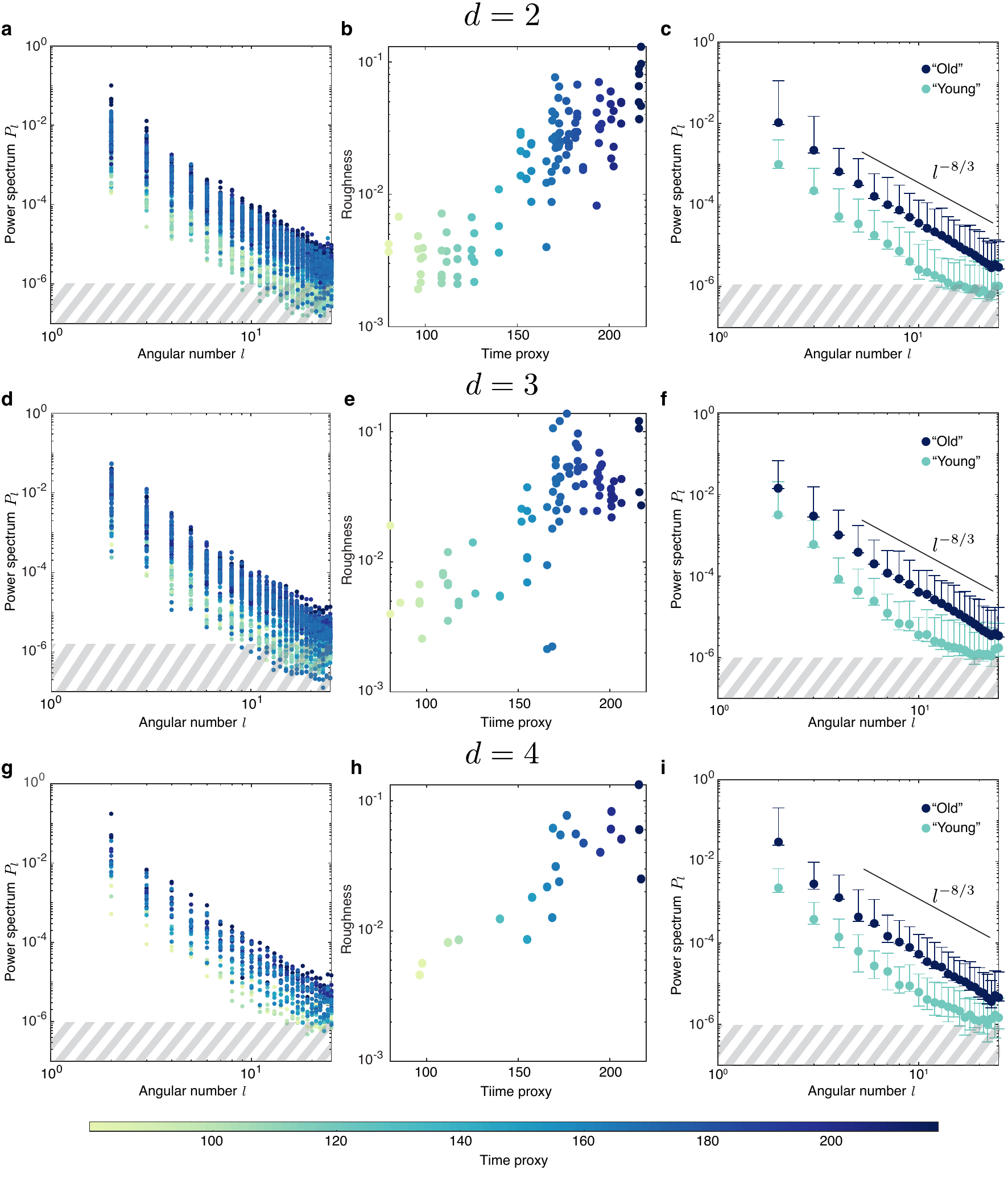}
    \caption{\textbf{The analyses used for Fig.~\ref{fig:fig1}e,f and Fig.~\ref{fig:fig2}c for different sets of nurse cell nuclei yield qualitatively similar results}. For nuclei~$d$ cells away from the oocyte, the same analysis as performed for cells directly connected to the oocyte is repeated with the same color bar throughout. \textbf{a,d,g} Power spectra $P_l$ colored by time proxy, as in Fig.~\ref{fig:fig1}e. \textbf{b,e,h} Roughness as in Fig.~\ref{fig:fig1}f. \textbf{c,f,i} Binned average of ``old'' nuclei as in Fig.~\ref{fig:fig2}c reveals the same scaling behavior. ``Young'' nuclei have time proxy between $80-140$; ``Old'' nuclei have time proxy between $160-220$. Intervals in panels c,f,i represent extremal values. $N=115$ nuclei for $d=2$  (29 young and 72 old), from 41 egg chambers, $N=86$ nuclei for $d=3$ (17 young and 59 old), from 39 egg chambers, $N=23$ nuclei for $d=4$ (4 young and 16 old), from 23 egg chambers.}
    \label{fig:sfig_layers}
\end{figure*}

\subsection{Lamin C measurements} \label{sec:si_lamin}
\textbf{Lamin C intensity over time.} To extract relative Lamin C intensities across egg chambers, it was necessary to compare endogenously-tagged Lamin C::GFP levels to Nup107::RFP levels, yielding a GFP:RFP ratio. Although the same laser settings and gains were used for all measurements, the use of this ratio helps avoid artifacts resulting from illumination differences between samples based on distance from the objective or other factors, and it helps reduce artifacts from intensity decrease stemming from the varying amount of tissue between the objective and nucleus in question. A simpler intensity correction based on depth was not possible here due to the curved shape of egg chambers. To further account for intensity decrease with imaging depth, especially given the variation in egg chamber size across stages, the GFP:RFP ratio in nurse cell NEs in each frame was normalized by the same ratio from follicle cell NEs. Importantly, Nup107::RFP and Lamin C::GFP levels were approximately constant over developmental time in follicle cells, and Nup107::RFP levels were also approximately constant over time in nurse cell NEs (Supp. Fig.~\ref{fig:sfig_lamC}a). To measure this ratio of ratios, the following steps were performed: 
\begin{enumerate}
    \item Nurse cell NEs were segmented as described above.
    \item A rough convex hull of the egg chamber was generated in Matlab using the Lamin C channel. A second hull containing just the nurse cells was generated from the nurse cell NE segmentation.
    \item These hulls were used to generate masks for the follicle cell region on a slice-by-slice basis: follicle cell regions were defined as any pixels inside the egg chamber convex hull but outside a dilated version of the nurse cell convex hull. Nurse cell masks were simply the output of the watershed algorithm in step 1. Pixels in which there was autofluorescence in the oocyte due to yolk granules were removed by hand if necessary and excluded from analysis.
    \item Individual nurse cell nuclei were identified by connected-component labeling.
    \item The intensities of Lamin C::GFP and Nup107::RFP were used to construct the GFP:RFP ratio, and outliers (defined as any ratio $>50$x the mean ratio per slice) were removed to account for division by near-zero pixel values.
    \item For each nurse cell nucleus, the sum of all ratios was divided by the sum of all ratios in follicle cells in the corresponding $z$-slices to generate the `ratio of ratios':
\begin{equation}
Ratio = \frac{\sum_{slices}(I_{LamC,NC}/I_{Nup,NC})}{\sum_{slices}(I_{LamC,FC}/I_{Nup,FC})},
\end{equation}
where $I_{LamC,NC}$ is the intensity of Lamin C in the nurse cells, $I_{Nup,NC}$ is that of Nup107 in the nurse cells, and so on for the follicle cells. The summation index \textit{slices} refers to every slice containing a portion of the nurse cell in question. This ratio was calculated separately for each nurse cell, requiring a different set of slices for each. Plots of the mean Lamin C::GFP intensity and of the ratio of Lamin C::GFP intensity in nurse cells to that in follicle cells also show a decrease with increasing time proxy (Supp. Fig.~\ref{fig:sfig_lamC}b,c). Furthermore, the ratio of Nup107::RFP in nurse cells to that in follicle cells is roughly constant (Supp. Fig.~\ref{fig:sfig_lamC}d), in accordance with both live and fixed images.
\end{enumerate}

\textbf{Overlap between Lamin C and Nup107 in wrinkles.} Wrinkles were identified by eye from maximum intensity projections of egg chambers from flies expressing both labels (Supp. Fig.~\ref{fig:sfig_lamC}e-g). A line of ~2-4 microns in width was drawn perpendicular to the wrinkles and intensities for both labels measured using FIJI. Intensities were plotted in Matlab after smoothing with a sliding window average along the line length using a window of 0.3-0.7 microns (Supp. Fig.~\ref{fig:sfig_lamC}h). To generate the average wrinkle profile (Supp. Fig.~\ref{fig:sfig_lamC}i), intensity peaks were located manually and a region of the peak full-width at half-maximum to either side of the peak was extracted. Each identified `wrinkle' signal was normalized by intensity and length (such that the start is 0 and the end 1), then the 13 wrinkles from 8 nuclei were averaged (data came from 5 egg chambers; nuclei ranged from 13 to 20 $\mu$m in radius).

\begin{figure}
    \centering
    \includegraphics{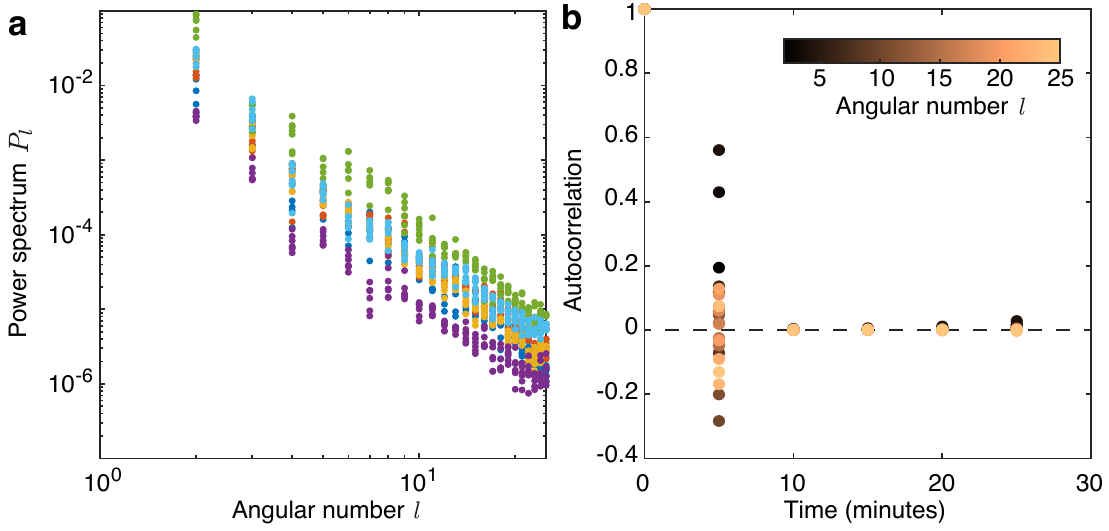}
    \caption{\textbf{3D high-resolution imaging reveals the fluctuating nature of NE wrinkles.} \textbf{a,} Power spectrum $P_l$ of different frames of a time series of nuclei 3d snapshots do not overlap, indicating that NE morphology is fluctuating in time. Each color corresponds to a different nucleus, with identically-colored points corresponding to different frames. \textbf{b,} Time autocorrelation $A_l(t) = \langle  \sum_m f_{lm}(t)f_{lm}(0) \rangle_Z / \langle  \sum_m f_{lm}(0)^2 \rangle_Z$ for $2\leq l \leq 25$ where the ensemble averages $\langle . \rangle_Z$ are estimated over the $7$ different experimental runs presented in~\textbf{a}. The decay of autocorrelation suggests that the correlation time for all observed modes is less than $10$ minutes. Number of samples contributing to $A_l(t)$ at each time point (from $t=0$ to $25$ min): $n= 30, 25, 20, 16, 12, 8$. Only data with $n>5$ are shown.}
    \label{fig:sfig_fluct}
\end{figure}

\begin{figure}
    \centering
    \includegraphics[scale=0.92]{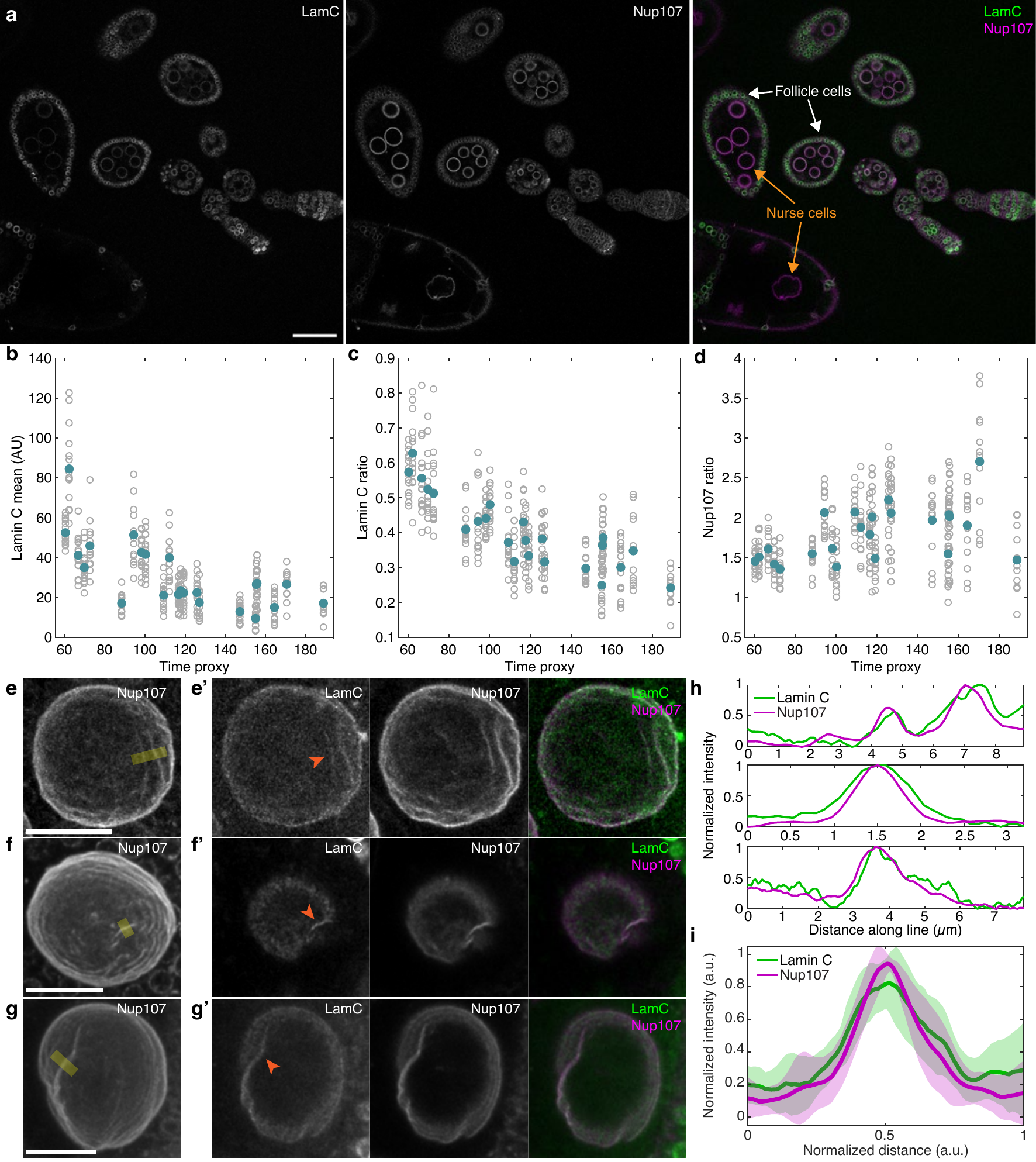}
    \caption{\textbf{Images and intensity measurements of live nurse cells show Lamin C decrease over time.} \textbf{a,} Single optical section of Lamin C::GFP, Nup107::RFP, and merged image for several egg chambers, showing decrease in Lamin C intensity with age while Nup107 intensity is relatively constant. In contrast to the nurse cells, Lamin C intensity remains roughly constant in the follicle cells, as shown by the continuous green outline in the merged image (white arrows; compare to nurse cells, orange arrows). Curved arrow denotes increasing age; scale bar 50\,$\mu$m. \textbf{b,} Mean Lamin C intensity (arbitrary units, AU) versus time proxy for each nucleus, with raw data in gray circles and the mean for each egg chamber in blue. \textbf{c,} Ratio of Lamin C intensity in nurse cells to that of follicle cells, calculated as described in Sec.~\ref{sec:si_lamin}. \textbf{d,} Ratio of Nup107 intensity in nurse cells to that in follicle cells, showing no obvious trend with age. Plots in \textbf{b-d} are of the same 337 nuclei from 23 egg chambers as in Fig.~\ref{fig:fig2}e. \textbf{e,} Maximum-intensity projection of Nup107 signal through 30 $\mu$m in depth of a nucleus (time proxy 148). The three panels in \textbf{e'} show the Lamin C signal, the Nup107 signal, and the merge for a projection through $\sim$11µm, showing the presence of both signals in a pair of wrinkles (orange arrowhead). \textbf{f,} As in \textbf{e}; the left image shows a projection through 44 $\mu$m (time proxy 172) and \textbf{f'} shows a single slice. \textbf{g,} As in \textbf{e}; the left image shows a projection through 34 $\mu$m (time proxy 166) and \textbf{g'} shows projections through 6 $\mu$m. Scale bars for \textbf{e-g}: 20 $\mu$m. Concentric rings in the left images are projection artifacts from discrete z-spacing during imaging. \textbf{h,} Intensity profiles of Lamin C and Nup107 across lines shown in  \textbf{e} (top) \textbf{f} (middle), and \textbf{g} (bottom). \textbf{i,} Average intensity profile for 11 wrinkles from 8 nuclei (5 egg chambers), showing overlap between the two signals (envelope represents the standard deviation).}
    \label{fig:sfig_lamC}
\end{figure}

\subsection{Excess perimeter measurements for LINC complex RNAi experiments}
In order to use RNA interference to knock down Klarsicht or Klaroid in the egg chamber, we were required to use a different fluorescently-tagged version of Nup107 that was less bright and less resistant to photobleaching. As a result, acquiring images of comparable resolution to those used for the reconstructions and roughness measurements in the main text was not possible. Nonetheless, we were able to use fractional excess perimeter as a less-detailed two-dimensional analogue of roughness to compare the degree of NE `wrinkliness' between wild-type and LINC complex-knockdown egg chambers (Supp. Fig.~\ref{fig:si_chromatinLINC}j):
\begin{enumerate}
    \item Nurse cells were imaged with a pixel size of about 100-200 nm instead of the 60-70 nm used previously.
    \item The midplane slice was determined manually and the NE was segmented as before using MorphoLibJ.
    \item Fractional excess perimeter was calculated as:
    $$x = \frac{P}{2\sqrt{\pi A}} - 1,$$
    where A and P are the area and perimeter of the midplane, a value of zero denotes a perfect circle, and higher values indicate increasing wrinkling/invagination of the NE.
\end{enumerate}

We compared this value between three conditions: 1) A subset of the 13 egg chambers from the  wild-type data used for the roughness analysis, 2) 9 egg chambers from the Klarsicht-RNAi condition, and 3) 10 egg chambers from the Klaroid-RNAi condition. Egg chambers were divided into `young' (approximately pre-wrinkling) egg chambers of time proxy $<150$ and `old' (generally post-wrinkling) egg chambers of time proxy $\geq 150$, similarly to Fig.~\ref{fig:fig2}c of the main text. The mean value of fractional excess perimeter per egg chamber was then compared between wild type and each of the RNAi conditions using a two-tailed Welch’s t-test for each pair of conditions and in each age range. For each of the four comparisons, the resulting p-value was at least 0.36.

\begin{figure}
    \centering
    \includegraphics{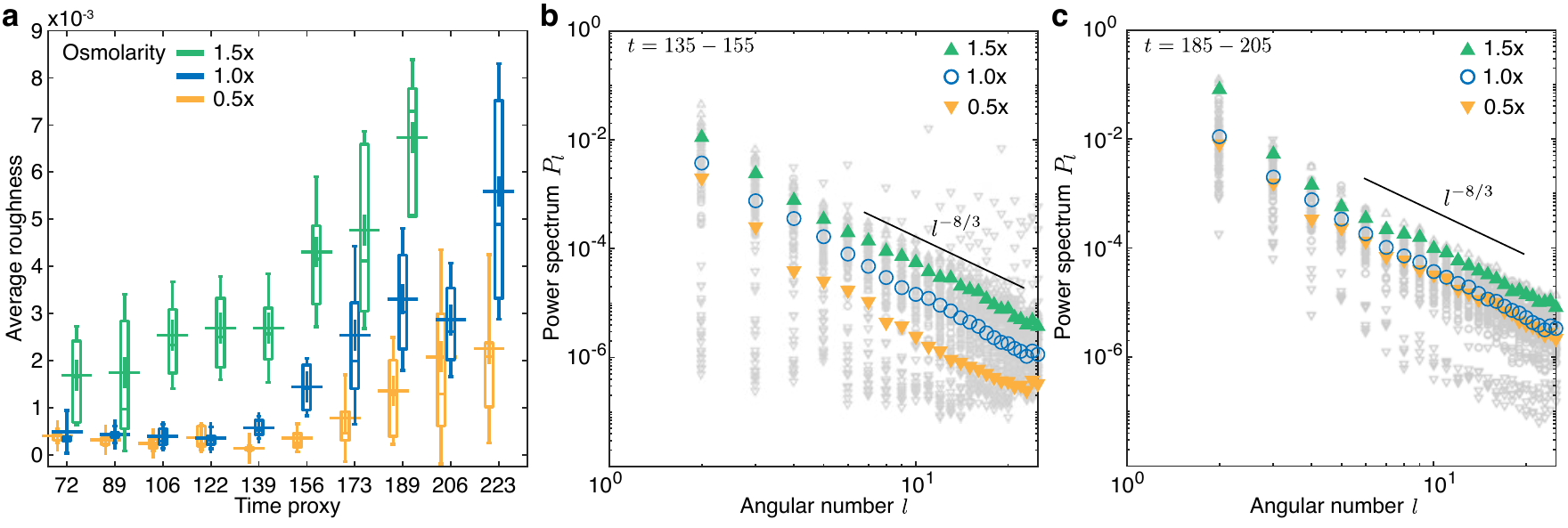}
    \caption{\textbf{Osmotic shock experiments robustly induce roughness changes across developmental time.} Plot of binned roughness values for different values of time proxy for hypertonic shock (green), wild type (blue), and hypotonic shock (orange); n = 254 nuclei from 21 chambers, 302 nuclei from 44 chambers, and 164 nuclei from 23 chambers, respectively. Plus sign, horizontal bar, and whiskers in box plots correspond to mean, median, and 9 to 91\% of the data range. respectively. Because the purpose of hypertonic shock experiments was to determine whether increased osmolarity causes increased wrinkling, hypertonic shock was not applied to older egg chambers (the maximum time proxy for hypertonic shock conditions is 187), although time proxies for hypotonic shock conditions go up to 223. Bins contain 5-85 nuclei each from 1-11 egg chambers; all nuclei are shown in the plot, regardless of their distance from the oocyte. \textbf{b,c,} Power spectra at different times for different osmotic conditions, showing the robustness of our results from Fig.~\ref{fig:fig3}e. For \textbf{b}, time proxies ranged from 135 to 155. 25 hypertonic, 27 control and 24 hypotonic nuclei. For \textbf{c}, time proxies ranged from 185 to 205. 5 hypertonic, 54 control and 19 hypotonic samples. Nuclei are included regardless of their distance from  the oocyte.}
    \label{fig:sfig_osmotic}
\end{figure}

\subsection{Heterochromatin intensity measurements}
Fixed images of wild-type egg chambers stained with Hoechst and an antibody against trimethylated histone H3K9 (abbreviated H3K9me3 from here on) were acquired. The optical section corresponding to the midplane of each nucleus was estimated and extracted manually for 132 nuclei from 11 egg chambers. For each image, the mean ratio of H3K9me3 to Hoechst (the ratio was used to account for differing intensity reductions resulting from the different depths of each nucleus) in the nucleus was obtained via a Matlab script implementing the following procedure. First, pixels corresponding to the nucleus were determined by performing background subtraction, followed by a median filter and binarization using Matlab's imbinarize function. For 13 nuclei, the chromatin signal was lobe-like and not sufficiently continuous (see Supp. Fig.~\ref{fig:si_chromatinLINC}b, leftmost egg chambers), so the pixel mask was made by manually outlining the nucleus in FIJI, then imported to Matlab. Second, for pixels inside the pixel mask, H3K9me3 signal was binarized and the mean intensity of all foreground pixels was calculated. Third, the same process was applied for the Hoechst signal, using a new binarization based on that signal instead. Finally, the mean intensity value of all H3K9me3 pixels was divided by that of all Hoechst pixels to yield the final H3K9me3:Hoechst intensity ratio.

\subsection{Error estimate for spectra} 
After segmentation, the NE surface has a multiple-pixel thickness that leads to an absolute standard deviation error of $\sim$200\,nm on the membrane position. This noise appears in $P_l$ as a constant value that can become dominant at high angular numbers $l$. For NEs of radius 5\,$\mu$m, the absolute error translates to a 5\,\% relative standard deviation, which gives an error on $P_l$ of about $10^{-6}$, setting the approximate level of the noise floor shown in figures as a shaded gray area. For larger nuclei, the noise is less of a concern.

\section{Elastic membrane theory and modeling} \label{sec:si_theory}
In this section, we describe the theoretical framework used to rationalize the nuclear membrane dynamics observed in experiments. Specifically, we introduce the mechanical model used in the main text, discuss the surface fluctuation power spectrum expected from its linear response, and derive the scaling results described in the main text when nonlinear contributions become important. After presenting details of the numerical simulation approach and its validation via analytic results, we provide arguments supporting that a fluid membrane model is not sufficient to explain the experimental and numerical observations.

\subsection{Theory of shallow elastic shells}

\subsubsection{Deformations of surfaces and shallow shell approximation}
We describe the nuclear envelope as a two-dimensional surface $\Gamma\subset\mathbb{R}^3$ with an undeformed reference shape parameterized by $\mathbf{r}(s^1,s^2)$, where $s^1,s^2$ are generalized coordinates. On this surface, tangent and normal vectors are defined by $\mathbf{e}_i=\partial_i\mathbf{X}$ ($\partial_i:=\partial/\partial s^i$, $i=1,2$) and $\mathbf{n}=\mathbf{e}_1\times\mathbf{e}_2/|\mathbf{e}_1\times\mathbf{e}_2|$, respectively. The metric tensor and the curvature tensor (the first and second fundamental forms) of the reference surface are given by $g_{ij}=\mathbf{e}_i\cdot\mathbf{e}_j$ and $C_{ij}=\mathbf{n}\cdot\partial_i\partial_j\mathbf{r}$. 

When the surface continuously deforms, a point $\mathbf{r}(s^1,s^2)$ moves to a new position $\mathbf{r}'=\mathbf{r}(s^1,s^2)+\boldsymbol{\xi}(s^1,s^2,t)$, where $\boldsymbol{\xi}(s^1,s^2,t)$ describes the displacement and we drop the arguments in the following for brevity. The general strain tensor $u_{ij}$ describing the relative deformations of the surface is defined by $(\mathrm{d}\mathbf{r}')^2-(\mathrm{d}\mathbf{r})^2=2u_{ij}\mathrm{d}s^i\mathrm{d}s^j$. Using $d\mathbf{r}'=(\mathbf{e}_i+\partial_i\boldsymbol{\xi})\mathrm{d}s^i$ and $\mathrm{d}\mathbf{r}^2=g_{ij}\mathrm{d}s^i\mathrm{d}s^j$, the strain tensor takes the form
\begin{align}
u_{ij}&=\frac{1}{2}\left[\mathbf{e}_j\cdot\partial_i\boldsymbol{\xi}+\mathbf{e}_i\cdot\partial_j\boldsymbol{\xi}+(\partial_i\boldsymbol{\xi})\cdot(\partial_j\boldsymbol{\xi})\right]\nonumber\\
&=\frac{1}{2}\left(\nabla_iu_j+\nabla_ju_i\right)-C_{ij}f+\frac{1}{2}(\nabla_iu_k-C_{ik}f)(\nabla_ju^k-C_j^{\,k}f)+\frac{1}{2}(C_i^{\,k}u_k+\partial_if)(C_j^{\,l}u_l+\partial_jf),\label{eq:st_full_si}
\end{align}
where we have split the deformation field $\boldsymbol{\xi}=\mathbf{u}+f\mathbf{n}$ into tangential $\mathbf{u}=u^i\mathbf{e}_i$ and normal $f\mathbf{n}$ contributions. 

To capture the geometric effects of bending deformations, we additionally discuss how surface deformations $\boldsymbol{\xi}$ change the mean curvature $H=C_i^{\;i}$ of an arbitrary surface. The mean curvature $H'$ of a general deformed surface, parameterized on the reference surface with curvature $H$, is to first order in the deformation field given by~\cite{salbreux17}:
\begin{equation}
H'=H+\nabla^2f+C_{ij}C^{ij}f+u^k\nabla_kH+O(\boldsymbol{\xi}^2),\label{eq:mc_si}
\end{equation}
where $\nabla^2f:=g^{ij}\nabla_i\nabla_jf$ denotes the Laplace-Beltrami operator on the surface.\\

The strain tensor in Eq.~(\ref{eq:st_full_si}) and changes of mean curvature Eq.~(\ref{eq:mc_si}) can be further simplified by taking into account the slenderness of the nuclear envelope. First, for envelope thickness~$h$ and deformation length scale $L$, plate and shell theory predict in-plane deformations $u_i$ to be of order $O(f\frac{h}{L})$~\cite{audoly_elasticity_2010}. As a second simplifying assumption, we consider a shell with small and slowly varying curvature, such that $h C_{ij} \ll 1$. Thus, in thin shells with $h/L\ll1$, in-plane deformations are expected to be significantly smaller than out-of-plane deformations $f$ and all higher order terms involving $u_i$ and/or the curvature $C_{ij}$ in Eq.~(\ref{eq:st_full_si}) are subdominant. In this shallow-shell approximation, we describe the undeformed reference shape as a spherical surface of radius $R$, which is oriented such that the normal $\mathbf{n}$ points towards the inside of the sphere. In this case, $C_{ij} = g^{\text{s}}_{ij}/ R$, where $g^{\text{s}}_{ij}$ denotes the metric tensor on the unit sphere and~\cite{audoly_elasticity_2010,paulose_fluctuating_2012,van2009w} 
\begin{equation}
u_{ij}\approx\frac{1}{2}\left[\nabla_i u_j+\nabla_j u_i+(\partial_i f)(\partial_j f)\right]-g^{\text{s}}_{ij}\frac{f}{R}.\label{eq:epsR}
\end{equation}

Similarly, changes in the mean curvature up to linear order in curvature and deformations follow from Eq.~(\ref{eq:mc_si}) as
\begin{equation}
H'\approx H+\nabla^2f\label{eq:mc_si_simp}.
\end{equation}
For a spherical reference surface of radius $R$, we have $H=2/R$.

\subsubsection{Free energy of shallow elastic shells}

Previous work investigating the mechanical properties of nuclear envelopes has demonstrated that their main structural constituents -- the double lipid bilayer and the associated lamin filament meshwork -- collectively give rise to an effective elasticity and bending rigidity~\cite{kim_volume_2015,funkhouser_mechanical_2013}. We thus describe the nuclear envelope in this work as a thin elastic membrane whose elasticity and bending rigidity lead to a resistance against stretching and bending, respectively. Such properties are captured by the free energy
\begin{equation}
F=\int\mathrm{d}\mathbf{r}^2\left[\frac{\kappa}{2} (H'-H_0)^2 + \frac{1}{2}C_{ijkl}(u^{ij}-u^{\ ij}_0)(u^{kl}-u^{\ kl}_0)- pf\right]\label{eq:FEgen},        
\end{equation}
where $\kappa$ denotes the bending rigidity and we use an elastic modulus tensor \hbox{$C_{ijkl}=\lambda g_{ij}g_{kl}+\mu(g_{ik}g_{jl}+g_{jk}g_{il})$} with Lam\'e parameters $\lambda, \mu$ to describe an isotropic elastic response of the surface. For a 2D material, these parameters are related to the 2D Young modulus $Y$ and the Poisson ratio $\nu$ by $\lambda = \nu Y/(1+\nu)$ and $2\mu=Y/(1+\nu)$. Additionally, we have introduced a spontaneous curvature $H_0$. We note that the spontaneous curvature $H_0$ adds an effective surface tension to the spherical reference surface that vanishes for $H_0=2/R$. As we will show below (SI~Sec.~\ref{sec:si_pressure}) the spectral response of the shell is for sufficiently small curvature variations and mean radius changes equivalently affected by pressure and surface tension. Without loss of generality, we therefore only consider the effects of pressure and set $H_0=2/R$ in the free energy Eq.~(\ref{eq:FEgen}). The spontaneous strain~$(u_{0})_{ij}$ in the energy Eq.~(\ref{eq:FEgen}) will be specified below. Furthermore, we have introduced in Eq.~(\ref{eq:FEgen}) an external normal load that is dimensionally equivalent to a pressure $p(\mathbf{x},t)$, where $p>0$ ($p<0$) correspond to normal forces pointing inwards (outwards). This pressure collects contributions from thermodynamic and osmotic pressure differences across the nuclear envelope, as well as from other mechanical interactions with the complex environment. The effects of a non-zero mean pressure are discussed in Sec.~\ref{sec:si_pressure} below. 

We also note that the NE area is constantly increasing as development progresses: with a roughly $2.3$-fold area increase in the 6 hours of stage 9, we find an area growth rate $r = \frac{1}{A}\frac{\Delta A}{\Delta t}\approx \frac{\ln 2.3}{6h}$ of about $r=4\cdot10^{-5}$ per second on average. To determine whether this growth dynamics can impact the dynamic properties of wrinkles, we compare the growth rate $r$ with the average wrinkle lifetime $T$, or wrinkle turnover rate $1/T$. Specifically, if $r\ll 1/T$, the effects of wrinkle turnover dynamics are expected to be the dominant factor in modulating statistical properties of the surface geometry, and the impact of growth dynamics would be negligible. Empirically, we find that the time-autocorrelation of mode amplitudes vanishes in less than 10\,min for all angular numbers $l$, such that $rT < 0.05 \ll 1$ (Fig.~\ref{fig:sfig_fluct}b). Therefore, area growth is very slow compared to the timescale of shape fluctuation, and in what follows, it is neglected as we consider a quasistatic approximation.

We allow for material properties that are described by the framework of non-Euclidean shells~\cite{efrati2009}. Specifically, the effective reference surface, with respect to which the energetic cost of stretching deformations is determined, may not actually be physically realizable in 3D Euclidean space. Such incompatibilities can arise from inelastic effects in the microscopic structure, such as growth, shrinkage or plastic cross-linking~\cite{pezzulla_curvature-driven_2017}, but they can also be thought of as a way to effectively account for the presence of metastable states with low energy barriers in the free energy landscape. In this work, we found that an isotropic preferred strain of the form
\begin{equation}
    (u_0)_{ij} = g_{ij}^{\text{s}} \left(\frac{1}{R_c} -  \frac{1}{R}\right) f,\label{eq:spontstr_si}
\end{equation}
is required to explain the experimental observations. The spontaneous strain given in Eq.~(\ref{eq:spontstr_si}) gives rise to incompatibilities, which can be seen as follows: If $R_c = R$, the energetic cost of surface stretching is entirely determined by deformations away from a spherical surface of radius $R$, compatible with the mean curvature $H_0=2/R$. If instead $R_c \ne R$, the effective reference surface for evaluating the energetic cost of out-of-plane deformations becomes instead a spherical surface of radius~$R$ with preferred principal curvatures $1/R_c$ everywhere. 

With the spontaneous curvature and strain discussed above, and taking into account the shallow shell approximations for strains and curvature changes given in Eqs.~(\ref{eq:epsR}) and (\ref{eq:mc_si_simp}), we arrive at the free energy used in the main text. Specifically, Eq.~(\ref{eq:FEgen}) becomes
\begin{equation}
    F = \int\mathrm{d}\mathbf{r}^2 \left[ \frac{\kappa}{2} (\nabla^2 f)^2 + \frac{\lambda}{2}\epsilon_i^{\;i} \epsilon_j^{\;j} + \mu\, \epsilon_{ij}\epsilon^{\,ij}  - pf\right],\label{eq:si_free_energy}
\end{equation}
where $\epsilon_{ij} = u_{ij} - (u_0)_{ij}$ denotes the effective strain tensor with components
\begin{equation}
\epsilon_{ij}=\frac{1}{2}\left[\nabla_i u_j + \nabla_j u_i + (\partial_i f)(\partial_j f)\right] - g^{\text{s}}_{ij}\frac{f}{R_c}.\label{eq:finstr_si}
\end{equation}
From the strain given in Eq.~(\ref{eq:finstr_si}), we see that in the limit of $R_c \rightarrow \infty$ the shell equations reduce to the conventional plate theory.\\

For vanishing pressure $p=0$, three non-dimensional numbers govern the behavior of a shell described by Eq.~(\ref{eq:si_free_energy}) \cite{paulose_fluctuating_2012,kosmrlj_statistical_2017}: the F\"oppl-von K\'arm\'an (FvK) number $\gamma = Y R^2/\kappa \approx 10^4-10^6$ (see main text, \cite{funkhouser_mechanical_2013,kim_volume_2015}), the bending rigidity relative to fluctuation intensity $\kappa / kT_\text{eff}$, and the curvature incompatibility $R/R_c$. If $p\neq 0$, a natural characteristic pressure is the critical buckling pressure $p_c = 4 \sqrt{\kappa Y}/R^2$, above which a spherical shell with $R_c=R$ classically buckles; a fourth non-dimensional number $p/ p_c$ then becomes relevant to characterize the dynamics of the shell~\cite{paulose_fluctuating_2012}.

Finally, we note that by construction, the quality of the small-strain and shallow-shell approximation considered here degrades when used to describe large deformations, and deformations at  large scales $\sim R$. Specifically, the geometric approximations made to arrive at Eq.~(\ref{eq:si_free_energy}) will not capture the increasing importance of nonlinear contributions when  deviations from the mean radius $\approx |f|/R$, become large. We also note that this free energy does not account for metastable states which commonly arise in shells; we however expect those metastable states to be less relevant when $R_c > R$ and the shell behaves closer to a `spherical plate'.

\subsection{Scaling predictions}
In this section we establish a range of linear and nonlinear results that can be obtained from the energy Eq.~\eqref{eq:si_free_energy}. After discussing the basic linear response, we describe how to obtain the schematic scaling contributions introduced in the main text and combine the latter with insights from previous renormalization analysis of similar models.

Throughout this section, we consider Gaussian fluctuations with an effective temperature $T_\text{eff}$ to account for both passive and active fluctuations.

\subsubsection{Linear response to fluctuations} 
\label{sec:si_linresponse}

We first derive the equilibrium power spectrum $P_l$ of shells described by Eq.~\eqref{eq:si_free_energy} in the linear response regime by extending a result from reference \cite{paulose_fluctuating_2012} to the case of $R_c \neq R$. For an external uniform pressure $p=0$, when shell fluctuations are small enough for the system to be in the linear regime, the free energy Eq.~\eqref{eq:si_free_energy} is given to quadratic order in the fields by
\begin{equation}
    F^{\text{lin}} = \frac{1}{2}\int \mathrm{d}^2\mathbf{r} \left[\kappa (\nabla_\mathcal{S}^2 f)^2 + \frac{4(\lambda + \mu)}{R_c^2}f^2 + \lambda (\nabla_\mathcal{S} \cdot \mathbf{u})^2 + \frac{\mu}{4}\left(\nabla_i u_j+\nabla_j u_i\right)\left(\nabla^iu^j+\nabla^ju^i\right) + \frac{2(\lambda+\mu)}{R_c} f (\nabla_\mathcal{S}\cdot \mathbf{u})\right],\label{eq:FElin}
\end{equation}
where we have introduced the spherical in-plane gradient $\nabla_\mathcal{S} = R^{-1} ( \mathbf{e}_\theta\partial_\theta  + \mathbf{e}_\phi\sin{\theta}^{-1} \partial_\phi )$. We expand the components~$f$ of normal displacements and in-plane displacements $\mathbf{u}=u^i\mathbf{e}_i$ using real scalar and vector spherical harmonics, respectively, such that 
\begin{subequations}
\begin{align}
    f(\mathbf{r}) & = \sum_{l=0}^{\infty}\sum_{m=-l}^{m=l} f_{lm} Y_{lm}(\theta,\phi) \\
    \mathbf{u}(\mathbf{r})  & = \sum_{l=1}^{\infty}\sum_{m=-l}^{m=l} \left( u_{lm}^{(1)} \boldsymbol{\Psi}_{lm}(\theta,\phi)  + u_{lm}^{(2)} \boldsymbol{\Phi}_{lm}(\theta,\phi)  \right),
\end{align}
\end{subequations}
where $\boldsymbol{\Psi}_{lm} = \nabla_\mathcal{S} Y_{lm}$ and $\boldsymbol{\Phi}_{lm} = \mathbf{e}_r \times \nabla_\mathcal{S} Y_{lm}$. Note that with this convention, $u_{lm}^{(1)}$ and $u_{lm}^{(2)}$ have dimensions of~(length)$^2$. Using the identities 
\begin{subequations}
\begin{align}
\int \mathrm{d}^2\mathbf{r} f(\nabla_\mathcal{S} \cdot \mathbf{u}) & =  -\frac{1}{R}\sum_{l,m} l(l+1)u_{lm}^{(1)} f_{lm} \\
\int \mathrm{d}^2\mathbf{r} (\nabla_\mathcal{S} \cdot \mathbf{u})^2 &  =  \frac{1}{R^2}\sum_{l,m} l^2(l+1)^2\left(u_{lm}^{(1)}\right)^2 \\
\frac{1}{4}\int \mathrm{d}^2\mathbf{r} \left(\nabla_iu_j+\nabla_ju_i\right)\left(\nabla^iu^j+\nabla^ju^i\right) &  = \frac{1}{R^2}\sum_{l,m} l(l+1)\left( l(l+1)-1\right)\left(u_{lm}^{(1)}\right)^2 \nonumber\\
& + \frac{1}{2R^2}\sum_{l,m} l(l+1)\left( l(l+1)-2\right)\left(u_{lm}^{(2)}\right)^2,
\end{align}
\end{subequations}
which follow from standard properties of scalar and vector-valued spherical harmonics \cite{mietke_dynamics_2018}, we can expand the linearized free energy Eq.~(\ref{eq:FElin}) in terms of the spherical harmonic basis as
\begin{align}
F^{\text{lin}} = \sum_{l,m} & \frac{1}{2}\left[\kappa(l-1)^2(l+2)^2 +  4 (\lambda+\mu)\left(\frac{R}{R_c}\right)^2R^2 \right] \left(\frac{f_{lm}}{R}\right)^2 \nonumber\\
& + \frac{1}{2}l(l+1)\left[ (\lambda+2\mu)l(l+1) - 2\mu \right]\left(u^{(1)}_{lm}\right)^2  + \frac{\mu}{2}l(l+1)\left( l(l+1)-2\right)\left(u_{lm}^{(2)}\right)^2\\ & - 2 \frac{\lambda +\mu}{R_c}R l(l+1) u^{(1)}_{lm} f_{lm}.\nonumber
\end{align}

Within our Gaussian fluctuation  assumption, we can functionally integrate out the $u^{(1)}_{lm}$ fields to obtain an effective free energy in terms of $f_{lm}$ only
\begin{align}
    F^{\text{lin}}_{\text{eff}} = \sum_{lm} \frac{1}{2}\left[\kappa(l-1)^2(l+2)^2 +  4 (\lambda+\mu)\left(\frac{R}{R_c}\right)^2R^2\left( 1 - \frac{(\lambda +\mu) l(l+1)}{(\lambda+2\mu)l(l+1) - 2\mu}\right)\right] \left(\frac{f_{lm}}{R}\right)^2
\end{align}
This quadratic free energy then lends itself to the equipartition theorem, and  we finally find that the spherical harmonic power spectrum $P_l = (2l+1)^{-1}\sum_m (f_{lm}/R)^2 $ follows
\begin{align}
    kT_\text{eff} P_l^{-1} = \kappa (l+2)^2(l-1)^2 + 4\mu \left(\frac{R}{R_c} \right)^2 R^2 \frac{(\lambda + \mu)(l^2+l-2)}{(\lambda+2\mu)l(l+1) - 2\mu}.\label{eq:Plfin}
\end{align}
This result agrees with Ref.~\cite{paulose_fluctuating_2012} for $R=R_c$ and serves as a benchmark to validate our numerical simulations~(Supp. Fig.~\ref{fig:sfig_sims}a,b, see Sec.~\ref{sec:si_simulations}).
From Eq.~(\ref{eq:Plfin}) it follows that $P_l$ is dominated by the effects of bending at high angular number, where it behaves as $P_l \propto l^{-4}$. At low angular number $P_l$ is dominated by the harmonic `confining' elastic term and approaches a constant of the order $P_l \sim 1/Y$ with the 2D Young modulus $Y\sim\lambda, \mu$ \cite{thorpe_new_1992}. The crossover between those two regimes is expected to happen at angular number \smash{$l= l_{\text{el}} \equiv \gamma^{1/4}\sqrt{R/R_c}$}, where $\gamma = Y R^2/\kappa$ is the FvK number. 

A criterion that strictly ensures the validity of conclusions drawn from this linear analysis is given by \hbox{$(kT_\text{eff}/\kappa) \sqrt{\gamma} \ll 1$} \cite{paulose_fluctuating_2012,kosmrlj_statistical_2017}. For FvK number values of $\gamma > 10^4$ considered here, this criterion then approximately yields the condition $kT_\text{eff}/\kappa \ll 10^{-2}$. However, even for a typical lipid bilayer bending rigidity $\kappa \sim 10 kT_\text{eq}$ with room temperature~$T_\text{eq}$ this condition is not expected to be satisfied. Hence, while this linear analysis provides important insights into the effects of the curvature mismatch $R/R_c$ and helps validating the numerical approach, further analysis presented below is needed the explain the scaling observed in experiments and in numerical simulations for strong fluctuations.

\subsubsection{Effective radial free energy}\label{sec:SI_EffFE}
In the next step, we explain how to obtain the scaling form for the free energy presented in Eq.~\eqref{eq:scaling_scheme} of the main text. Assuming Gaussian fluctuations of an elastic shell described by Eq.~\eqref{eq:si_free_energy}, it is possible to integrate out the in-plane displacements $\mathbf{u}=u^i\mathbf{e}_i$  and the mean radial displacement $f_0 = (4\pi R^2)^{-1}\int \mathrm{d}^2\mathbf{x} \, f$~\cite{sachdev_crystalline_1984, nelson_statistical_1989,paulose_fluctuating_2012,kosmrlj_statistical_2017}. This procedure yields an effective free energy written purely in terms of the normal displacements. For convenience, we collect in the following several results distributed across the above references into a single explicit derivation.

We start from the elastic free energy in Eq.~\eqref{eq:si_free_energy}, rewritten in terms of the isotropic elastic tensor $C_{ijkl} = \mu (\delta_{ik}\delta_{jl} + \delta_{il}\delta_{jk}) +\lambda \delta_{ij}\delta_{kl}$ and dropping the explicit pressure term for now.
We write
\begin{equation}
F = F^{\text{bend}}+F^{\text{stretch}}:=\int \mathrm{d}^2\mathbf{x} \left[ \frac{\kappa}{2}(\nabla^2 f)^2 + \frac{1}{2}C_{ijkl}\epsilon_{ij}\epsilon_{kl} \right].
\end{equation}
To use the  shallow-shell approach, we consider here a  portion of a shallow shell  of area $A$, such that we can use a two-dimensional Cartesian coordinate system and Fourier transforms to describe our fields.
As found in Eq.~(\ref{eq:finstr_si}) by considering in the shallow-shell regime a Cartesian metric $g_{ij} = \delta_{ij}$, the strain tensor is $\epsilon_{ij} = u_{ij} + \frac{1}{2}\partial_i f \partial_j f - \delta_{ij} f/R_c$, where we denote in this Sec.~\ref{sec:SI_EffFE} by $u_{ij} = \frac{1}{2}[\partial_i u_j + \partial_j u_i]$ only the \textit{in-plane} contributions of the strain tensor. To simplify the integration procedure, we decompose the symmetric tensor $A_{ij} = \frac{1}{2}\partial_i f \partial_j f$ into a longitudinal and transverse part \cite{sachdev_crystalline_1984, nelson_statistical_1989}
\begin{equation}
\frac{1}{2}\partial_i f \partial_j f  = \frac{1}{2}[\partial_i v_j + \partial_j v_i] + P_{ij}^T h,
\end{equation}
with a vector field $v_i$, a scalar field $h$, and the transverse projection operator $P_{ij}^T = (\delta_{ij} - \partial_i\partial_j/\nabla^2)$. Applying the latter to both sides of this equation, we find  $h(\mathbf{x}) = \frac{1}{2}P_{ij}^T \partial_i f \partial_j f$. We will also separate the radial displacement $f(\mathbf{x}) = f_0 + f'(\mathbf{x})$ into its uniform and spatially-dependent parts $f_0$ and $f'(\mathbf{x})$, respectively.

Here, we use the Fourier convention of the Supplementary Information of Ref.~\cite{paulose_fluctuating_2012}, where the direct transform is $f(\mathbf{q}) = (1/A)\int\mathrm{d}^2\mathbf{x}\, f(\mathbf{x}) e^{i\mathbf{q}\cdot \mathbf{x}}$  and the inverse is given by $f(\mathbf{x}) =  \sum_{\mathbf{q}} f(\mathbf{q})e^{-i\mathbf{q}\cdot \mathbf{x}}$. To keep the notation compact, we use the same symbols for functions in real and Fourier space and indicate their dependence by explicitly writing the argument. Finally, we note that by reality of the displacements fields, each field $f(\mathbf{x}), \ h(\mathbf{x})$ and $\mathbf{u}(\mathbf{x})$ satisfies $f(\mathbf{q}) = f(-\mathbf{q})^*$, with $f(\mathbf{q})^*$ denoting the complex conjugate of $f(\mathbf{q})$.\\

\textit{Fourier representation of the free energy:} We can now rewrite the free energy in terms  of the displacement fields mode  amplitudes. The bending energy can be expressed as $F^{\text{bend}}=(\kappa/2)\sum_\mathbf{q} q^4 |f(\mathbf{q})|^2$. In the following, we focus on the stretching part $F^{\text{stretch}}=\frac{1}{2}\int \mathrm{d}^2\mathbf{x}\, C_{ijkl}\epsilon_{ij}\epsilon_{kl}$ of the free energy. By defining the shifted variable
\begin{equation}
\tilde{u}_i(\mathbf{q}) = u_i(\mathbf{q}) + v_i(\mathbf{q}),    
\end{equation}
we can express the contributions from $u_i(\mathbf{q})$ and $v_i(\mathbf{q})$ in terms of the quadratic form $C_{ij}(\mathbf{q}) = \mu q^2(\delta_{ij} - q_i q_j / q^2) + (\lambda+2\mu)q_i q_j$. We then find a free energy that is quadratic in $\tilde{u}_i(\mathbf{q})$, $h(\mathbf{q})$ and $f(\mathbf{q})$, and given by
\begin{align}
F^{\text{stretch}} =\ & AC_{ijkl}\left(u_{ij}^0 + A^0_{ij}- \delta_{ij} f_0/R_c\right) \left( u_{kl}^0 + A^0_{kl}- \delta_{kl} f_0/R_c\right) \nonumber \\
& + \sum_{\mathbf{q}\neq 0} \tilde{u}_i(\mathbf{q}) C_{ij}(\mathbf{q}) \tilde{u}_j(-\mathbf{q}) + \tilde{u}_i(\mathbf{q}) B_i(-\mathbf{q}) + B_i(\mathbf{q}) \tilde{u}_i(-\mathbf{q}) \nonumber\\&+\sum_{\mathbf{q}\neq 0}  (\lambda + 2\mu) |h(\mathbf{q})|^2 + 4(\lambda+\mu)\frac{|f(\mathbf{q})|^2}{R_c^2} - \frac{2}{R_c}(\lambda+\mu)\left[f(\mathbf{q}) h(-\mathbf{q}) + f(-\mathbf{q}) h(\mathbf{q})\right].\label{eq:FTel}
\end{align}
Here, we denoted for convenience the uniform modes by
\begin{equation}
u^0_{ij}:=u_{ij}(\mathbf{q}=0)\quad\text{and}\quad A^0_{ij}:=A_{ij}(\mathbf{q}=0),
\end{equation}
and introduced $h(\mathbf{q}) = (\delta_{ij} - q_i q_j/q^2) A_{ij}(\mathbf{q})$, as well as
\begin{equation}
B_i(\mathbf{q}) = -i \lambda q_i h(\mathbf{q}) + i \frac{2}{R_c}(\lambda + \mu) q_i f(\mathbf{q}).
\end{equation}
To proceed, we distinguish between the spatially uniform modes ($\mathbf{q}=0$) and spatially non-uniform modes~($\mathbf{q} \neq 0$).\\

\textit{Spatially non-uniform modes} ($\mathbf{q}\neq 0$): We can observe from Eq.~(\ref{eq:FTel}) that spatially non-uniform modes contribute quadratically in $\tilde{\mathbf{u}}(\mathbf{q}\neq 0)$. These modes can therefore be integrated out using the Gaussian Path integral \cite{kardar_statistical_2007}
\begin{equation}
F_\text{eff}[f_0, f'] = -k T_\text{eff} \ln \left( \int \prod \mathcal{D}u_{ij}^0 \prod_{\mathbf{q}\neq0} \mathcal{D}\tilde{\mathbf{u}}(\mathbf{q}) \exp\left(-F/(kT_\text{eff})\right) \right).
\end{equation}
The contributions from the in-plane degrees of freedom given in the second line of Eq.~(\ref{eq:FTel}), once integrated out, yield a contribution to the effective free energy that is given by
\begin{align}
\sum_{\mathbf{q}\neq 0} -C_{ij}^{-1}(\mathbf{q})B_i(\mathbf{q})B_j(-\mathbf{q}),
\end{align}
with the inverse of the elastic quadratic form $C_{ij}^{-1}(\mathbf{q}) = \left(\delta_{ij} -q_i q_j/q^2\right)/(\mu q^2) + (\lambda +2\mu)^{-1} (q_i q_j/ q^4)$. 
Taken together, the spatially non-uniform contributions in the free energy Eq.~(\ref{eq:FTel}) yield an effective free energy
\begin{align}
F_\text{eff}^{\text{stretch}, \mathbf{q}\neq 0} =\ & \frac{1}{2}\sum_{\mathbf{q}\neq 0} \frac{1}{\lambda + 2\mu} \left( - \lambda^2|h(\mathbf{q})|^2  - 4(\lambda+\mu)^2\frac{f^2}{R_c^2} + 2\frac{\lambda}{R_c}(\lambda+\mu)\left(f(\mathbf{q}) h(-\mathbf{q}) + f(-\mathbf{q}) h(\mathbf{q})\right) \right) \nonumber\\
&+ (\lambda + 2\mu) |h(\mathbf{q})|^2 + 4(\lambda+\mu)\frac{f^2}{R_c^2} - \frac{2}{R_c}(\lambda+\mu)\left(f(\mathbf{q}) h(-\mathbf{q}) + f(-\mathbf{q}) h(\mathbf{q})\right).
\end{align}
This expression can be simplified using that $h(\mathbf{\mathbf{x}})$ and $f(\mathbf{\mathbf{x}})$ are real: Defining the 2D Young modulus $Y = 4\mu (\lambda +\mu)/(\lambda+2\mu)$, the contributions from the spatially-varying modes can be written as
\begin{align}
F_\text{eff}^{\text{stretch}, \mathbf{q}\neq 0}  & = \frac{1}{2}\sum_{\mathbf{q}\neq 0} Y|h(\mathbf{q})|^2  +Y\frac{|f(\mathbf{q})|^2}{R_c^2} - \frac{Y}{R_c}\left(f(\mathbf{q}) h(-\mathbf{q}) + f(-\mathbf{q}) h(\mathbf{q})\right) \nonumber\\
& = \sum_{\mathbf{q}\neq 0} \frac{Y}{2}\left| h(\mathbf{q}) - \frac{f(\mathbf{q})}{R_c} \right|^2 \nonumber\\
& = \int \mathrm{d}^2\mathbf{x} \,\frac{Y}{2}\left( \frac{1}{2}P_{ij}^T (\partial_i f \partial_j f) - \frac{f'}{R} \right)^2,
\end{align}
where the last line follows from Parseval's theorem. In the case of a flat sheet with  $R=\infty$, we recover the classical result presented in Ref.~\cite{nelson_statistical_1989}. In particular, one can understand  the role of the term $\frac{1}{2}P_{ij}^T (\partial_i f \partial_j f)$ through the geometric identity~\cite{nelson_statistical_1989}
\begin{equation}
    \nabla^2 \left[ \frac{1}{2}P_{ij}^T (\partial_i f \partial_j f) \right] = - K(\mathbf{x}),
\end{equation}
where $K(\mathbf{x})$ is the Gaussian curvature of the deformed sheet: In a similar fashion as the Airy stress function formalism of plate mechanics \cite{audoly_elasticity_2010}, the integration of in-plane degrees of freedom leads to a situation in which Gaussian curvature acts as a charge density for a stress-generating `curvature potential'~\cite{mitchell_fracture_2017}.\\

\textit{Spatially uniform modes} ($\mathbf{q}= 0$): The zero-wavenumber contributions to the strain tensor, $\epsilon_{ij}^0:=\epsilon_{ij}(\mathbf{q}=0)$, in Fourier space can be written as
\begin{subequations}
\begin{align}
\epsilon_{11}^0 & = u_{11}^0 + A_{11}^0 - f_0/R_c \\
\epsilon_{22}^0 & = u_{22}^0 + A_{22}^0 - f_0/R_c\\
\epsilon_{12}^0 & = u_{12}^0 + A_{12}^0,
\end{align}
\end{subequations}
where we recall that $u_{ij}^0$ refers to the modes corresponding to uniform in-plane strains. Of the  three of these, only two are independent since $u_{11}^0 +  u_{22}^0 = 0$, as otherwise this would indicate a uniform stretching of the shell which would lead to a variation in mean radius. We can then define $\Delta u^0 = u_{11}^0 - u_{22}^0$,  and write $u_{11}^0  = - u_{22}^0 = \Delta u^0/2$. We can thus rewrite the spatially uniform contributions to the free energy (first line in Eq.~(\ref{eq:FTel})) as
\begin{align}
F^\text{stretch}_0 = \frac{A}{2}C_{ijkl}  \epsilon_{ij}^0 \epsilon_{kl}^0 = & \frac{A}{2}\left[ \lambda\left( A_{11}^0 + A_{22}^0 - 2 \frac{f_0}{R_c}\right)^2 + 4\mu \left( u_{12}^0+ A_{12}^0\right)^2 \right. \\ & \left. +2\mu\left(\frac{\Delta u^0}{2}+ A_{11}^0 - \frac{f_0}{R_c}\right)^2 + 2\mu\left(-\frac{\Delta u^0}{2}+ A_{22}^0 - \frac{f_0}{R_c}\right)^2 \right]. \nonumber
\end{align}
We can now directly integrate out the $u_{12}^0$ field, and by shifting $\Delta u^0  \leftarrow \Delta u^0 + A_{11}^0 + A_{22}^0$ and integrating the field $\Delta u^0$, we finally find
\begin{align}
F^\text{stretch}_{0,\text{eff}} = & \frac{A}{2}(\lambda +\mu)\left( A_{11}^0 + A_{22}^0 - 2 \frac{f_0}{R_c}\right)^2. 
\end{align}
By using $ A_{11}^0 + A_{22}^0  = \frac{1}{A}\int \mathrm{d}^2 \mathbf{x} |\nabla f|^2$ and including the pressure term, we hence have the final, total effective free energy as 
\begin{align}
F_\text{eff}[f',f_0] = & \int \mathrm{d}^2 \mathbf{x} \left[ \frac{\kappa}{2}(\nabla^2 f)^2 + \frac{Y}{2}\left(\frac{1}{2}P_{ij}^T (\partial_i f \partial_j f) - \frac{f'}{R_c} \right)^2 + \frac{\lambda + \mu}{2}\left( \frac{1}{A}\int \mathrm{d}^2 \mathbf{x} |\nabla f|^2 - 2\frac{f_0}{R_c}\right)^2  \right] \nonumber \\ &- \int \mathrm{d}^2 \mathbf{x} \, p(\mathbf{x},t) f(\mathbf{x},t). \label{eq:si_Feff_fpf0}
\end{align}

\textit{Effect of pressure:} If $p(\mathbf{x},t) = p_0$, then $\int \mathrm{d}^2 \mathbf{x} \, p(\mathbf{x},t) f(\mathbf{x},t) = \int \mathrm{d}^2 \mathbf{x}\, p_0 f_0$, and after completing the square we can integrate out the average value of $f_0$. We now find
\begin{align}
F_\text{eff}[f'] = & \int \mathrm{d}^2 \mathbf{x} \left[ \frac{\kappa}{2}(\nabla^2 f)^2 - \frac{p_0R_c}{2}|\nabla f'|^2 + \frac{Y}{2}\left(\frac{1}{2}P_{ij}^T (\partial_i f \partial_j f) - \frac{f'}{R_c} \right)^2 \right].
\end{align}
 Denoting the spatial average by $\langle . \rangle$, the mean value of $f_0$ is given by
\begin{align}
f_0 = \frac{p_0 R_c^2}{4(\lambda + \mu)} + \frac{R_c}{4}\langle  |\nabla f'|^2 \rangle. \label{eq:f0_p0}
\end{align}
This equation reflects average area conservation: as the surface stretches very little for the large FvK numbers we consider, the appearance of wrinkles and folds must be accompanied by a reduction in mean radius $\Delta R \sim  - f_0$.
If we now consider inhomogeneous contributions to the pressure and write $p(\mathbf{x},t) = p_0+p'(\mathbf{x},t)$ with \hbox{$\int \mathrm{d}^2\mathbf{x}\,p'(\mathbf{x},t)=0$}, we find an effective free energy in terms of $f'$ that reads
\begin{align}
F_\text{eff}[f', p'] = & \int \mathrm{d}^2 \mathbf{x} \left[ \frac{\kappa}{2}(\nabla^2 f)^2 - \frac{p_0R_c}{2}|\nabla f'|^2 + \frac{Y}{2}\left(\frac{1}{2}P_{ij}^T (\partial_i f \partial_j f) - \frac{f'}{R_c} \right)^2  - p'f'\right].  \label{eq:si_Feff_fprime}
\end{align}
After expanding the square, this yields
\begin{align}
F_\text{eff}  = & \frac{1}{2}\int \mathrm{d}^2\mathbf{r} \left[ \kappa (\nabla^2 f)^2 -p_0 R_c (\nabla f)^2 + \frac{Y}{R_c^2}f^2 \right] \nonumber\\   & + \frac{1}{2}\int \mathrm{d}^2\mathbf{r} \left[\frac{Y}{R_c}f P_{ij}^T \partial_i f \partial_j f \right] \nonumber \\ & +\int \mathrm{d}^2\mathbf{r} \left[\frac{Y}{8} P_{ij}^T \partial_i f \partial_j f P_{kl}^T \partial_k f \partial_l f  - p'f'\right] \label{eq:si_effectiveF}.
\end{align}

\textit{Scaling form:} If we can assume $p_0 = 0$ (see SI Sec.~\ref{sec:si_pressure}), we can consider shape variations on a length scale $L$, and read off from Eq.~(\ref{eq:si_effectiveF}) the expected relative contributions of the different terms to the free energy density
\begin{equation}
    \delta F_\text{eff} \sim \kappa \left( \frac{f}{L^2}\right)^2 + Y \left(\frac{f}{R_c} \right)^2 + \frac{Y}{R_c} \left(\frac{f^3}{L^2} \right) + Y  \left(\frac{f}{L} \right)^4 - p f,\label{eq:scalSI}
\end{equation}
where $F_\text{eff} = \int \mathrm{d}^2\mathbf{r} \,\delta F_\text{eff}$. This model forms the basis for the scaling analysis presented in the main text Eq.~\eqref{eq:scaling_scheme}. If $p_0 \neq 0$, we expect generically from Eq.~\eqref{eq:si_effectiveF} the presence of an additional surface tension term $\sim \nabla^2 f$. However, we find that `crumpled' shells have vanishing surface tension,  under conditions detailed in Sec.~\ref{sec:si_pressure}. This scaling analysis thus holds for the rough surfaces we consider in experiments, with $p$ replaced by an effective pressure $p_\text{eff}$ to accommodate the effects of geometric confinement (Sec.~\ref{sec:si_pressure}).

\subsubsection{Renormalization of mechanical surface properties and comparison with simulations}
\label{sec:si_renorm}

\begin{figure}
    \centering
    \includegraphics[scale=0.95]{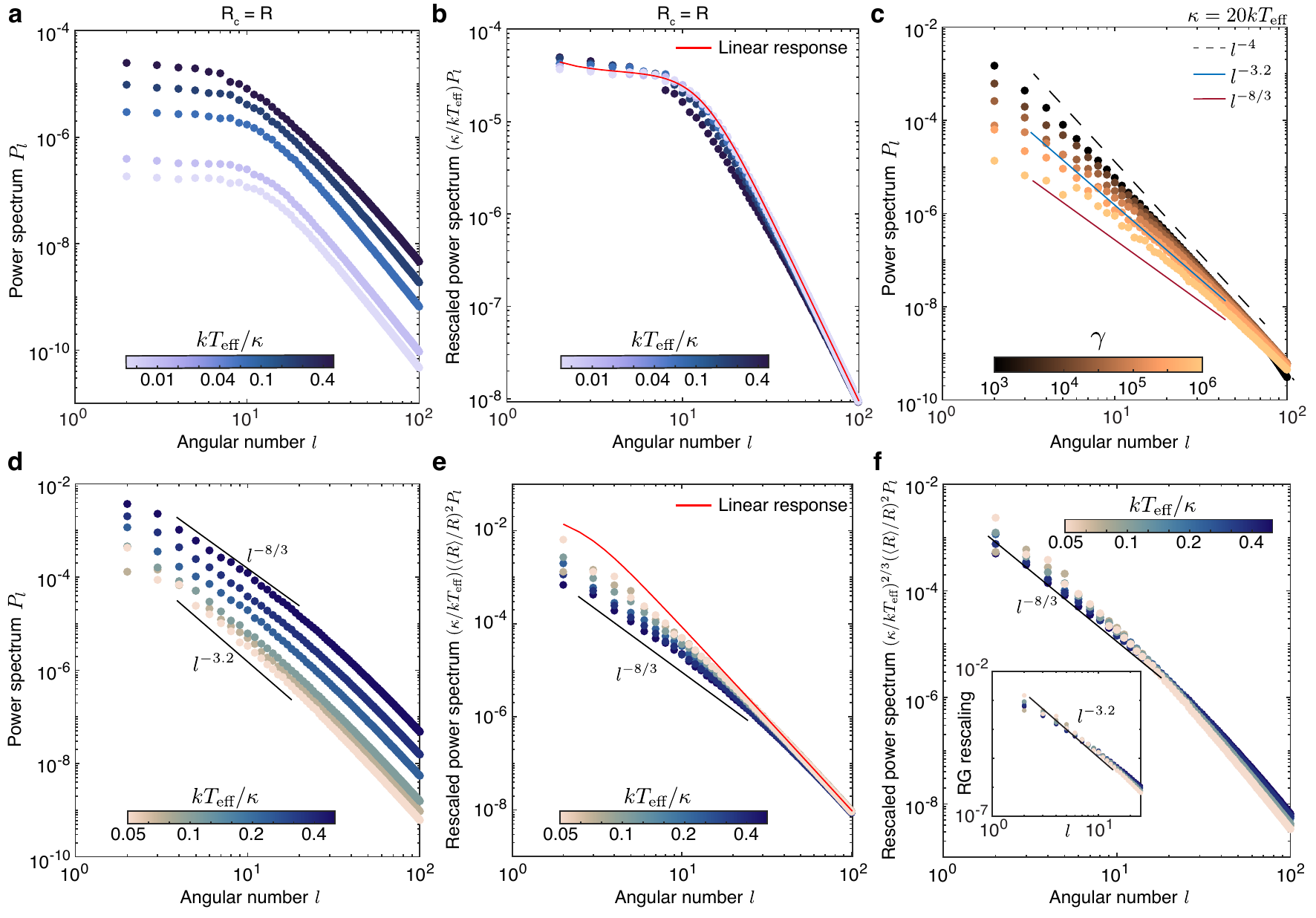}
    \caption{\textbf{Fluctuating elastic shells.} \textbf{a,}~Fluctuating elastic shell power spectra for $R_c = R$, $\gamma = 3\times 10^4$ for a variable elastic modulus controlled through $kT_\text{eff}/\kappa$.  \textbf{b,}~Rescaled power spectra from panel \textbf{a} show the validity of the linear regime for stiff shells, and the departure of the shell response from linear response (Eq.~(\ref{eq:Plfin}), red solid line) with decreasing rigidity (see Sec.~\ref{sec:si_linresponse}). \textbf{c,}~Fluctuating elastic shell power spectra for $R_c = 20R$ and variable FvK number $\gamma$ for constant bending modulus controlled through $\kappa = 20 kT_\text{eff}$. With increasing $\gamma$, nonlinear contributions to the shell elasticity modulate the power spectrum of radial deformations, consistent with predictions from renormalization and scaling arguments (see Sec.~\ref{sec:si_renorm}). \textbf{d,}~Full power spectrum for the simulations presented in Fig.~\ref{fig:fig2}a,b for fixed FvK number $\gamma = 3\times 10^4$ and $R_c = 20R$. \textbf{e,}~Rescaling the power spectra from \textbf{d} by $(\kappa/kT_\text{eff}) (\langle R\rangle/R)^2$ shows the departure from linear response Eq.~(\ref{eq:Plfin}) (see Sec.~\ref{sec:si_Pl_large}). \textbf{f,}~Rescaling the power spectra from \textbf{d} by $(\kappa/kT_\text{eff})^{2/3}(\langle R\rangle/R)^2$ shows a collapse to a master curve at intermediate wavenumbers, in accordance with the scaling analysis Eq.~(\ref{eq:Plsi_meanR_b})(see Sec.~\ref{sec:si_Pl_large}). Inset: Rescaling by $(\kappa/kT_\text{eff})^{0.6}(\langle R\rangle/R)^2$  as suggested by the RG scaling prediction Eq.~(\ref{eq:Plsi_meanR_RG}) leads to a similar collapse, confirming the similarity between both viewpoints. All simulations were performed using $T_\text{eff} = 10 T_\text{eq}$ with $T_\text{eq}= 300$\,K.}
    \label{fig:sfig_sims}
\end{figure}

We now refine Eq.~(\ref{eq:scalSI}) by adapting results from renormalization studies of curved elastic surfaces in order to explain in more detail the observed deviations from linear response in the simulations for soft shells (Supp. Fig.~\ref{fig:sfig_sims}).

Generally, it is expected that thermal fluctuations at length scales $L \sim R/l$ contribute most significantly to the effective mechanical properties of the surface if $L>L_{\text{th}}$, where \smash{$L_{\text{th}} = \sqrt{ \kappa^2/(kT_\text{eff} Y)}$} is the characteristic thermal length scale of the shell. Indeed, it was shown in Refs.~\cite{paulose_fluctuating_2012,kosmrlj_statistical_2017} that fluctuations in the presence of strong geometrical nonlinearities give rise to length scale-dependent elastic constants and an effective pressure. Specifically, the bending modulus~$\kappa$, the 2D Young modulus~$Y$, as well as a fluctuation-induced pressure behave in weakly nonlinear regimes as~\cite{kosmrlj_statistical_2017} 
\begin{subequations}\label{eq:scalAll}
\begin{align}
    \kappa_R(L) = & \kappa(L/L_{\text{th}})^\eta\label{eq:kappaR}  \\
    Y_R(L) = & Y(L/L_{\text{th}})^{-\eta_u}\\
    p_R(L) -p_0= & p_c (L_{\text{th}}/L_\text{el})(L/L_{\text{th}})^{2\eta}\label{eq:pR}
\end{align}
\end{subequations}
with exponents set by $\eta \approx 0.8$ and $\eta_u + 2\eta = 2$ and $L_\text{el} = (\kappa R^2 / Y)^{1/4}=R\gamma^{-1/4}$ is the characteristic elastic length scale beyond which stretching effects dominate over bending for $R=R_c$. Note that except for $p_c$, all the quantities and scaling exponents in Eqs.~(\ref{eq:scalAll}) are independent of the radius $R$ and hence do not depend on the presence of the curvature mismatch $R_c > R$. Far in the bending response-dominated regime at high angular numbers, we then expect from Eqs.~(\ref{eq:kappaR}) and Eq.~(\ref{eq:scalSI}) \smash{$P_l\sim(kT_\text{eff}/\kappa_R)l^{-4}\sim l^{-3.2}$}. Thus, as observed in our simulations (Supp. Fig.~\ref{fig:sfig_sims}c), an increasing FvK number $\gamma$ -- corresponding to an increasing significance of nonlinear contributions to the surface mechanics -- gradually changes angular number power laws describing the shell response from $l^{-4}$ to $l^{-3.2}$, before reaching a $l^{-8/3}$-\,scaling [see main text Eq.~(\ref{eq:scaling})] in the strongly nonlinear regime.

To better understand the limits of the renormalization approach, it is helpful to remember that the above scaling relations Eq.~(\ref{eq:scalAll}) are derived from the integration of the RG flow equations: as contributions from rapidly-varying modes are integrated during the renormalization procedure, the effective scale-dependent elastic parameters are  found to satisfy a set of  ordinary differential equations in the the parameter $s = \ln(L/a)$ known as the RG flow equations
\begin{subequations}
\begin{align}
    \frac{\mathrm{d}\kappa_R}{\mathrm{d}s} & = \beta_\kappa(\kappa_R, p_R, Y_R, R_R, \Lambda) \\
    \frac{\mathrm{d}p_R}{\mathrm{d}s} & = \beta_p(\kappa_R, p_R, Y_R, R_R, \Lambda) \\
    \frac{\mathrm{d}Y_R}{\mathrm{d}s} & = \beta_Y(\kappa_R, p_R, Y_R, R_R, \Lambda) \\
        \frac{\mathrm{d}R_R}{\mathrm{d}s} & = \beta_R(\kappa_R, p_R, Y_R, R_R, \Lambda),
\end{align} \label{eq:si_rgflowAll}
\end{subequations}
where $a = 1/\Lambda \ll R$ is a  microscopic lengthscale that  does not affect the integration results as long  as it is chosen to be small compared  to all other lengthscales involved.  The integration of  these equations from initial conditions $(\kappa, p_0, Y, R)$ then leads  to the scale-dependent parameters $\kappa_R(L), p_R(L), Y_R(L)$, where  the lengthscale relates to the wavenumber through $L = \pi/q$ (Fig.~\ref{fig:sfig_pressure}a,c). The precise form of the $\beta$-functions is available  in Appendix A of Ref.~\cite{kosmrlj_statistical_2017}.

\subsubsection{Power spectrum scaling for large fluctuations} \label{sec:si_Pl_large}

A surprising result from renormalization analysis~\cite{kosmrlj_statistical_2017} indicated by Eq.~(\ref{eq:pR}) is that fluctuations induce an effective pressure acting on the surface which can cause shells to spontaneously buckle. For the purpose of the following discussion, we consider the renormalized pressure $p_R$ in Eq.~(\ref{eq:pR}) at long wavelengths~\cite{paulose_fluctuating_2012}. In this regime, the renormalized pressure is given~by
\begin{align}
    p_R \sim p_c \left(\frac{kT_\text{eff}}{\kappa}\right) \sqrt{\frac{YR^2}{\kappa}} \sim \frac{Y}{R} \left(\frac{kT_\text{eff}}{\kappa}\right), \label{eq:pressure_scaling}
\end{align}
where $p_c = 4 \sqrt{\kappa Y}/R^2$ is the classical critical buckling pressure determined from a linear stability analysis of an elastic shell~\cite{paulose_fluctuating_2012,kosmrlj_statistical_2017}. Consequently, when $p =p_R + p_0 > p_c$ the elastic shell is expected to buckle. For $p_0=0$ and $R_c = R$, Ref.~\cite{kosmrlj_statistical_2017} finds from integrating the renormalization group flow that buckling should therefore occur around $p_R/p_c=(kT_\text{eff}/\kappa) \sqrt{\gamma}\gtrapprox 160$. When $R_c > R$, the added tendency for the shell to locally flatten will further modify the buckling pressure, with a new rescaled critical pressure given by $\bar{p}_c = (R/R_c)^2 p_c$. In general, by showing the instability of the  ground state of the elastic free energy Eq.~(\ref{eq:si_effectiveF}), this signals the breakdown of the validity of the renormalization group analysis from Ref.~\cite{kosmrlj_statistical_2017}. However one could expect the scaling form of $p_R$ to stay valid in the post-buckling regime of interest to us. Under this assumption, and if $p_0=0$, using the scaling result $P_l \sim (pR/Y)^{2/3} l^{-8/3}$ from the large fluctuation regime with $p\rightarrow p_R$ we find
\begin{equation}
    P_l \sim \left(\frac{kT_\text{eff}}{\kappa}\right)^{2/3} l^{-8/3},\label{eq:Plsi_R}
\end{equation}
independently of the stretching modulus. However, we remark that when the mean radius of the shell $\langle R \rangle = R - (4\pi)^{-1}\int \mathrm{d}\Omega \, f$ deviates from the undeformed radius $R$ (especially when we let $R_c > R$) this scaling has to be corrected as follows. 

In the asymptotic regime where we expect the $l^{-8/3}$ scaling to hold, the balance of the nonlinear contribution to surface elasticity with the pressure in Eqs.~(\ref{eq:scaling_scheme})~and~(\ref{eq:scaling_scheme_l}) in the main text still leads to
\begin{equation}
    \frac{Y}{R^4} l^4 f_l^3 \sim p,\label{eq:NonLineBalsi}
\end{equation}
where the spatial length scale $L=R/l$ is related to the undeformed sphere radius $R$ in our simulations.
Letting $\langle R \rangle \neq R$, our definition of $P_l \sim (f_l/\langle R \rangle)^2$ in Eq.~(\ref{eq:roughness}) corresponds to a normalization with respect to the \emph{deformed} average shell radius, and we have with Eq.~(\ref{eq:NonLineBalsi}) that
\begin{equation}
 P_l \sim \left( \frac{p R^4}{Y \langle R \rangle^3}\right)^{2/3}l^{-8/3}. \label{eq:Plsi_meanR}
\end{equation}
We then observe that the result of the scaling of the pressure in Eq.~(\ref{eq:pressure_scaling}) does not depend on the mean radius $\langle R\rangle$ of the shell, as the outcome of an analysis where the radius of the shell $R$ only plays a role as a material parameter. With this observation, we find from Eq.~(\ref{eq:Plsi_meanR})
\begin{equation}
    P_l \sim \left(\frac{kT_\text{eff}}{\kappa}\right)^{2/3} \left(\frac{R}{\langle R \rangle }\right)^2 l^{-8/3}, \label{eq:Plsi_meanR_b}
\end{equation}
which is equivalent to Eq.~(\ref{eq:Plsi_R}) for $R/\langle R \rangle\approx 1$.

A similar argument stands for comparisons to theoretical linear regime predictions Eq.~(\ref{eq:Plfin}), where a rescaling of~$P_l$ by $(\kappa/kT_\text{eff}) (\langle R \rangle / R)^2$ should lead to a curve collapse even when $R_c > R$ and the mean radius deviates significantly from~$R$. However, we find that for the simulations presented in Fig.~\ref{fig:fig2}a,b (main text) and Supp. Fig.~\ref{fig:sfig_sims}d, a rescaling by $(\kappa/kT_\text{eff}) (\langle R \rangle / R)^2$ does not lead to a curve collapse, confirming the departure from linear response for those simulation parameters (Supp. Fig.~\ref{fig:sfig_sims}e). Instead, rescaling $P_l$ by $(\kappa/kT)^{2/3} (\langle R \rangle / R)^2$, as suggested by Eq.~(\ref{eq:Plsi_meanR_b}), does lead to a collapse of simulation results onto a master curve for corresponding angular numbers $l$ (Supp. Fig.~\ref{fig:sfig_sims}f), further highlighting the importance of contributions from nonlinear surface elasticity for explaining our observations in simulations and experiments.

We note that these scaling results are very similar to RG scaling predictions: in a regime where the shell response is dominated by the renormalized bending rigidity, then
\begin{equation}
    P_l \sim  \left(\frac{kT_\text{eff}}{\kappa}\right)^{1-\eta/2} \gamma^{-\eta/2} \left(\frac{R}{\langle R \rangle }\right)^2 l^{-4+\eta} \approx  \left(\frac{kT_\text{eff}}{\kappa}\right)^{0.6} \gamma^{-0.4} \left(\frac{R}{\langle R \rangle }\right)^2 l^{-3.2} , \label{eq:Plsi_meanR_RG}
\end{equation}
with $\eta \approx 0.8$. The corresponding curve collapse is shown in the inset of Fig.~\ref{fig:sfig_sims}f. This result serves as confirmation that both approaches, either using perturbative corrections to the linear response or using dimensional arguments to obtain a scaling form for the nonlinear response, lead to reasonable predictions \cite{goldenfeld_intermediate_1989}.

\subsection{Effects of non-zero uniform pressure}
\label{sec:si_pressure}

Our  results above assumed  that  the  system is in an environment where $p_0=0$. Here, we revisit this assumption on  the basis of RG predictions and simulations, and provide justification for an experimentally-relevant regime in which the effective surface tension of the  shell is $0$.
Integrating  out the  mean radial  dynamics in Eq.~(\ref{eq:si_Feff_fpf0}) leads to an effective  surface tension in Eq.~(\ref{eq:si_Feff_fprime}) with value  $\sigma_{\text{eff}} = - p_0R$. The physical effects due to this tension strongly depend on  the sign of  $p_0$:

When the  pressure is oriented outwards ($p_0 <0$) it tends to stabilize the shell, akin to an inflated rubber balloon. Consequently, fluctuations are suppressed, leading to the power spectrum scaling as $l^{-2}$ due to the gradient-squared term in the free energy Eq.~(\ref{eq:si_effectiveF}).
However, when the pressure forces are oriented inwards ($p_0 > 0$) they tend to buckle the shell, with the shell eventually becoming unstable. Its stabilization can then only be realized through higher-order geometric nonlinearities. The resulting states are characterized by `ruffled surfaces' with excess area available to compensate for surface  tension.

\begin{figure}
    \centering
    \includegraphics{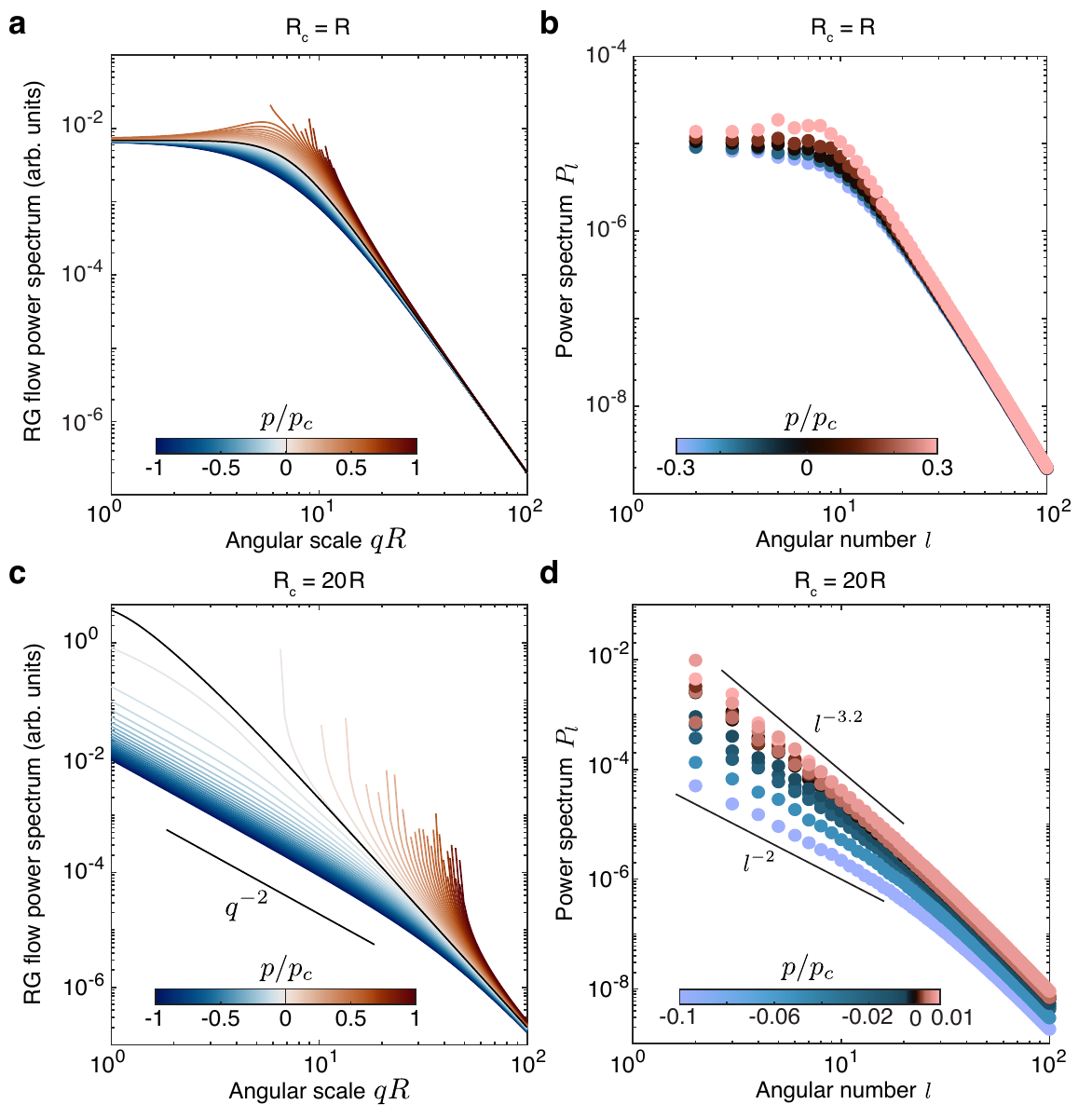}
    \caption{\textbf{The effects of non-zero pressure.} \textbf{a,} Effective  power  spectrum obtained by integrating the renormalization group (RG) flow near the linear regime for  varying pressure.  High-enough positive pressures lead to a divergent RG flow, indicated by  interrupted lines. In  these cases, RG predictions are unreliable. \textbf{b,} Simulation data of a shell with $\gamma = 3\cdot10^4$, $\kappa = 5 kT_{\text{eff}}$ and $R_c = R$ for varying pressure. In this smaller-deformation regime,  data qualitatively agrees with RG. \textbf{c,} RG flow for varying pressure for $R_c  =20 R$. Negative  pressures lead  to a $q^{-2}$ scaling  response  at long-wavelengths, while even small positive  pressures lead to divergent responses within 1-loop perturbative RG. \textbf{d,}  Simulation data of a shell with $\gamma = 3\cdot10^4$, $\kappa = 5 kT_{\text{eff}}$ and $R_c = 20 R$ for varying pressure. Negative pressures, which  stabilize the  shell, lead  to a  $l^{-2}$ response in  accordance with RG. Positive pressures are stabilized by nonlinear effects.}
    \label{fig:sfig_pressure}
\end{figure}

\begin{figure}
    \centering
    \includegraphics{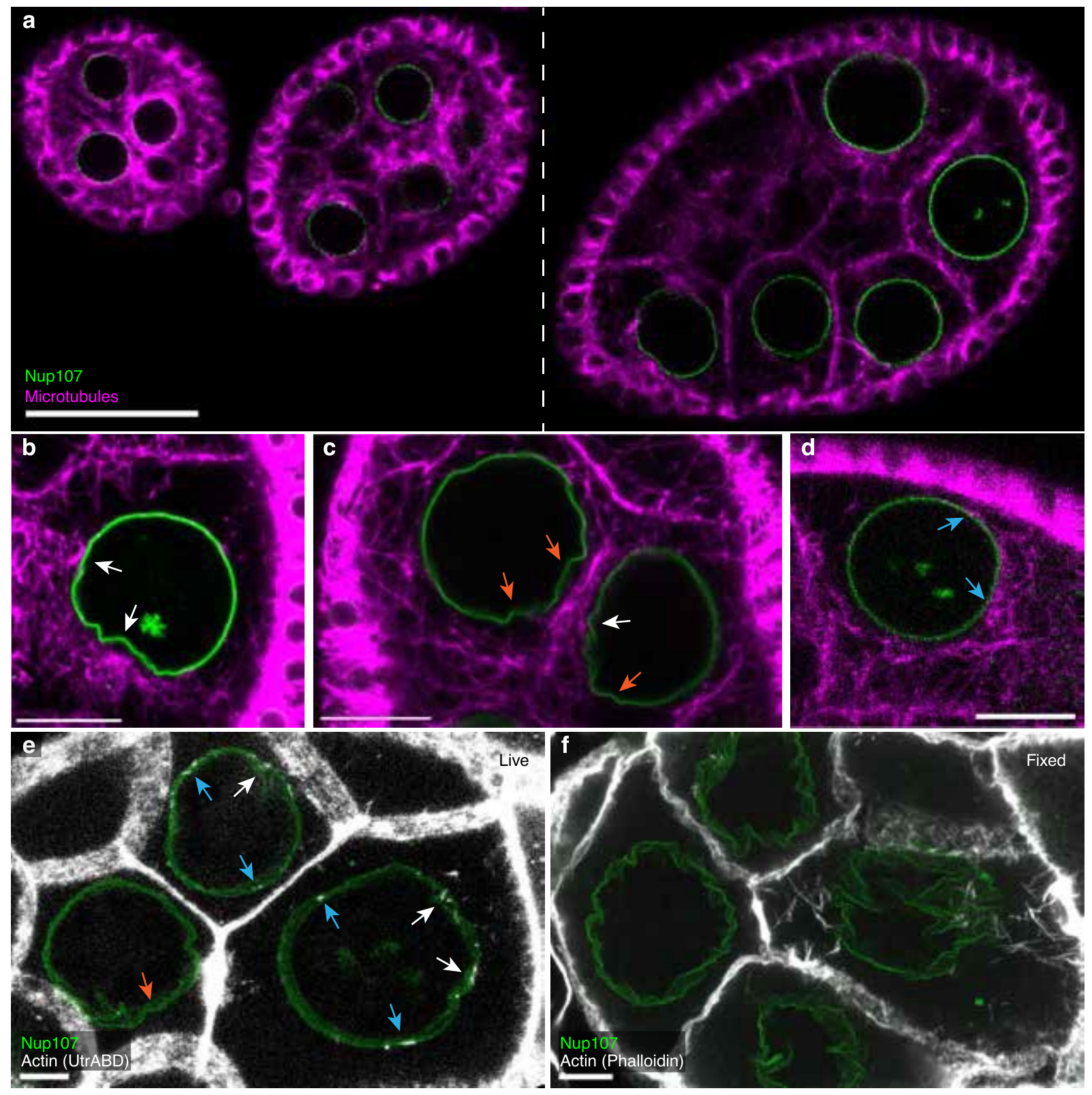}
    \caption{\textbf{Stable microtubule contact or impingement by actin are not sufficient to explain the observed patterns of NE wrinkling.} \textbf{a,} Single optical section through three egg chambers of time proxies 57, 76, and 122 from left to right, showing NEs in green and microtubules in magenta. Microtubules are present around the nuclei in all three egg chambers of various ages prior to wrinkle appearance, suggesting microtubule appearance at a particular time does not cause wrinkling. Rightmost egg chamber (separated by dashed line) is from a different ovariole than the first two. \textbf{b,} Single section of a nucleus (time proxy 149) showing microtubules correlated with the location of wrinkling or deformation. \textbf{c,} Section through two nuclei (time proxy 169) that show wrinkles without obvious colocalization of microtubules. \textbf{d,} Section through another nucleus from the egg chamber in \textbf{c}, showing no wrinkles despite surrounding microtubules. \textbf{e,} Projection through $\sim$7.5\,$\mu$m of an egg chamber (time proxy 184) expressing Nup107::RFP and a GFP-tagged actin-binding domain (ABD) of Utrophin. \textbf{f,} Projection through $\sim$3.5\,$\mu$m of a fixed egg chamber showing Nup107 in green and phalloidin-stained F-actin in white. F-actin is visible in some but not all cells and does not appear to correlate strongly with NE deformation. For all panels, arrows point to regions with cytoskeletal signal adjacent to wrinkles (white), wrinkles not near cytoskeletal signal (orange), or signal not near wrinkles (blue). Scale bars: 50\,$\mu$m (\textbf{a}), 20\,$\mu$m (\textbf{b-d}), 10\,$\mu$m (\textbf{e,f}).}
    \label{fig:sfig_cytoskeleton}
\end{figure}

\begin{figure}
    \centering
    \includegraphics[scale=0.95]{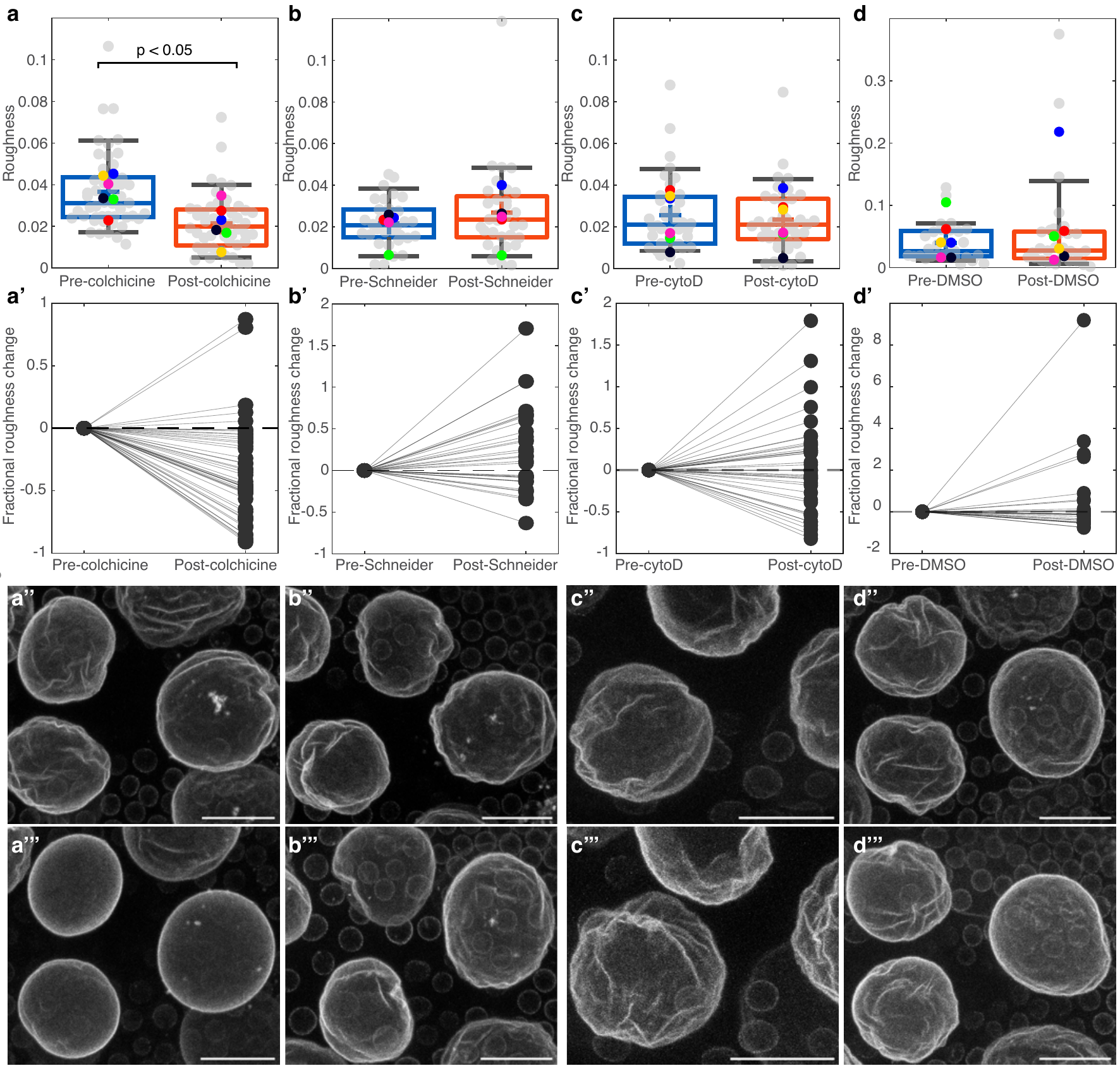}
    \caption{\textbf{Disruption of microtubules but not actin decreases NE wrinkling.} \textbf{a-d,} Box plots of roughness before (left) and after (right) addition of (\textbf{a}) colchicine dissolved in Schneider's medium, (\textbf{b}) Schneider's medium alone (control), (\textbf{c}) cytochalasin-D dissolved in DMSO, or (\textbf{d}) DMSO alone (control). Gray points represent individual nuclei, while colored points represent the mean for each egg chamber. Plus sign denotes the mean for all nuclei combined, middle line is the median, top and bottom edges of the box are the upper and lower quartiles, and whiskers span from 9\% to 91\% of the data range. \textbf{a’-d’,} Plots of fractional roughness change ($R_{v,f}-R_{v,i})/R_{v,i}$, where $R_{v,i}$ and $R_{v,f}$ are the initial and final effective radii, calculated from the measured volume) upon addition of drug or control medium. \textbf{a’’-d’’}, Representative images of egg chambers (time proxies 173, 171, 182, and 164, from \textbf{a’’} to \textbf{d’’}) before addition of drug or control medium. \textbf{a’’’-d’’’,} Same nuclei, but after $\sim$30 minutes incubation in the drug or control medium. Scale bars: 20\,$\mu$m. All comparisons: Student’s two-tailed t-test with unequal variance, p-values for \textbf{a}-\textbf{d}: 0.015, 0.57, 0.81, and 0.61, respectively. Sample sizes: 49 nuclei from 6 egg chambers (\textbf{a}), 32 nuclei from 5 egg chambers (\textbf{b}), 40 nuclei from 6 egg chambers (\textbf{c}), and 29 nuclei from 6 egg chambers (\textbf{d}). For t-tests, averages over all nuclei per egg chamber were compared.}
    \label{fig:sfig_drugaddition}
\end{figure}

\begin{figure}
    \centering
\includegraphics[scale=0.95]{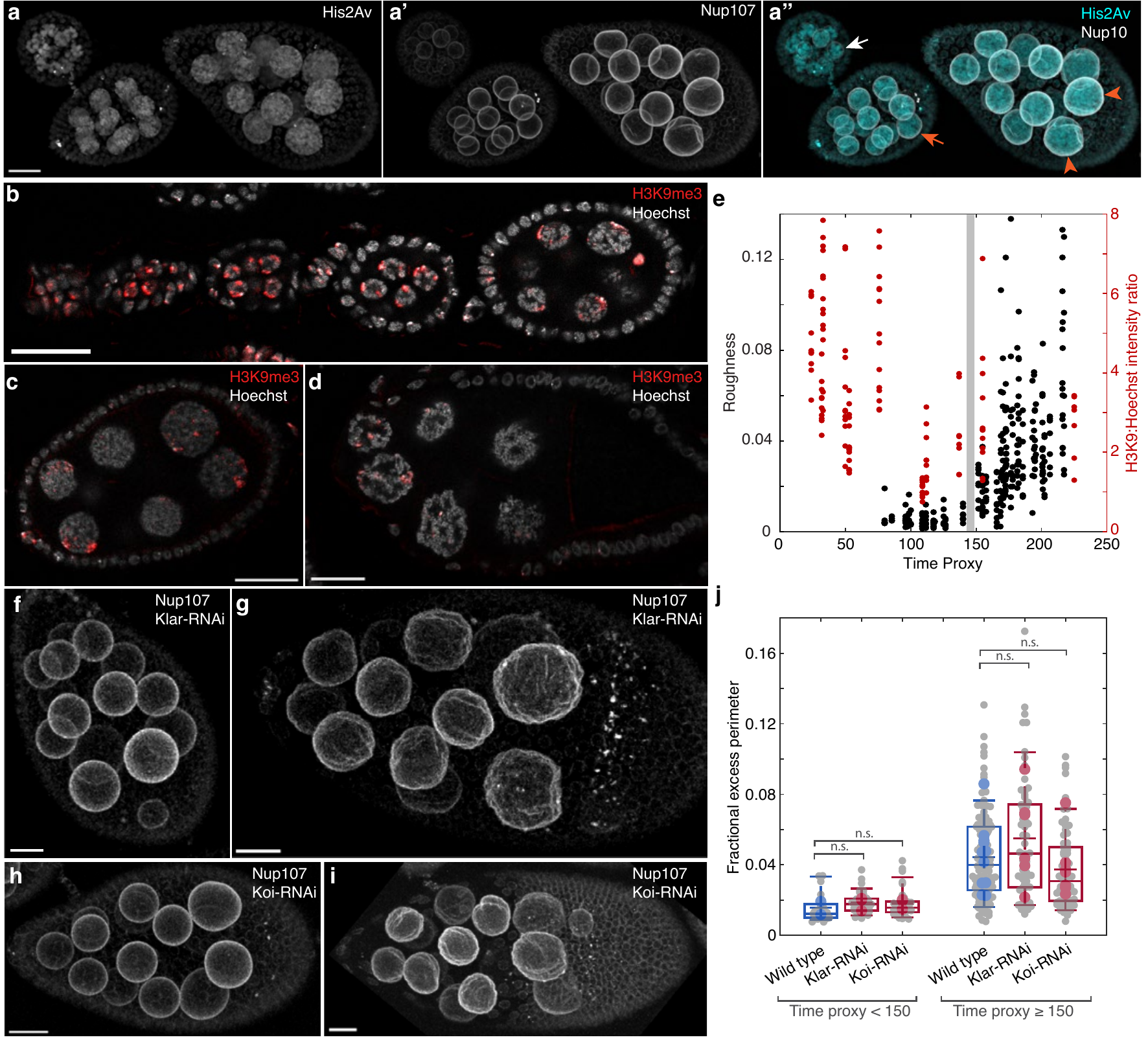}
    \caption{\textbf{Chromatin reorganization and the LINC complex do not explain appearance of wrinkles.} \textbf{a-a'',} Histone, nuclear membrane, and merged images of three egg chambers (time proxies 49, 77, and 122, from left to right). Chromatin reorganizes in egg chambers from a lobed structure (white arrow in merge) to a more diffuse structure (orange arrows) several hours prior to wrinkles first becoming visible (orange arrowheads). \textbf{b-d,} Immunofluorescence images showing trimethylated H3K9 (a marker for heterochromatin) and DNA across stages. \textbf{e,} Ratio of trimethylated H3K9 to Hoechst signal (arbitrary units, AU) in midplace slices of 132 nuclei from 11 egg chambers, plotted alongside roughness values for all 302 wild-type nuclei from 44 egg chambers used elsewhere in this study. H3K9 intensity decreases earlier in development than wrinkling becomes visible (gray line). \textbf{f,g,} Two Klar-knockdown egg chambers (time proxies 107 in \textbf{f} and 158 in \textbf{g'}). \textbf{h,i,} Two Koi-knockdown egg chambers (time proxies 120 in \textbf{h} and 175 in \textbf{i'}). \textbf{j,} Plot of fractional excess perimeter (an analog of roughness, defined in SI text) for wild-type, Klarsicht-RNAi, and Klaroid-RNAi egg chambers, split into younger chambers (time proxy $<$ 150) and older chambers (time proxy $\geq$ 150). Small gray points denote individual nuclei and larger colored points averages per egg chamber. Two-tailed Welch's t-tests, using average value per egg chamber as the samples, between wild type and each RNAi condition per age group resulted in p-values of at least 0.36. Sample sizes, from left to right: 25 nuclei / 2 chambers, 36 nuclei / 3 chambers, 34 nuclei / 3 chambers, 117 nuclei / 11 chambers, 62 nuclei / 6 chambers, 69 nuclei / 7 chambers. n.s. = not significant. All scale bars: 20\,$\mu$m.}
    \label{fig:si_chromatinLINC}
\end{figure}

To make those physical ideas precise, we integrate the RG flow equations Eqs.~(\ref{eq:si_rgflowAll}) and run numerical Langevin simulation for a range of different pressures (Fig.~\ref{fig:sfig_pressure}). In a low-deformation regime ($R_c = R$), simulation results agree with RG predictions (Fig.~\ref{fig:sfig_pressure}a,b):  Negative pressure stabilizes the shell, while positive pressure leads to a resonant wavelength. Note, however, that larger positive pressures lead to divergent responses in the RG prediction: To 1-loop perturbative order, the RG flow equations diverge as nonlinear contributions to the free energy dominate the linear effects \cite{kosmrlj_statistical_2017}. 
In a high-deformation regime ($R_c =20R$), large negative  pressure indeed leads to the expected  $l^{-2}$, which corresponds to a tension-dominated regime. However, positive pressure always leads to divergences; the corresponding  simulations show a  qualitatively similar response between $p_0=0$ and $p_0>0$ curves, with a scaling exponent that corresponds to the effective bending response (Fig.~\ref{fig:sfig_pressure}c,d). It is interesting to note that even the  $p_0=0$ response diverges at long-wavelength (not shown), which in the simulated system is limited by the infrared cutoff $1/R$.

Up to the limit where very large inward pressure forces lead to self-contact of shells, our results suggest that ruffled surfaces with wrinkles available to provide excess area have responses that are rather insensitive to pressure variation. This finding complements previous work that has experimentally found that the mean radius and volume of nuclear envelope in yeast cells are well-described by membranes with vanishing surface tension, which is explained by the presence of excess area buffering tension \cite{lemiere_control_2022,deviri_balance_2022}. This suggests that the dominant effect of osmotic pressure changes is an increase or decrease in the amount of excess area, while surface tension remains negligible. Therefore, we expect the $P_l \sim l^{-8/3}$ scaling result or its RG equivalent, which were obtained neglecting surface tension, still hold in our system even when osmotic pressure differences $p_0$ are non-zero.

To make the connection between $p_0$ introduced in Eq.~(\ref{eq:si_Feff_fprime}) with the pressure $p_\text{eff}$ appearing in Eq.~\eqref{eq:scaling} (main text) explicit, consider the scaling law obeyed by the deformed nucleus in Eq.~\eqref{eq:Plsi_meanR},
\begin{equation}
    P_l = \left(\frac{p'R^4}{Y\langle R\rangle^3}\right)^{2/3}l^{-8/3} \equiv \left(\frac{p_\text{eff}R}{Y}\right)^{2/3}l^{-8/3},
\end{equation}
where the effective pressure is given by 
\begin{equation}\label{eq:peff}
p_\text{eff} = p'\left(\frac{R}{\langle R\rangle}\right)^3,    
\end{equation}
with $\langle p_\text{eff}\rangle = 0$. This equation holds in our system, assuming that the crumpled surface leads to a vanishing surface tension. The effective pressure $p_\text{eff}$ captures the effect of excess area in the power spectrum $P_l$, which in turn changes the mean radius $\langle R \rangle$. To connect the osmotic pressure difference~$p_0$ to the crumpled shell's mean radius $\langle R\rangle$, we rewrite Eq.~\eqref{eq:f0_p0} in terms of the power spectrum $P_l$ as
\begin{equation}\label{eq:f0bar}
    f_0 = \frac{p_0 R_c}{4(\lambda+\mu)}+\frac{R_c}{4}\sum_l P_l (2l+1)l(l+1),
\end{equation}
where $\langle R\rangle=R-\langle f_0 \rangle$ by construction of the Fourier transform in Sec.~\ref{sec:SI_EffFE} (note that $f_0$ is distinct frorm the spherical harmonic $f_{00}$). Note that as $P_l\sim l^{-4}$ for  large $l$,  the sum is logarithmically divergent. However, for a stabilizing outward pressure  $p_0<0$, a 1-loop perturbative RG expansion reveals a well-defined nonlinear relationship between $\langle R\rangle$ and $p_0$~\cite{kosmrlj_statistical_2017}. For $p_0>0$, we can use a similar approach to the one in Ref.~\cite{milner_dynamical_1987} to obtain an expression for the excess area by introducing a cut-off angular number $l_\xi < \gamma^{1/2}=R/h$. The resulting,  cut-off-dependent estimate of $\langle R\rangle$ then follows from Eqs.~\eqref{eq:Plsi_meanR}  and \eqref{eq:f0bar} as the solution of
\begin{equation}
    \langle R \rangle^3 + \left(\frac{p_0 R_c}{4(\lambda +\mu)} -R\right)\langle R\rangle^2+ \frac{R_c}{4}\left(\frac{p' R^4}{Y} \right)^{2/3}\xi = 0,\label{eq:meanR_p_sum}
\end{equation}
where $\xi = \sum_{l=1}^{l_\xi} (2l+1)l(l+1)l^{-8/3}$. Eq.~\eqref{eq:meanR_p_sum} then leads to a nonlinear relationship between $p_\text{eff}$ and $p_0$ in the presence of inhomogeneous loads $p'$. Our previous scaling results can thus be applied when $p_0 \neq 0 $ by replacing $p'$ with the effective load $p_\text{eff}$ given in Eq.~(\ref{eq:peff}).

\subsection{Numerical simulations}
\label{sec:si_simulations}

In this section, we detail our simulation approach for results shown in Fig.~\ref{fig:fig2}a,b (main text) and in Supp. Fig.~\ref{fig:sfig_sims}.

We directly simulate the stochastic PDE derived from the elastic energy $F$ in Eq.~(\ref{eq:si_free_energy}) using a spherical harmonic variant of Fourier space Brownian dynamics~\cite{lin_brownian_2004}. In the absence of more detailed noise data, we assume that the fluctuations follow the fluctuation-dissipation theorem, with Gaussian fluctuations at an effective temperature $T_\text{eff}$. With this assumption, our results for the steady-state power spectra do not depend on the choice of damping function; we thus authorize ourselves a simplified description of the fluid environment of the nucleus. For simplicity, we assume that both the inner and outer surroundings of the nuclear envelope can be described as a viscous low-Reynolds number fluid~\cite{imran_alsous_dynamics_2021}. To keep the computational cost feasible, we approximate the corresponding non-local hydrodynamic coupling of the shell with the surrounding. To this end, we neglect radial shape variations and follow Refs.~\cite{milner_dynamical_1987,turlier_equilibrium_2016} by considering the first order effect of viscous damping experienced by a spherical surface that pushes via a deformation into the surrounding fluid. In harmonic mode space, such a damping is described by~\cite{lamb1932hydrodynamics} \hbox{$\tilde{\Lambda}_{lm}^{-1} = \eta (2l+1)(2l^2+2l-1 + 2 \delta_{l,0})/[l(l+1)R] \sim \eta l / R$}. For simplicity, we additionally assume that the damping of tangential modes is the same as for the normal modes. We then find Langevin equations for the displacement fields $f(\mathbf{r},t)$ and $\mathbf{u}(\mathbf{r},t)$ in terms of the harmonic modes $f_{lm}$ and $\mathbf{u}_{lm}=[u^{(1)}_{lm}, u^{(2)}_{lm}]^\top$~as
\begin{subequations}
\begin{align}
    \partial_t f_{lm}(t) &= \tilde{\Lambda}_{lm} \left( - \left[\frac{\delta F}{\delta f}\right]_{lm}\hspace{-0.3cm}(t) + \zeta_{f,lm}(t) \right)\label{eq:Lang1}\\
    \partial_t \mathbf{u}_{lm} &=  \tilde{\Lambda}_{lm}  \left( - \left[\frac{\delta F}{\delta\mathbf{u}}\right]_{lm}\hspace{-0.3cm}(t) + \boldsymbol{\zeta}_{u,lm}(t) \right).\label{eq:Lang2}
\end{align}
\label{eq:oseen_sh}
\end{subequations}
The Gaussian white noise components $\zeta_{f,lm}(t)$ and $\boldsymbol{\zeta}_{u,lm}(t)$ have zero mean and satisfy \smash{$\langle \zeta_{f,lm}(t) \zeta_{f,l'm'}(t') \rangle = 2kT_{\text{eff}} R^{-2} \tilde{\Lambda}_{lm}^{-1} \delta_{mm'} \delta_{ll'} \delta(t-t')$} and \smash{$\langle \boldsymbol{\zeta}^\alpha_{u,lm}(t) \boldsymbol{\zeta}^\beta_{u,l'm'}(t') \rangle = 2kT_{\text{eff}} R^{-2} \tilde{\Lambda}_{lm}^{-1} \delta_{mm'} \delta_{ll'} \delta(t-t') \delta_{\alpha\beta}$}, where $\alpha = 1, 2$ labels the two tangential components. As noted above, this noise satisfies the fluctuation-dissipation theorem. The functional derivatives of the free energy Eq.~(\ref{eq:si_free_energy}) used in the Langevin Eqs.~(\ref{eq:Lang1}), (\ref{eq:Lang2}) are given in real space by
\begin{subequations}
\begin{align}
\frac{\delta F}{\delta f} &= \kappa \nabla^2\nabla^2 f -p + 4\frac{\lambda+\mu}{R_c^2}f - \frac{2}{R_c}(\lambda + \mu) \nabla \cdot \mathbf{u} +\frac{1}{R_c}(\lambda+\mu)\left[(\nabla f)^2 + 2 f \nabla^2 f \right] \nonumber \\
&   - \mu \nabla \cdot ([ \nabla \mathbf{u} + \nabla \mathbf{u}^\top ]\cdot \nabla f ) - \left(\frac{\lambda}{2}+\mu\right)\nabla \cdot [ (\nabla f)^2 \nabla f] - \lambda \nabla \cdot [ (\nabla \cdot \mathbf{u})\nabla f]\\
\frac{\delta F}{\delta \mathbf{u}} &=  - (\lambda + \mu) \nabla(\nabla \cdot \mathbf{u}) - \mu \nabla^2 \mathbf{u} + (\lambda + \mu) \frac{2}{R_c}\nabla f - \frac{\lambda}{2} \nabla [ (\nabla f)^2 ]- \mu \left[ \nabla^2 f \nabla f + \nabla \nabla f \cdot \nabla f\right].
\end{align}
\end{subequations}

We assume for all simulations that the Poisson modulus is constant with an intermediate value of $\nu = 0.3$. The 2D Lam\'e parameters are then related to the 2D Young modulus $Y$ by \cite{thorpe_new_1992}
\begin{align}
    \lambda = \frac{\nu Y}{1-\nu^2} & & \mu = \frac{Y}{2(1+\nu)}.
\end{align}
All simulations are performed using $\eta = 4$\,Pa$\cdot$s. For simulations with fixed FvK number $\gamma$, we set $R$, $R_c$ and $kT_\text{\text{eff}} \geq kT_\text{eq}$, and vary $\kappa$ and $Y = \gamma \kappa / R^2$. To stay within reasonable physical regimes ($\kappa\sim 25 kT_\text{eq}\sim 10^{-19}$\,J for lipid bilayers), and $\gamma \sim 10^4-10^6$ using $T_\text{\text{eff}} = 10 T_\text{eq}$, we vary $\kappa/kT_\text{eff}$ in the range $2-20$. In the absence of  structural information on the preferred curvature of the nuclear envelope, and motivated by the experimental observation that the qualitative response of the shell does not  depend on  the size of the nuclei, we choose $R_c = 20 R$, such that the crossover lengthscale $L_{\text{el}}=R \gamma^{-1/4} \sqrt{R_c/R}$ where the linear response function `falls off' into the bending-dominated regime is approximately equal to $R/2$ for $\gamma=10^4$. This choice both suppresses the appearance of a plateau region in $P_l$ (Sec.~{\ref{sec:si_linresponse}}, Supp.~Fig.~\ref{fig:sfig_sims}), and lowers the energy barrier to larger deformations which, while allowed by the non-convex elastic free energy in finite deformation regimes, are not allowed in our free energy based on a shallow-shell assumption.

We use the pseudo-spectral solver Dedalus 3 \cite{burns_dedalus_2020} to solve the Langevin equation on the surface of the sphere. The equation is spatially discretized with spherical harmonics up to degree $L_\mathrm{max} = N-1$, resolving scales down to $\delta = \pi R / N$ on the surface of the sphere. The nonlinear terms are computed on the Gaussian quadrature grid using a dealiasing factor of 2. The system is temporally integrated using the Euler-Mayurama stochastic integrator. We choose a resolution of $N=256$ to ensure we sufficiently resolve the elastic scale $L_{\text{el}} = R \gamma^{-1/4}$ for FvK number $\gamma =10^6$. We rescale all lengths by the radius $R$ and time by $\tau = 4\eta R/Y$, which is the characteristic timescale of the dynamics induced by the cubic terms in the PDEs. The simulations are performed for a duration of $3000\tau$ with time steps of $\Delta t = 10^{-2} \tau$, where each simulation ran on 32 Intel Xeon Platinum 8260 cores on the MIT Supercloud cluster~\cite{Supercloud}, totalling approximately 360 hours of CPU time per simulation for the parameter values considered.

\textit{A posteriori} checks for convergence are performed by monitoring the dynamics of the mean shell radius. We also verify that the shallow-shell approximation remains valid throughout simulations by monitoring the average norm of the in-plane displacements $\langle |\mathbf{u}| \rangle$. Specifically, we check that in-plane displacements remain small relative to the shell radius and relative to radial deformations, i.e. $\langle |\mathbf{u}| \rangle < 10^{-2} R < \langle |f| \rangle$ for results in Fig.~\ref{fig:fig2}a,b. We note that the mean radius variation can reach large fractions of the radius for very soft nuclei, limiting the validity of the free energy in Eq.~(\ref{eq:si_free_energy}). 

\subsection{Limitations of fluid membrane models}
\label{sec:si_fluid}

In this section, we discuss the linear response behaviour of fluctuating fluid membranes. Specifically, we point out several discrepancies between our experiments, estimates of material parameters from the literature and theory, which indicates the need for the more complex shell models used throughout this work.

Biological membranes are often modeled as fluid membranes with properties described by the Helfrich-Canham Hamiltonian \cite{zhong-can_instability_1987}
\begin{equation}
    H =\int \mathrm{d}\mathbf{r}^2\,\frac{\kappa}{2}(\nabla^2 f)^2 + \sigma A,\label{eq:Hamil}
\end{equation}
with bending modulus $\kappa$, effective surface tension $\sigma$~\cite{gueguen_fluctuation_2017} and total surface area $A$. Because under some small-deformations assumptions the Hamiltonian in Eq.~(\ref{eq:Hamil}) is quadratic in normal displacements $f$, the equipartition theorem can be readily applied to determine the linear response power-spectrum of radial out-of-plane deformations. For a fluctuating equilibrium steady state about a spherical surface, this well-known power spectrum is given by~\cite{milner_dynamical_1987,gueguen_fluctuation_2017}
\begin{equation}
    P_l = \frac{kT}{(l-1)(l+2)[\kappa l(l+1) + \sigma R^2]}.
\end{equation}
The behavior of $P_l$ can be schematized as $P_l \sim l^{-2}$ for $l  \ll l_c$, where $l_c = \sqrt{\sigma R^2/\kappa}$ is the critical wave number, and $P_l \sim l^{-4}$ for $l \gg l_c$.  Using experimental measurements of mechanical nuclear membranes properties ($\kappa \approx 10^{-18}$\,J and $\sigma \approx 0.1-100$\,mN/m, \cite{guilak_viscoelastic_2000,funkhouser_mechanical_2013,kim_volume_2015}), we can estimate for a nucleus of radius $R=5$\,$\mu$m a \textit{lower} bound $l_c^*\approx50$ for the critical wave number. Under the reasonable assumption that the experimentally observed reduction of lamin concentration reduces the rigidity of the NE, the developmental progression of growing nuclei will most likely push~$l_c^*$ to even larger values. For the length scale and wave number regime $l\in [2, 25]$ to which we have experimental access, the fluid membrane model would therefore imply a dominant $l^{-2}$ scaling of the power spectrum, in stark contrast to our observations (Fig.~\ref{fig:fig2}c, Supp. Fig.~\ref{fig:sfig_layers}).

This contradiction, together with the experimental observation of sharp surface creases that suggest nonlinear stress focusing, lends further support for the nonlinear elastic surface model employed in this work.

\section{Supplementary video captions}

\paragraph*{\textbf{Supplementary Video 1}} Three z-stacks of Nup107 signal through individual nurse cell nuclei from egg chambers of time proxies 118, 157, and 216 from left to right, showing progression of NE shapes from unwrinkled (left) to heavily wrinkled (right) during development. Scale bars: 5\,$\mu$m; 10\,fps.

\paragraph*{\textbf{Supplementary Video 2}} Two time-lapses of maximum-intensity projections of two nurse cell nuclei each from egg chambers expressing Nup107::GFP, showing fluctuations in NE wrinkling over time. First movie: lower-resolution movie acquired at 1 frame per 40 seconds (time proxy 160). Second movie: higher-resolution movie acquired at 1 frame every 5 minutes (time proxy 200). Scale bars: 10\,$\mu$m. 10\,fps; second movie has had each frame duplicated 4 times for an effective 2\,fps.

\paragraph*{\textbf{Supplementary Video 3}} Z-stack through an egg chamber (time proxy 166) expressing Nup107::GFP (green) and stained with SPY555-tubulin to visualize microtubules (magenta). White box highlights region corresponding to the zoomed-in z-stack in the latter part of the movie. Microtubules are present around the nurse cell nuclei, but their location does not seem to correlate with the location of NE wrinkles. Orange arrows point to surface fluctuations without nearby tubulin signal, white arrow to surface fluctuations with nearby tubulin signal, and cyan arrow to tubulin signal near undeformed surface. Scale bars: 20\,$\mu$m; 15\,fps.

\paragraph*{\textbf{Supplementary Video 4}} Egg chamber (time proxy 172) expressing Nup107::GFP before and after addition of 9\,mg/mL colchicine to disrupt microtubules. After drug addition, nurse cell NE wrinkles decrease markedly in amplitude over the next 30 minutes. This effect is not simply due to addition of fresh medium (see Supplementary Video 5). White box in second and third frames indicates the rough border of the zoom-in shown for the rest of the movie. Scale bars: 20\,$\mu$m; 5 fps.

\paragraph*{\textbf{Supplementary Video 5}} Egg chamber (time proxy 163) expressing Nup107::GFP before and after addition of fresh, identical culture medium, used as a control for colchicine addition. After medium addition, the NE wrinkles show no major change. White box in second and third frames indicates the rough border of the zoom-in shown for the rest of the movie. Scale bars: 20\,$\mu$m; 5\,fps.

\paragraph*{\textbf{Supplementary Video 6}} Two timelapses showing single optical sections from nuclear envelopes (Nup107, green), cell membranes (magenta), and motion of cytoplasmic components (reflected light, gray) in egg chambers under normal conditions (no inhibitors added). First timelapse is from an egg chamber of time proxy 190; second is of time proxy 173. Scale bars: 20\,$\mu$m; 30\,fps.

\paragraph*{\textbf{Supplementary Video 7}} Three reflection microscopy timelapses showing effects of colchicine addition on intracellular motion. First timelapse: maximum intensity projection through 4\,$\mu$m near the top of a nucleus in one nurse cell; the NE appears as a large, bright object deforming and rotating alongside fluctuations of small objects in the cytoplasm (white dots). Alternating dark and light regions on the right side of the NE are fluctuating wrinkles. Second timelapse: projection through 8\,$\mu$m of the same egg chamber after addition of 9\,mg/mL colchicine. Boxed nucleus corresponds to the nucleus shown in first portion of the movie. Reduction in NE roughness can be seen by the reduction in linear patterns of dark and light regions. Reduction in cytoplasmic motion becomes more easily apparent around 10 minutes in. Third timelapse: projection through 6.5\,$\mu$m of a nucleus from a different egg chamber, $>$30 minutes after colchicine was added, to ensure slowing in the second timelapse did not result from photodamage. Here, too, the NE is smoother than expected given the egg chamber age, and small cytoplasmic objects are substantially less mobile. Scale bars: 10\,$\mu$m; 20\,fps.

\paragraph*{\textbf{Supplementary Video 8}} Egg chamber (time proxy 178) expressing Nup107::GFP before and after addition of 10\,$\mu$g/mL cytochalasin-D to disrupt F-actin. After drug addition, nurse cell NE wrinkles show no substantial change. Addition of DMSO alone also causes no major change (see Supplementary Video 9). White box in second and third frames indicates the rough border of the zoom-in shown for the rest of the movie. Scale bars: 20\,$\mu$m; 5\,fps.

\paragraph*{\textbf{Supplementary Video 9}} Egg chamber (time proxy 170) expressing Nup107::GFP before and after addition of DMSO, used as a control for cytochalasin-D addition. After DMSO addition, the NE wrinkles show no major change. White box in second and third frames indicates the rough border of the zoom-in shown for the rest of the movie. Scale bars: 20\,$\mu$m; 5\,fps.

\paragraph*{\textbf{Supplementary Video 10}} Three z-stacks through individual nurse cell nuclei (from Supplementary Video 1) from egg chambers of time proxy 118, 157, and 216 from left to right. The Nup107 signal is shown in gray following the background subtraction and Gaussian blur used to preprocess before segmentation; the segmentation output is shown in red. The rightmost nucleus is near the oldest of the nuclei segmented and hence effectively represents a lower bound on the accuracy achieved for segmentations used in reconstructions. Scale bars: 5\,$\mu$m; 10\,fps.

\ \\
\let\oldaddcontentsline\addcontentsline
\renewcommand{\addcontentsline}[3]{}
\putbib[bibliography]
\let\addcontentsline\oldaddcontentsline
\end{bibunit}


\end{document}